\documentclass[12pt]{article}
\usepackage{amsmath}
\usepackage{graphicx}
\usepackage{enumerate}
\usepackage{natbib}
\usepackage{url} 
\usepackage[ruled,vlined]{algorithm2e}

\usepackage{amsthm,amsmath,amsfonts,amssymb}
\usepackage[colorlinks,citecolor=black,urlcolor=black]{hyperref}
\usepackage{graphicx}
\usepackage[font=footnotesize]{caption}

\newcommand{\blind}{1}

\usepackage{mathtools}
\usepackage{amsmath}
\DeclareMathOperator*{\argmax}{arg\,max}

\newcommand{\REV}{\textcolor{black}}

\newcommand{\REVV}{\textcolor{black}}

\usepackage{hyperref}
\usepackage{graphicx}
\usepackage{subcaption}
\usepackage{bm}

\usepackage{gensymb}
\usepackage{multirow}

\SetKwInput{KwInput}{Input}
\SetKwInput{KwOutput}{Output}

\begin{document}

\def\spacingset#1{\renewcommand{\baselinestretch}%
{#1}\small\normalsize} \spacingset{1}


\if1\blind
{
  \title{\bf Space-time extremes of severe US thunderstorm environments}
  \author{Jonathan Koh\thanks{
    The authors gratefully acknowledge the Swiss National Science Foundation (project $200021\_178824$) for funding, and Olivia Martius and Daniela Domeisen for meteorological help.}\hspace{.2cm}\\
    and \\
    Erwan Koch\\
    and \\
    Anthony C.~Davison\\ \\
    Institute of Mathematics,\\
    Ecole Polytechnique F\'ed\'erale de Lausanne (EPFL),\\
    1015 Lausanne, Switzerland}
  \maketitle
} \fi

\if0\blind
{
  \bigskip
  \bigskip
  \bigskip
  \begin{center}
    \LARGE\bf Space-time extremes of severe US thunderstorm environments
\end{center}
  \medskip
} \fi

\bigskip
\begin{abstract}
Severe thunderstorms cause substantial economic and human losses in the United States. Simultaneous high values of convective available potential energy (CAPE) and storm relative helicity (SRH) are favorable to severe weather, and both they and the composite variable $\mathrm{PROD}=\sqrt{\mathrm{CAPE}} \times \mathrm{SRH}$ can be used as indicators of severe thunderstorm activity. Their extremal spatial dependence exhibits temporal non-stationarity due to seasonality and large-scale atmospheric signals such as El Ni\~no-Southern Oscillation (ENSO). In order to investigate this, we introduce a space-time model based on a max-stable, Brown--Resnick, field whose range depends on ENSO and on time through a tensor product spline.  We also propose a max-stability test based on empirical likelihood and the bootstrap. The marginal and dependence parameters must be estimated separately owing to the complexity of the model, and we develop a bootstrap-based model selection criterion that accounts for the marginal uncertainty when choosing the  dependence model. In the case study, the out-sample performance of our model is good.  We find that extremes of PROD, CAPE and SRH are generally more localized in summer and, in some regions, less localized during El Ni\~no and La Ni\~na events, and give meteorological interpretations of these phenomena.
\end{abstract}

\vfill

\newpage
\spacingset{1.7} 

\section{Introduction}

Severe thunderstorms cause a substantial fraction of the economic and human losses due to natural disasters in the United States (US), such as those due to the tornadoes experienced across six states on 10--11 December 2021, so it is imperative to have a good understanding of their origins. A severe US thunderstorm is defined as one that produces tornadoes, hailstones over one inch ($2.54$ cm) in diameter, or wind gusts in excess of 50~kts ($1$ knot, abbreviated to kt,  corresponds to approximately $0.51$~m~s$^{-1}$). Supercells, which are thunderstorms with a deep and persistent rotating updraft, are responsible for many severe thunderstorm reports \citep[79\% of tornadoes, for example, according to][]{trapp2005tornadoes}, even though only about 10\% of thunderstorms are supercells \citep{Doswell2015Meso}.  

The available thunderstorm data record is compromised by issues, such as observational bias, that complicate its use for modelling \citep{Verbout2006,Allen:Hail:2015,Edwards2018:wind}, so it is worthwhile to consider meteorological environments that are conducive to severe thunderstorms. Such storms, especially supercell storms, are more probable in the presence of elevated values of convective available potential energy (CAPE) and of certain measures of vertical wind shear \citep[e.g.,][]{Brooks2003, brooks2013severe} such as storm relative helicity (SRH), which have been used by weather forecasters and climatologists for more than two decades. High values of the combined variable $\mathrm{PROD}=\sqrt{\mathrm{CAPE}} \times \mathrm{SRH}$ are favorable to severe thunderstorms, and PROD has been used as a proxy of severe thunderstorm activity \citep[by, e.g.,][]{TippettCohen:ExtremeTornado,koch2020trendPROD}; see for example \citet[][Equation (1)]{Brooks2003} and \citet[Section~1]{koch2020trendPROD} for justification for this. In addition to the absence of observational bias, an advantage of using thunderstorm environments as a proxy for thunderstorm reports is that their reanalysis values are available at high and regular spatio-temporal resolution (typically 1$^\circ$ longitude and 1$^\circ$ latitude every hour or three hours), so techniques from extreme-value theory and geostatistics can be applied. Among papers that have studied the temporal evolution of the extremes of a quantity similar to PROD over part of the contiguous US are \cite{gilleland2013spatial}, who applied the conditional extreme-value model of \cite{heffernan2004conditional} to $\mathrm{WS} \times W_{\mathrm{max}}$, where WS is a measure of wind shear and $W_{\mathrm{max}} = \sqrt{2 \times \mathrm{CAPE}}$, \cite{mannshardt2013extremes}, who fitted the generalized extreme-value (GEV) distribution to the annual maxima of $\mathrm{WS} \times W_{\mathrm{max}}$, and \cite{heaton2011spatio}, who considered, for the same variable, a hierarchical extreme-value model based on a Poisson point process. Other studies have investigated the link between ENSO and seasonal or monthly means of environments during winter and spring in the US \citep{Allen:ENSO2014, Lepore:ENSO:2017}, or between ENSO and monthly maxima of environments such as PROD, CAPE and SRH \citep{koch2020trendPROD}. \cite{gilleland2013spatial} studied the time evolution of spatial patterns of rather extreme events, but did not include the effect of time in their model or consider pointwise maxima. The other papers mentioned above focus on the influence of time or ENSO at the marginal level (grid point by grid point) only, and, to the best of our knowledge, no article has yet considered the potential influence of these or other time-varying covariates on the spatial dependence of maxima. This issue is nevertheless prominent for risk analysis, and both data and physical arguments suggest that large-scale signals such as ENSO or seasonality may affect spatial  extremal dependence, for instance through the characteristic dimension of individual extreme events.  

In this paper we address this question by using a space-time model that consists, at each time point, of a max-stable field whose dependence structure can evolve as a function of large-scale atmospheric signals or seasonal effects. We use reanalysis data from the North American Regional Reanalysis (NARR), and focus on PROD, CAPE and SRH from 1979 to 2015 over a large rectangle of the contiguous US that contains Tornado Alley, its riskiest region. Max-stable fields \citep[e.g.,][]{haan1984spectral, de2007extreme, davison2012} provide a natural extension of multivariate extreme-value distributions to the infinite-dimensional setting. They are well-suited to model spatial extremes, as they arise as the only possible non-degenerate limiting random fields of appropriately rescaled pointwise maxima of independent replications of a field. The most commonly used parametric max-stable models are the \citet{Smith}, \citet{Schlather2002}, Brown--Resnick \citep{brown1977extreme, kabluchko2009stationary} and extremal-$t$ \citep{opitz13} models. Although many models for time-varying marginal (GEV) parameters have been proposed \citep[e.g.,][and references therein]{davison.et.al.2013}, to our knowledge, no model for time-varying dependence of spatial extremes has yet been developed. Thus, our model fills a gap in the literature while allowing us to address a significant practical problem. In the multivariate setting, this problem was tackled by \citet{mhalla.2017}, who introduce covariates into the \citet{pickands1991}  dependence function  of a max-stable random vector using a generalized additive model. The modelling of spatial non-stationarity (as opposed to temporal non-stationarity) in the dependence structure of max-stable fields has received slightly more attention. \cite{Smith2009} allowed the covariance matrix of the Smith model to vary across space, whereas \cite{Huser2016nonstat} extended the extremal-$t$ model by taking non-stationary correlation functions for the underlying Gaussian random field, and proposed a max-mixture of max-stable models with spatially dependent weights. Space-varying covariates such as longitude, latitude and elevation, can be incorporated into the correlation functions and the weights. 

In our space-time model, when the Brown–Resnick or extremal-t fields are used, we
propose to let the parameters (range, smoothness and possible anisotropy parameters) of
the variogram or correlation function of their underlying Gaussian field depend on covariates
through a general regression function that may involve splines or wavelets. In our application, the time step corresponds to one month and we use a fractional Brown--Resnick field whose range depends on ENSO and month through a tensor product spline. Such a spline basis captures the interactions between the covariates and allows for monthly variation in the ENSO effect. We fit our model using pairwise likelihood \citep[e.g.,][]{padoan2010likelihood} and show by simulation that its parameters can be estimated rather accurately. Furthermore, the out-sample performance of our model is good and our findings have broad meteorological explanations. 
\REV{The range parameter tends to be lower in summer for all regions and variables, although the situation is more subtle for CAPE. The ENSO effect mostly occurs in late winter and spring. The range for PROD is higher during both El Ni\~no \REVV{(EN)} and La Ni\~na \REVV{(LN)} phases in the North-East and during \REVV{the EN phase} in the South-East. The range is higher during both phases in \REVV{the East} for CAPE, and in the \REVV{North during EN} for SRH. The extremes of PROD, CAPE and SRH are thus more spatially extensive during both phases in some regions.}

We also contribute new methods for inference on max-stable fields. In environmental applications, data may exhibit asymptotic independence \REV{and/or weakening dependence}, in which case max-stable models are unsuitable, and several subasymptotic models have been proposed to alleviate this \citep[e.g.,][]{huser.wadsworth.18, huser.2021}. One should always assess the validity of max-stable models in applications, and \citet{gabda.2012} and \citet{Buhl.2016} proposed graphical diagnostics for data with standardized margins. Here we develop a max-stability test that accounts for the unknown margins encountered in practice by approximating the distribution of a specific test statistic under the null hypothesis of max-stability using the \REV{bootstrap.}
Owing to the complexity of our model and the large number of grid points considered, simultaneous estimation of the marginal (GEV) parameters and the dependence parameters of the max-stable field is computationally too intensive, so we first fit the marginal model by estimating the GEV parameters, and then transform the data to  standard Fréchet margins and fit the dependence parameters using pairwise likelihood. We demonstrate that, following a two-step procedure, the \REVV{estimated \citet{Huber:1964} sandwich covariance matrix, henceforth the sandwich matrix for brevity,} gives poor confidence intervals, and the non-parametric bootstrap gives better coverage. We also show that  in such a context model selection using the composite likelihood information criterion \citep{padoan2010likelihood} is sub-optimal and propose a better criterion using a bootstrap-based estimator of the non-normalized composite Kullback--Leibler divergence.

\REV{The test mentioned above casts no doubt on max-stability as a working hypothesis for our data. Nor does the $\chi$ function \citep{coles1999dependence}, and plots of conditional exceedance probabilities do not exhibit weakening dependence with increasing levels, though these diagnostics are rather variable. 
Thunderstorms tend to be most intense when they take the form of supercells or occur within organized systems such as multicellular thunderstorms, squall lines, or mesoscale convective complexes; see Chapters~8 and~9 of \cite{markowski2011mesoscale}, respectively, for more information. Although supercells, for instance, represent only about 10\% of thunderstorms \citep{Doswell2015Meso}, they lead to many severe thunderstorm reports \citep[79\% of tornadoes according to][]{trapp2005tornadoes}. All these systems are much more spatially extended than single-cell thunderstorms, indicating that severer thunderstorms are not necessarily more localized. 
Moreover, here we model not severe thunderstorms but environments conducive to them, and a weakening of dependence is less expected for such large-scale variables than for the hazards they cause. 
}

The rest of the paper is organized as follows. In Section~\ref{Sec_Preliminaries}, \REVV{we outline max-stable fields 
and present the data and some exploratory analyses.} Section~\ref{Sec_Methodology} details our main methodological contributions: the model, the max-stability test, and the bootstrap-based model selection criterion. Section~\ref{Sec_CaseStudy} is dedicated to the case study: we apply the model and methodologies developed in Section~\ref{Sec_Methodology} to the thunderstorm environment data. Section~\ref{Sec_Discussion} summarizes our main contributions and findings and gives some future perspectives. Throughout the paper, $\mathcal{X}$ denotes a subset of~$\mathbb{R}^2$, and $\overset{d}{=}$ and $\overset{d}{\rightarrow}$ denote equality and convergence in distribution, respectively; in the case of random fields, these should be understood as applying to all finite-dimensional distributions. 

\section{Preliminaries}
\label{Sec_Preliminaries}

\subsection{Max-stable random fields}
\label{Subsec_MaxStableFields}

A random field $\{ G(\bm{s}) : \bm{s} \in \mathcal{X} \}$ is said to be max-stable if there exist sequences of functions 
$\{ a_n(\bm{s}), \bm{s} \in \mathcal{X}\}_{n \geq 1}> 0$ and $\{ b_n(\bm{s}), \bm{s} \in \mathcal{X}\}_{n \geq 1} \in \mathbb{R}$
such that, for any $n \geq 1$,
$$
\left \{ \frac{ \max_{i=1}^{n} G_i(\bm{s}) -b_n(\bm{s} )}{a_n(\bm{s} )} : \bm{s} \in \mathcal{X}  \right \}  \overset{d}{=} \left \{ G(\bm{s}) : \bm{s} \in \mathcal{X} \right \},
$$
where $G_1,\dots,G_n$ are independent replicates of $G$. Let $ \tilde{T}_1, \dots, \tilde{T}_n$, be independent replications of a random field $\{ \tilde{T}(\bm{s}) : \bm{s} \in \mathcal{X} \}$. Let 
$\left\{ c_n(\bm{s}), \bm{s} \in \mathcal{X} \right\}_{n \geq 1}$ and $\left\{ d_n(\bm{s}), \bm{s} \in \mathcal{X} \right\}_{n \geq 1}$ be sequences of functions that respectively take values in the strictly positive and real numbers. If there exists a non-degenerate random field $\{ G(\bm{s}) : \bm{s} \in \mathcal{X} \}$ such that
\begin{equation}\label{eq:maxima_convergence}
 \left \{ \frac{\max_{i=1}^n  \tilde{T}_i(\bm{s}) -d_n(\bm{s})}{c_n(\bm{s})} : \bm{s} \in \mathcal{X} \right \} \overset{d}{\rightarrow}  \left \{ G(\bm{s}) : \bm{s} \in \mathcal{X} \right \}, \quad n \to \infty,   
\end{equation}
then $G$ must be max-stable \citep{haan1984spectral}, and this explains the relevance of max-stable fields as models for the pointwise maxima of random fields. If $\{ G(\bm{s}) : \bm{s} \in \mathcal{X} \}$ is a \REVV{non-degenerate} max-stable field, then, for any $\bm{s} \in \mathcal{X}$, $G(\bm{s})$ has a GEV distribution \citep[e.g.,][Section~3.1]{Coles:2001} with location, scale and shape parameters $\eta_{\bm{s}}$, $\tau_{\bm{s}}$ and $\xi_{\bm{s}}$. The transformed variable $Z(\bm s) = \left[1 + \xi_{\bm s} \{G(\bm s) - \eta_{\bm{s}}\}/{\tau_{\bm{s}}} \right]^{1/\xi_{\bm{s}}}$ is standard Fréchet distributed, i.e.,  $\mathbb{P}\{Z(\bm{s}) \leq z\} = \exp(-1/z)$ for $z>0$; max-stable fields having standard Fréchet margins are said to be simple. Max-stable fields are sometimes instead standardized to have Gumbel margins; the Gumbel distribution function with location parameter $\mu$ is $\exp[-\exp\{-(x - \mu)\}]$, $x \in \mathbb{R}$, and the standard Gumbel distribution appears when $\mu=0$.

Any simple max-stable field can be represented as \citep[][]{haan1984spectral}
\begin{equation}\label{eq2}
Z(\bm{s}) = \max_{i=1}^{\infty} R_i U_i(\bm{s}), \quad \bm{s} \in \mathcal{X},
\end{equation}
where the $(R_i)_{i \geq 1}$ are the points of a Poisson point process on $(0, \infty)$ with intensity function $r^{-2} \mathrm{d}r$ and the $(U_i)_{i \geq 1}$ are independent replicates of a non-negative random field $\{U(\bm{s}), \bm{s} \in \mathcal{X}\}$ such that $\text{E}\{U(\bm{s})\}=1$ for any $\bm{s} \in \mathcal{X}$. Any field defined by~\eqref{eq2} is simple max-stable, moreover, and this allows parametric max-stable fields to be constructed, such as the \citet{Smith},  \citet{Schlather2002}, Brown--Resnick \citep{brown1977extreme,kabluchko2009stationary}, and extremal-$t$ \citep{opitz13} models. The last two are flexible and have been found to capture extremes well.  In Section~\ref{Sec_CaseStudy} we use the Brown--Resnick model. 

\REV{Let $\{ \varepsilon(\bm{s}) : \bm{s} \in \mathcal{X} \}$ be a {centred} Gaussian random field with stationary increments and semivariogram $\gamma$. Using $U(\bm{s}) = \exp \left[ \varepsilon(\bm{s}) - \mathrm{Var}\{\varepsilon(\bm{s})\}/2 \right]$  in~\eqref{eq2}, where $\mathrm{Var}$ denotes variance, leads to the Brown--Resnick random field associated with  $\gamma$.}
A popular isotropic semivariogram is
$ \gamma(\bm{s}) = \left( \| \bm{s} \|/\rho \right)^{\alpha}$, $\bm{s} \in \mathcal{X}$, where $\rho > 0$ and $\alpha \in (0, 2]$ are the range and smoothness parameters, respectively, and $\|\cdot\|$ is the Euclidean distance. Theorem~3.1 of \citet{kabluchko2010ergodic} implies that an unbounded semivariogram such as this yields a field  that is mixing, which is appropriate if the extreme events are spatially localized. As pointwise maxima typically arise from several individual events, the spatial scale of \REV{dependence reflects the extent} of the individual extreme events \citep[e.g.,][]{Dombry2018}. Thus the range parameter \REV{influences} the characteristic size of individual extreme events, whereas the smoothness parameter controls the regularity of the field's sample paths. We will account for possible geometric anisotropy by using the semivariogram
\begin{equation}
\label{Eq_ExpressionAnisotropicVariogram}
\gamma(\bm{s}) = \left( \| A \bm{s} \|/\rho \right)^{\alpha}, \quad \bm{s} \in \mathcal{X},
\end{equation}
 where
\begin{equation}\label{eq_aniso}
A=
\begin{pmatrix}
\cos \kappa & -\sin \kappa  \\
r \sin \kappa & r \cos \kappa
\end{pmatrix},
\end{equation}
with scaling and rotation parameters respectively $r >0$ and $\kappa \in [0, \pi]$, which \cite{blanchet2011} used to introduce anisotropy into the Schlather model. 

The Schlather and extremal-$t$ models are also typically parametrized by a range parameter $\rho > 0$ and smoothness parameter $\alpha \in (0, 2]$ through the isotropic correlation function $C$ of their underlying standard Gaussian field; the powered exponential, Cauchy and Whittle--Matérn correlation functions are often used in applications. In anisotropic cases we consider $C(\| A \bm{s} \|)$ instead of $C(\|\bm{s} \|)$; then the correlations also depend on $r$ and~$\kappa$.

For any simple max-stable field, we have, for $\bm{s}_1, \ldots, \bm{s}_D \in \mathcal{X}$ and $z_1, \ldots, z_D >0$,
\begin{equation}
\label{Eq_MultivariateDfMaxStable}
    \mathbb{P}\{Z(\bm{s}_1) \leq z_1, \ldots, Z(\bm{s}_D) \leq z_D\}= \exp \left\{- V_{\bm{s}_1, \ldots, \bm{s}_D}(z_1, \ldots, z_D) \right\},
\end{equation}
with \citep[][]{pickands1991}
$$ 
V_{\bm{s}_1, \ldots, \bm{s}_D}(z_1, \ldots, z_D) = \int_{\mathcal{S}_D} \max \left \{ \frac{w_1}{z_1}, \ldots, \frac{w_D}{z_D} \right \} \mathrm{d}M_{\bm{s}_1, \ldots, \bm{s}_D}(w_1, \ldots, w_D),
$$
where $M_{\bm{s}_1, \ldots, \bm{s}_D}$ is a measure on the $D$-dimensional simplex $\mathcal{S}_D$ satisfying 
$$
\int w_d \,\mathrm{d}M_{\bm{s}_1, \ldots, \bm{s}_D}(w_1, \ldots, w_D) = 1
$$ 
for each $d \in \{1, \ldots, D \}$.
The function $V_{\bm{s}_1, \ldots, \bm{s}_D}$, called the exponent measure of the max-stable random vector $(Z(\bm{s}_1), \ldots, Z(\bm{s}_D))^{\prime}$, entirely characterizes its dependence and is homogeneous of order $-1$; the~ \REV{``$^{\prime}$''} denotes transposition of a vector. Dependence measures proposed for max-stable fields/vectors include the extremal coefficient \citep{schlather2003dependence}. If $Z$ is a simple max-stable field, then the bivariate distribution function satisfies 
\begin{equation}\label{eq:extcoeff}
    \mathbb{P}(Z(\bm{s}_1) \leq u, Z(\bm{s}_2) \leq u) = \exp \left\{- \frac{\theta(\bm{s}_1, \bm{s}_2)}{u} \right\}, \quad \bm{s}_1, \bm{s}_2 \in \mathcal{X},
\end{equation}
where $u>0$ and $\theta(\bm{s}_1, \bm{s}_2)$ is the bivariate extremal coefficient. By homogeneity of the exponent measure, $\theta(\bm{s}_1, \bm{s}_2) = V_{\bm{s}_1, \bm{s}_2}(1, 1)$. Furthermore, $\theta(\bm{s}_1, \bm{s}_2) \in [1, 2]$ for any $\bm{s}_1, \bm{s}_2 \in \mathcal{X}$, with values $1$ and $2$ indicating perfect dependence and independence, respectively. The lower the value of $\theta(\bm{s}_1, \bm{s}_2)$, the higher the dependence. The pairwise extremal coefficient has a one-to-one relation with the F-madogram \citep{Cooley2006} and this allows it to be estimated non-parametrically. If $Z$ in~\eqref{eq:extcoeff} is a Brown--Resnick field associated with the semivariogram $\gamma$, then \citep[e.g.,][]{davison2012}
\begin{equation}
\label{Eq_ExtrCoeffBRField}
\theta(\bm{s}_1, \bm{s}_2)= 2 \Phi \left\{ \sqrt{\gamma(\bm{s}_2 - \bm{s}_1)}/2 \right\}, \quad \bm{s}_1, \bm{s}_2 \in \mathcal{X},
\end{equation}
where $\Phi$ denotes the standard univariate Gaussian distribution function.

If $\{ Z(\bm{s}) : \bm{s} \in \mathcal{X} \}$ is a simple max-stable field, then \citep[e.g.,][]{davison2012}
\begin{equation}
\label{eq:chi}
\lim_{z\rightarrow \infty} \mathrm{Pr}\{Z(\bm s_1) > z \mid Z(\bm s_2) > z \} = 2 - \theta(\bm{s}_1, \bm{s}_2), \quad \bm{s}_1, \bm{s}_2 \in \mathcal{X}.
\end{equation}
Thus, unless $Z$ is standard Fréchet white noise, there exist $\bm{s}_1, \bm{s}_2$ such that the limit in~\eqref{eq:chi} is strictly positive and therefore $Z$ is asymptotically dependent. One can assess the suitability of max-stable models by checking whether the left hand-side of~\eqref{eq:chi} vanishes for different pairs of grid points using tests \REV{such as those of \citet{huser.wadsworth.18} or those} reviewed by \citet{miguel12}; see for instance \citet{Bacro2010}. However, evidence of asymptotic dependence does not entail suitability of max-stable models, as data may show asymptotic dependence without being max-stable. \REV{Moreover max-stable models cannot capture weakening dependence at increasing intensity levels, 
which is sometimes encountered in finite datasets.} In this paper, we \REV{use diagnostic plots to check for this instability, while also} explicitly testing the null hypothesis of max-stability; see Sections~\ref{sec:ms_test} and \ref{Subsec_sub-region}. \REV{Sub-asymptotic models for extremes \citep{Zhang.etal.2022, huser.wadsworth.18, huser17gaussian} might be used if this limiting assumption appears implausible.}


Let $\bm{s}_1, \ldots, \bm{s}_D \in \mathcal{X}$ denote grid points at which we regularly observe a field of pointwise maxima, which, based on~\eqref{eq:maxima_convergence}, we model by a max-stable field. \REV{One way to model the marginal parameters is to borrow strength across space, as in the Bayesian hierarchical approach used by \citet{davison2012}.} In Section~\ref{sec:model:case_study} we shall \REV{instead} justify modelling the margin at grid point $\bm{s}_d$ by a GEV distribution with location, scale and shape parameters $\eta_{\bm{s}_d}$, $\tau_{\bm{s}_d}$ and $\xi_{\bm{s}_d}$, \REVV{$d=1, \ldots, D$}.  In our setting this entails the estimation of $3\times D$, i.e., $1857$, parameters. Computational considerations make it usual to first estimate these marginal parameters by maximum likelihood, and then to fix them and estimate the dependence parameters of the max-stable field by maximizing a composite log-likelihood. \REV{The latter approach, and more specifically the maximum truncated pairwise likelihood estimation used in this paper\REVV{,} are described in Section~\ref{sec:composite} of the Supplementary Material.}

\subsection{Data and exploratory analysis}
\label{sec:data_explore}

The data we study were used in \cite{koch2020trendPROD} and constitute a coarse version of reanalysis data from the North American Regional Reanalysis (NARR). They consist of three-hourly time-series of 0--180 hPa CAPE (J kg$^{-1}$) and 0--3 km SRH (m$^2$ s$^{-2}$) from 1 January 1979 at 00:00 Coordinated Universal Time (UTC) to 31 December 2015 at 21:00 UTC. For consistency, we dropped any data for February 29; this does not impact our findings. The domain considered is a rectangle  over the contiguous US from~$-110^\circ$~to~$-80^\circ$~longitude and~$30^\circ$~to~$50^\circ$~latitude \REV{(see Figure~\ref{fig:regions})} containing Tornado Alley (Texas, Oklahoma, Kansas, Nebraska, Iowa and South Dakota), the riskiest region for  severe thunderstorms. The resolution is 1$^\circ$ longitude and 1$^\circ$ latitude, leading to 651 grid points in our region; no data are available for 32 grid points over water. We use the time series of CAPE and SRH to build three-hourly time series of $\mathrm{PROD}=\sqrt{\mathrm{CAPE}} \times \mathrm{SRH}$ (m$^3$ s$^{-3}$), \REVV{and then consider their monthly maxima.} As a measure of ENSO, we use monthly values of the Niño-3.4 index (${}^{\circ}\mathrm{C}$) from 1979 to 2015, taken from the ERSSTv5 data set publicly available from the National Oceanic and Atmospheric Administration (NOAA) Climate Prediction Center. \REVV{We project the original longitude-latitude coordinates using the Universal Transverse Mercator projection (with
zone 15 as a reference) so  Euclidean distances correspond to great circle distances (in 100 km) unless otherwise stated.} 

\begin{figure}[t]
\centering
  \begin{subfigure}[b]{.45\linewidth}
    \centering
    \includegraphics[width=.99\textwidth]{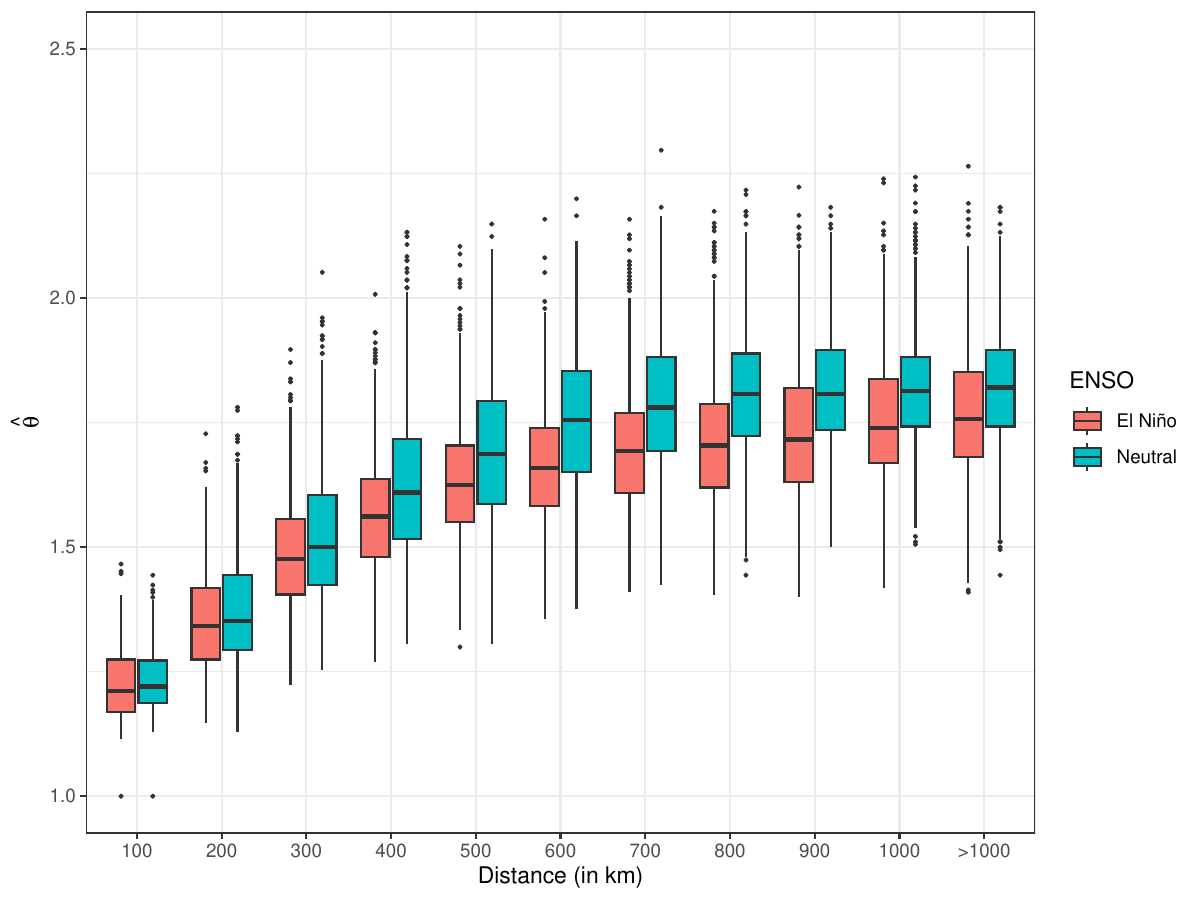}
  \end{subfigure}
  \begin{subfigure}[b]{.45\linewidth}
    \centering
    \includegraphics[width=.99\textwidth]{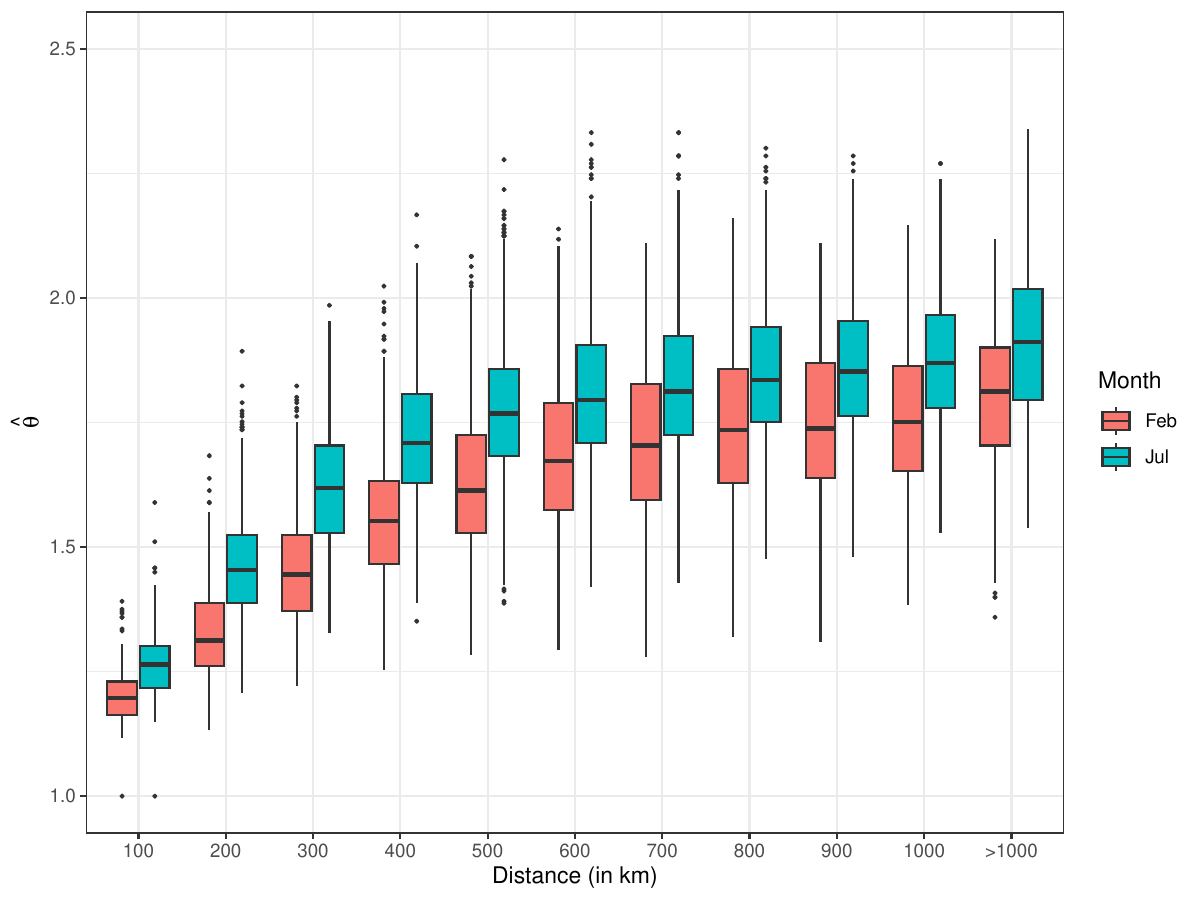}
  \end{subfigure} \\
  \caption{\REVV{Boxplots of the empirical pairwise extremal coefficients for SRH using all grid points in Region 2 of Figure \ref{fig:regions}. The left-hand panel concerns February, March and April maxima (pooled together) for $\text{ENSO}>+0.25^{\circ}$C (red) and $|\text{ENSO}|\leq 0.25^{\circ}$C (blue), and the right-hand panel concerns February (red) and July (blue) maxima.}}   
  \label{fig:explore:extcoeff}
\end{figure}

Figure~\ref{fig:explore:extcoeff} shows that for SRH maxima in \REVV{February, March and April}, the bivariate extremal coefficient tends to be lower during \REVV{EN} episodes than during periods with low absolute value of ENSO, indicating that the spatial dependence of pointwise maxima increases during \REV{EN events.} This may result from an increase of the spatial extent of individual SRH events. 
\REV{Similarly,} the extremal coefficients are lower in February than in July, \REV{suggesting wider SRH events, probably linked with the stronger and more organized jet stream in winter;
see Section~\ref{Subsec_MeteorologicalExpl}.}
These findings are true for regions \REVV{and variables} other than that used in  Figure~\ref{fig:explore:extcoeff} (not shown) and suggest the use of ENSO and month as covariates \REV{when modelling the extremal spatial dependence.} 

\section{Novel methodology}
\label{Sec_Methodology}

\subsection{Model}\label{sec:model_all_ms}

\subsubsection{General case}
\label{Subsubsec_GeneralVersion}

We propose a space-time model that is simple max-stable at each time point. Let $T$ be the number of time points and let $\bm{x}_t = (x_{1, t}, \ldots, x_{p, t})^{\prime}$ gather the values of $p$ covariates, such as large-scale atmospheric signals or the month, at time $t\in\{1, \ldots, T\}$, .
Our space-time model $\{ Z(\bm{s}, t) : \bm{s} \in \mathcal{X}, t  = 1, \ldots, T \}$ is then  defined as follows:
\begin{enumerate}
    \item for any $t$, the spatial field $\{ Z(\bm{s}, t) : \bm{s} \in \mathcal{X}\}$ is max-stable and its spatial dependence structure may depend on $\bm{x}_t$;
    \item the spatial fields $\{ Z(\bm{s}, 1) : \bm{s} \in \mathcal{X}\}, \ldots, \{ Z(\bm{s}, T) : \bm{s} \in \mathcal{X}\}$ are independent.
\end{enumerate}

As mentioned in Section~\ref{Subsec_MaxStableFields}, max-stable models 
typically depend on range and smoothness parameters $\rho > 0$ and $\alpha \in (0, 2]$, a scaling parameter $r>0$ and a rotation parameter $\kappa \in [0, \pi]$. \REVV{If $a_t$ denotes the value of $a \in \{\rho, \alpha, r, \kappa\}$  at time $t$, then $a_t$} may depend on $\bm{x}_t$ through the expression 
\begin{equation}
\label{EqLinearBasisExpansonRange}
g(a_t) = \sum_{k=1}^{K} \beta_k h_k(\bm{x}_t),
\end{equation}
where $g$ is a monotonic link function, $K \in \mathbb{N} \backslash \{0\}$, the $h_1,\dots,h_K$ are real-valued basis functions  and the $\beta_k$ are real numbers. A \REV{convenient} choice for $g$ in the case of $\rho$ is the logarithm, but other choices would be necessary for the other parameters. Depending on the choice of the $h_k$,~\eqref{EqLinearBasisExpansonRange} might represent a generalized linear model or a more flexible model (e.g., with splines or wavelets). 
Radial cubic splines are useful for continuous variables such as weather variables (e.g., ENSO), cyclic P-splines are appropriate when using the month as covariate, as they allow a smooth transition between the first and the last month, and tensor product splines can capture interactions between different covariates. The simultaneous modelling of several parameters with~\eqref{EqLinearBasisExpansonRange} requires enough data, as it can be hard to detect non-stationarity in the dependence structure. For instance, modelling both $\rho$ and $\alpha$ in this way is difficult because their effects may be difficult to distinguish. 

\subsubsection{Specification for the case study, and simulations}\label{sec:model:case_study}

Finding appropriate and parsimonious trend surfaces for the marginal parameters is challenging for large and meteorologically heterogeneous regions. Using ill-specified trend surfaces may bias our characterization of the dependence structure, so we model the monthly maxima by fitting GEV distributions separately at each grid point and then model the field obtained after using these separate fits to transform the data to standard Fr\'echet margins. Our space-time model for these transformed data lies in the class of max-stable models introduced in Section~\ref{Subsubsec_GeneralVersion}; we use the Brown--Resnick model with variogram~\eqref{Eq_ExpressionAnisotropicVariogram}, which often fits environmental extremes well \citep[e.g.,][]{davison2012}. Fitting a classical Brown--Resnick model for each month separately showed no evidence of month- or ENSO-specific smoothness, scaling and rotation parameters, for PROD, CAPE or SRH (not shown). Thus, based on our exploratory analysis (Figure~\ref{fig:explore:extcoeff}) and meteorological understanding, we model the range parameter $\rho$ as in~\eqref{EqLinearBasisExpansonRange} with ENSO and month as covariates, keep the parameters $\alpha$, $r$ and $\kappa$ constant, and take the time step to be one \REV{month. Our} vector of covariates at $t$ is therefore $\bm{x}_t = (\mathrm{ENSO}_t, t \ \mathrm{mod} \ 12)'$ and  $t=1, \ldots, 444$, 
where $\mathrm{ENSO}_t$ is the value of ENSO for the month associated with $t$, and mod denotes the modulo operation. The effect of these covariates on $\rho$ appears to be non-linear \citep[see, e.g.,][regarding the non-linearity of general responses to \REVV{EN} and \REVV{LN}]{hoerling1997nino}, and the covariates interact, so we choose the functions $h_k$ to be the components of a tensor product spline basis between a radial cubic spline basis in the ENSO direction and a cyclic P-spline basis in the month direction, allowing us to borrow strength from neighbouring ENSO phases and months. Thus our final model at time $t=1, \ldots, T$, for the standardized quantities is the Brown--Resnick field with semivariogram
\begin{equation}
\label{Eq_VariogramFinalModelCaseStudy}
\gamma(\bm{s}, t) = \left[ \frac{ \| A \bm{s} \|}{ \exp \left \{ \sum_{k=1}^{K} \beta_k h_k(\bm{x}_t) \right \} } \right]^{\alpha}, \quad \bm{s} \in \mathcal{X},
\end{equation}
where 
$A$ is the matrix (\ref{eq_aniso}) with $r >0$, $\kappa \in [0, \pi]$, \REV{and $K$ equals $1$ + number of ENSO knots $\times$ ($1 - $ number of month knots).} To choose the knots, we fit models with different numbers of knots positioned regularly in each direction and choose the best using the bootstrap-based selection criterion developed in Section~\ref{sec:bootstrap}. 

As detailed in the Supplementary Material,  we verified that the model parameters are identifiable and can be estimated adequately in a setting similar to the case study. This is especially important for the model in~\eqref{Eq_VariogramFinalModelCaseStudy} owing to its complexity, and as it is hard to detect non-stationarity in the extremal dependence structure with too few data. 

\subsection{A max-stability test 
with unknown margins}\label{sec:ms_test}

Assume that a \REV{simple} max-stable field $\{ Z(\bm{s}) : \bm{s} \in \mathcal{X} \}$ is observed in a testing region $\mathcal{T}\subseteq\mathcal{X}$ consisting of grid points $\{\bm s_1,\ldots,\bm s_D\}$, and let $\mathcal{S} = \{1,\ldots,D\}$. The homogeneity of order $-1$ of the exponent measure in~\eqref{Eq_MultivariateDfMaxStable} implies that  $Z_\mathcal{S} = \max_{d\in \mathcal{S}} \REV{\log } Z(\bm s_d)$ has a Gumbel distribution with location parameter $\mu_\mathcal{S} = \log V_{\bm{s}_d, d\in \mathcal{S}}\REVV{(1, \ldots, 1)}$, where $V_{\bm{s}_d, d\in \mathcal{S}}$ is the \REVV{exponent measure} of $\{Z( \bm s_d), d\in \mathcal{S}\}$. \citet{gabda.2012} and \citet{Buhl.2016} visually compared the empirical distribution of $Z_\mathcal{S}$ and a Gumbel distribution whose location $\hat{\mu}_\mathcal{S}$ has been estimated from simulated realizations of $Z_\mathcal{S}$. 

In applications we must transform data to \REV{standard Fr\'echet} margins before performing such a test. We use the Anderson--Darling \REV{\citep[A-D, ][]{Anderson.Darling.1954}} statistic to measure the distance between the empirical distribution of $Z_\mathcal{S}$ and a Gumbel distribution with location $\hat{\mu}_\mathcal{S}$ and thus to assess the null hypothesis that the multivariate distribution is max-stable. The null distribution should allow for the estimation of both $\mu_\mathcal{S}$ and the margins, and we use the bootstrap to do so.  

Our proposed bootstrap test is parametric for the margins and non-parametric for the dependence, on which we do not wish to impose a particular model. Let $({z}^\star_{1,m},\dots ,{z}^\star_{D,m})^\prime$ denote the vector of the $m$-th observed maxima at all grid points in $\mathcal{S}$, where $m\in\{ 1,\dots,M\}$. For each grid point $d\in \mathcal{S}$, we first fit the GEV distribution using ${z}^\star_{d,1},\dots, z^\star_{d,M}$ to obtain estimators $\hat{\eta}^\star_d$, $\hat{\tau}^\star_d$ and $\hat{\xi}^\star_d$. 
We approximate the distribution of the A-D statistic under the null hypothesis of max-stability by $B$ independent replications of the following procedure:
\begin{enumerate}
\item use the empirical likelihood-based approach outlined in the Supplementary Material to 
simulate $M$ replicates of a max-stable vector, denoted $(\tilde{z}_{1,1},\dots ,\tilde{z}_{D,1})^\prime,$ $\dots,$ $(\tilde{z}_{1,M},\dots ,\tilde{z}_{D,M})^\prime$, such that $\tilde{z}_{d,1},$ $\dots,$ $\tilde{z}_{d,M} $, $d\in \mathcal{S}$, are drawn from a GEV distribution with location, scale and shape parameters $\hat{\eta}^\star_d$, $\hat{\tau}^\star_d$ and $\hat{\xi}^\star_d$. \REV{This step uses the total number of underlying observations over all years, $n_M=n$ (block size) $\times M$ (number of years), to simulate $M$ replicates of a max-stable vector};
\item for each $d\in \mathcal{S}$, use a GEV distribution fitted to $\tilde{z}_{d,1}, \dots, \tilde{z}_{d,M}$ by maximum likelihood to transform the $\tilde z_{d,m}$ to approximately standard \REV{Fr\'echet}-distributed quantities ${z}_{d,1}, \dots, {z}_{d,M} $, yielding $({z}_{1,1},\dots ,{z}_{D,1})^\prime,$ $\dots,$ $({z}_{1,M},\dots ,{z}_{D,M})^\prime$; 
\item compute $z_{\mathcal{S},m} = \max_{d\in \mathcal{S}} \REV{\log}  z_{d,m}$($m=1,\dots,M$) and fit a Gumbel distribution to $z_{\mathcal{S},1},\dots, z_{\mathcal{S},M}$ using maximum likelihood, giving location parameter estimate $\hat{\mu}_\mathcal{S}$;
\item calculate the A-D statistic measuring the distance between the empirical distribution of the $z_{\mathcal{S},1},\dots, z_{\mathcal{S},M}$ and a Gumbel distribution with location $\hat{\mu}_\mathcal{S}$.
\end{enumerate}
\REV{It is critical to incorporate the marginal estimation uncertainty when approximating the distribution of the A-D statistic under the null hypothesis, which we do via Steps~1 and~2.} The same test might be applied with $\mathcal{S}$ any subset of $\{1,\dots,D\}$ of size two or more. \REV{To assess how our procedure works} we perform two experiments detailed in the Supplementary Material, with \REV{$D=100$}
 and $M=40$ block maxima of size $240$, \REV{so $n_M=40\times240$.} 
 
\subsection{Bootstrap-based uncertainty assessment and model selection}\label{sec:bootstrap}




Let $\mathcal{Y}=(\bm Y_1, \ldots, \bm Y_T)^{\prime}$ be a data matrix, where $\bm Y_1, \ldots, \bm Y_T$ are independent replicates of a $D$-dimensional random vector $\bm Y$, such as the vector of maxima at certain grid points at a given time, and suppose that the margins of $\bm Y$ have been estimated.  In this section we discuss uncertainty  quantification and model selection when using composite likelihood to estimate the dependence structure.

Suppose we have a parametric model for each margin of $\bm{Y}$ and that the marginal parameters of all components of $\bm{Y}$ are gathered in $\bm{\lambda}$. Below, we consider both the ideal situation where the exact marginal models and $\bm \lambda$ are known, and the more realistic situation where $\bm \lambda$ is estimated by $\hat{\bm \lambda}$ prior to dependence modelling. Let the function $t_{\bm{\lambda}}$ transform a data matrix to have known margins, let $\bm{Z}_1, \ldots, \bm{Z}_T$ and $\bm{Z}$ be transformations of $\bm Y_1, \ldots, \bm Y_T$ and $\bm Y$ to have known margins, and let $\mathcal{Z}= (\bm Z_1, \dots, \bm Z_T)^{\prime}$. Then $\mathcal{Z}=t_{\bm{\lambda}}(\mathcal{Y})$ if $\bm \lambda$ is known and $\mathcal{Z}=t_{\hat{\bm{\lambda}}}(\mathcal{Y})$ if $\bm \lambda$ is estimated by $\hat{\bm \lambda}$. 

Assume that we model $\mathcal{Z}$ using a family $\mathcal{F} = \{ f(\bm{z}, \bm{\psi}): \bm{z} \in \mathbb{R}^D, \bm{\psi} \in  \Psi \subseteq \mathbb{R}^p \}$ of density functions with known margins (typically standard Fréchet in the spatial extremes setting) and dependence parameter $\bm \psi$. The composite likelihood is
$L_C(\bm\psi ; \mathcal{Z})= \prod_{t=1}^{T} f_C(\bm \psi; \bm Z_t)$, where $f_C$ is defined through the density $f$ and characterizes the composite likelihood \citep[see, e.g.,][Definition 1]{varin.2005}, and $\hat{\bm\psi}$ denotes the maximum composite likelihood estimator. In the case of the truncated pairwise likelihood \REV{(see \eqref{Eq_TruncLogLik} in the Supplementary Material)}, $f_C$ is the sum over all pairs of tapered bivariate densities. 

When the true marginal models and $\bm \lambda$ are known, under mild regularity assumptions, $\hat{\bm \psi} \mathrel{\dot\sim} \mathrm{N}_p\{\bm \psi, \text{I}(\bm{\psi})^{-1}\}$ for $T$ large, where the \REV{\citet{Huber:1964}} sandwich information matrix  $\text{I}(\bm{\psi})$ equals $\text{H}(\bm{\psi}) \text{J}(\bm{\psi})^{-1} \text{H}(\bm{\psi})$ with $\text{H}(\bm{\psi})=\text{E}\{-\nabla^2_{\bm \psi} \log L_C(\bm \psi; \mathcal{Z})\}$ and $\text{J}(\bm{\psi})=\text{V}\{ \nabla_{\bm \psi} \log L_C(\bm \psi; \mathcal{Z}) \}$, $\nabla^2_{\bm \psi}$ and $\nabla_{\bm \psi}$ denote the Hessian and gradient operators with respect to $\bm \psi$, and $\text{V}$ indicates the covariance matrix operator. Confidence intervals can be based on the sandwich matrix $\hat{\text{H}}(\hat{\bm{\psi}})^{-1}\hat{\text{J}}(\hat{\bm{\psi}}) \hat{\text{H}}(\hat{\bm{\psi}})^{-1}$ \citep{padoan2010likelihood},
where
$$
\hat{\text{H}}(\hat{\boldsymbol{\psi}})=- \nabla_{\bm \psi}^2 \log  L_{C}(\hat{\bm\psi}; \mathcal{Z}), \quad 
\hat{\text{J}}(\hat{\boldsymbol{\psi}})=\sum_{t=1}^{T} \{ \nabla_{\bm \psi} \log  f_{C}(\hat{\bm\psi}; \bm Z_t) \} \{ \nabla_{\bm \psi} \log  f_{C}(\hat{\bm\psi}; \bm Z_t) \}^\prime.
$$
After fitting several models in $\mathcal{F}$ using composite likelihood, it is standard to select \REVV{the one} having the highest observed value of
$
\log L_C(\boldsymbol{\hat{\psi}}, \mathcal{Z} ) - \text{tr}\{ \hat{\text{J}}(\hat{ \boldsymbol{\psi}}) \hat{\text{H}}(\hat{ \boldsymbol{\psi}})^{-1}\}
$ \citep{varin.2005}, 
or equivalently the lowest observed value of 
\citep{padoan2010likelihood}
\begin{equation}\label{eq:CLIC}
\text{CLIC} = - 2\log L_C(\boldsymbol{\hat{\psi}}, \mathcal{Z} )+ 2\text{tr}\{ \hat{\text{J}}(\hat{ \boldsymbol{\psi}}) \hat{\text{H}}(\hat{ \boldsymbol{\psi}})^{-1}\}.
\end{equation}

If $\bm \lambda$ has been estimated in a first step, as is often the case in spatial extremal analysis, then use of the estimated covariance matrix and CLIC for uncertainty assessment of $\hat{\bm \psi}$ and model selection within $\mathcal{F}$ does not account for estimating the marginal parameters $\bm \lambda$. 
In spatial extremes, the non-parametric bootstrap is often used for uncertainty assessment \citep{davison.et.al.2013, davison18handbook, huser.wadsworth.18}, because researchers are aware of the shortcomings of using the sandwich matrix, but this cannot be said of the use of CLIC for model selection in a two-step setting. Many studies \citep[e.g.,][]{davison.et.al.2013, davison18handbook, Huser2016nonstat, huser.2021} do not allow for the estimation of the margins.

In order to account for how marginal estimation affects model selection when using composite likelihood, we propose bootstrap estimation of the non-normalized composite Kullback--Leibler divergence \citep{varin.2005}. In  Section~\ref{sec:known_margins} we suppose that the margins are known and extend the results of \cite{Shibata-1997} and \cite{Cavanaugh.1997} to the composite likelihood setting, and in Section~\ref{Subsubsec_CLICUnknownMargins} we  define a criterion to account for the marginal effects. Section~\ref{Subsubsec_SimStudyCoverageModelSelection} illustrates the resulting benefits through a simulation study.  
We suppress the dependence of $\mathcal{Y}$, $\mathcal{Z}$ and $\hat{\bm \psi}$ on $T$ throughout.  

\subsubsection{Known margins}\label{sec:known_margins}

We assume that $\bm \lambda$ in $t_{\bm{\lambda}}$ is known and we seek the best model for $\mathcal{Z}=t_{\bm{\lambda}}(\mathcal{Y})$ by  estimating the non-normalized composite Kullback--Leibler divergence from a model to the truth using a
non-parametric bootstrap, following what \citet{Cavanaugh.1997} and \citet{Shibata-1997} did for the non-normalized Kullback--Leibler divergence.

Let $g(\bm z)$, $\bm z\in\mathbb{R}^{D}$, be the true density of $\bm{Z}$.
The non-normalized composite Kullback--Leibler divergence for a model with density in $\mathcal{F}$ is
$ d_T(\bm\psi) = \text{E}\REVV{_g} \{ -\log L_C(\bm\psi; \mathcal{Z} )\}$, where $\text{E}\REVV{_g}$ is the expectation under $g$. The divergence of the model estimated by maximum composite likelihood (with $\hat{\bm\psi}$ as estimated parameter) to the truth is thus
\begin{equation}\label{eq:divergence}
d_T( \hat{\bm\psi}) = \text{E}\REVV{_g} \{ -\log L_C(\bm\psi ; \mathcal{Z} )\} \mid_{\bm\psi= \hat{\bm\psi}},
\end{equation}
but this cannot be computed unless we know $g$. \citet{varin.2005} showed that a biased estimator of (\ref{eq:divergence}) is $-\log L_C(\hat{\bm\psi} ; \mathcal{Z})$, and adjusted for the bias with a first-order correction. Now, suppose that $\hat{\bm\psi}^*$ is a bootstrap replicate of $\hat{\bm\psi}$, and let E$^*$ denote expectation with respect to the bootstrap distribution of $\hat{\bm\psi}$. Arguments similar to those in \citet{Cavanaugh.1997} imply under the usual regularity conditions that 
\begin{equation}\label{eq:bias_bootstrap}
\text{Bias}_T^*=-2\left[\text{E}^* \left \{ -\log L_C\left(\hat{\bm\psi}^* ; \mathcal{Z}\right) \right \} + \log L_C\left(\hat{\bm\psi}; \mathcal{Z}\right) \right]
\end{equation}
converges almost surely to the bias of $-\log L_C(\hat{\bm\psi} ; \mathcal{Z})$ as $T\rightarrow \infty$. \REVV{By the strong law of large numbers under dependence,} a Monte Carlo estimator from $B$ bootstrap replicates yields a strongly consistent estimator of $\text{Bias}_T^*$ as $B \to \infty$, 
\begin{equation}
\label{Eq_MCEstimatorBias}
\hat{\text{ Bias}_T^\star}=- \dfrac{2}{B} \sum_{b=1}^{B} \left[ - \log L_C \left \{\hat{\bm\psi}^*_b; t_{\bm \lambda}(\mathcal{Y}) \right \} + \log L_C\left \{\hat{\bm\psi}; t_{\bm \lambda}(\mathcal{Y}) \right \} \right].
\end{equation}
Thus, a natural estimator of twice the quantity in~\eqref{eq:divergence} is
\begin{equation}
\label{Eq_FirstExpreCLICb}
\quad -2\log L_C\left \{\hat{\bm\psi}; t_{\bm \lambda}(\mathcal{Y}) \right \} - 2\hat{\text{ Bias}_T^\star} 
= \dfrac{1}{B} \sum_{b=1}^{B} \left[ {2} \log L_C \left\{\hat{\bm\psi}; t_{\bm \lambda}(\mathcal{Y}) \right \} - {4} \log L_C\left \{\hat{\bm\psi}^*_b; t_{\bm \lambda}(\mathcal{Y}) \right \}\right].
\end{equation}
For $T$ and $B$ large enough, model selection based on CLIC and~\eqref{Eq_FirstExpreCLICb} should be equivalent.

\subsubsection{Unknown margins}
\label{Subsubsec_CLICUnknownMargins}

Suppose that $\bm \lambda$ in $t_{\bm{\lambda}}$ is estimated by $\hat{\bm \lambda}$, and we seek the best model  for $\mathcal{Z}=t_{\hat{\bm{\lambda}}}(\mathcal{Y})$ within $\mathcal{F}$. An attractive property of the bootstrap-based estimator of the non-normalized composite Kullback--Leibler divergence developed in Section~\ref{sec:known_margins} is that we can account for estimating the margins. In computing the maximum composite likelihood estimate of $\bm \psi$ for the $b$-th bootstrap replicate, we estimate the marginal parameters from the bootstrapped data, yielding an estimate $\hat{\bm \lambda}^*_b$. We make this explicit by writing the estimates $\hat{\bm \psi}$ and $\hat{\bm \psi}^*_b$ as functions of $\hat{\bm \lambda}$ and $\hat{\bm \lambda}^*_b$, respectively. The expectation $\text{E}^\star$ in (\ref{eq:bias_bootstrap}) with respect to the bootstrap distribution of $\hat{\bm\psi}$ takes the estimation of the margins into account.
Following~\eqref{Eq_FirstExpreCLICb}, we should consider choosing the model that minimises the criterion 
\begin{equation}\label{eq:CLICb}
\text{CLIC}^\text{b} = \dfrac{1}{B} \sum_{b=1}^{B} \left[ {2} \log L_C\left \{\hat{\bm\psi}\left(\hat{\bm \lambda} \right) ; t_{\hat{\bm \lambda}}(\mathcal{Y})\right\} - {4} \log L_C\left\{\hat{\bm\psi}^*_b \left(\hat{\bm \lambda}^*_b \right) ; t_{\hat{\bm \lambda}}(\mathcal{Y}) \right\} \right].
\end{equation}
A full likelihood is a composite likelihood, so this approach also applies more broadly.

The matrices $\hat{\text{H}}(\hat{ \bm \psi})$ and $\hat{\text{J}}(\hat{ \bm \psi})$ required for  confidence intervals or the $\mathrm{CLIC}$ are often cumbersome to compute, and careful application of pseudo-inverse procedures may be needed if $\hat{\text{H}}(\hat{ \bm \psi})$ is singular, especially for complex models with many parameters. This further supports the use of $\mathrm{CLIC^b}$, whose calculation costs the same as a bootstrap.

Expressions asymptotically equivalent to~\eqref{eq:bias_bootstrap} could be used, as in \citet[][Section~2]{Shibata-1997}, leading to different but asymptotically equivalent specifications of $\text{CLIC}^\text{b}$.

\subsubsection{Simulation study}
\label{Subsubsec_SimStudyCoverageModelSelection}

We perform two experiments with three procedures: $\mathrm{P^k}$, in which the correct margins are used when fitting the models and CLIC is used for selection; $\mathrm{P^u}$, in which the marginal distributions are supposed to be GEV and estimated in a first step, then transformed before fitting the dependence models and using CLIC for selection; and $\mathrm{P^b}$, which is like $\mathrm{P^u}$ but uses $\text{CLIC}^\text{b}$ in~\eqref{eq:CLICb} for model selection with a non-parametric block bootstrap ($B=200$), in which each replicate is a block. In each case the dependence models are fitted using the approach of Section~\ref{sec:composite}. The first procedure approximates the best that CLIC can do. 


In the first experiment, we generated $40$ independent replicates at $D \in \{25, 100, 225\}$ grid points of a Smith field \citep{Smith} with common standard Fr\'echet margins and twice the $2\times 2$ identity matrix as covariance matrix, and used $\mathrm{P^k}$, $\mathrm{P^u}$ and $\mathrm{P^b}$ to choose between an isotropic Smith model labelled SM$_0$ and a two-parameter Brown--Resnick model labelled BR$_1$; the latter is over-complex because the Smith field corresponds to the Brown-Resnick field with $\alpha=2$ \citep[e.g.,][]{HuserDavison}. Table~\ref{table:clic_simulation} shows that $\mathrm{P^k}$ correctly chooses SM$_0$  for any $D$ in around $92\%$ of 200 \REVV{repetitions}. This figure is much lower for $\mathrm{P^u}$ and drops to as low as $20\%$ when $D$ increases, due to unaccounted variation from the estimation of the margins, whereas $\mathrm{P^b}$ achieves performance close to that of $\mathrm{P^k}$. 

In a second experiment with a configuration that could be realistic in an environmental application, we generated $40$ independent replicates at $D \in \{25, 100, 225\}$ grid points of a Brown--Resnick field with common standard Fr\'echet margins, $\rho=2$ and $\alpha=1$, and used $\mathrm{P^k}$, $\mathrm{P^u}$ and $\mathrm{P^b}$ to choose between BR$_1$ and a simpler Brown--Resnick model labelled BR$_0$ with $\rho=2$ fixed and $\alpha$ estimated.  Table~\ref{table:clic_simulation} shows that $\mathrm{P^k}$ correctly chooses BR$_0$ with probability around $0.84$  for any $D$, as would be expected in the full likelihood setting with $n$ large.  
The frequency of true selection ranges from $28\%$ and $44\%$ with $\mathrm{P^u}$, and is much higher (between $76\%$ and $80\%$) for $\mathrm{P^b}$. 

The paired proportions test \citep{McNemar.1947} ascribes 5\% significance to all tests of differences between $\mathrm{P^u}$ and $\mathrm{P^b}$ for both experiments (not shown). Thus, if the marginal and dependence parameters are estimated in two distinct steps and if composite likelihood is used, we strongly advocate the use of $\mathrm{CLIC^b}$,~\eqref{eq:CLICb}, rather than CLIC,~\eqref{eq:CLIC}.

\begin{table}[t]
\centering
\begin{tabular}{lrrr}
  \hline
   & \multicolumn{3}{c}{P$^\mathrm{k}$/$\mathrm{P^u}$/P$^\mathrm{b}$} \\
\cline{2-4}
True/Alternative & $D=25$ & $D=100$ & $D=225$ \\ 
\hline
  \hline
 \multirow{1}{*}{SM$_0$/BR$_1$} 
& 93/82/89 & 90/54/90 & 94/20/81  \\ 
  \hline
     \multirow{1}{*}{BR$_0$/BR$_1$} 
   & 84/44/80 & 84/30/76 & 85/28/76  \\ 
     \hline
\end{tabular}
\caption{Frequency (in $\%$, computed over $200$ repetitions) of selection of the true (simpler) model for each experiment, procedure and value of $D$. }
\label{table:clic_simulation}
\end{table}

In the second experiment, we also compared $95\%$ confidence intervals for the range parameter estimates of BR$_1$ calculated using the sandwich matrix and the non-parametric block bootstrap \citep[][basic intervals in \S5.2]{davison.hinkley.1997} with logarithm as a variance-stabilizing transform; see Figure~\ref{fig:simulation:sandwichCI} for $D \in \{25, 225 \}$. The empirical coverages of the sandwich matrix-based intervals drop from $61\%$ to $39\%$ as $D$ increases from $25$ to $225$, whereas the corresponding values for the bootstrap intervals are $90\%$ and $85\%$, lower than the nominal coverage but not appalling so.

\section{Case study}
\label{Sec_CaseStudy}

\subsection{\REV{Choice of appropriate regions and max-stability}}\label{Subsec_sub-region}

\begin{figure}[t!]
\centering
  \begin{subfigure}[b]{.60\linewidth} 
    \centering
    \includegraphics[width=.99\textwidth]{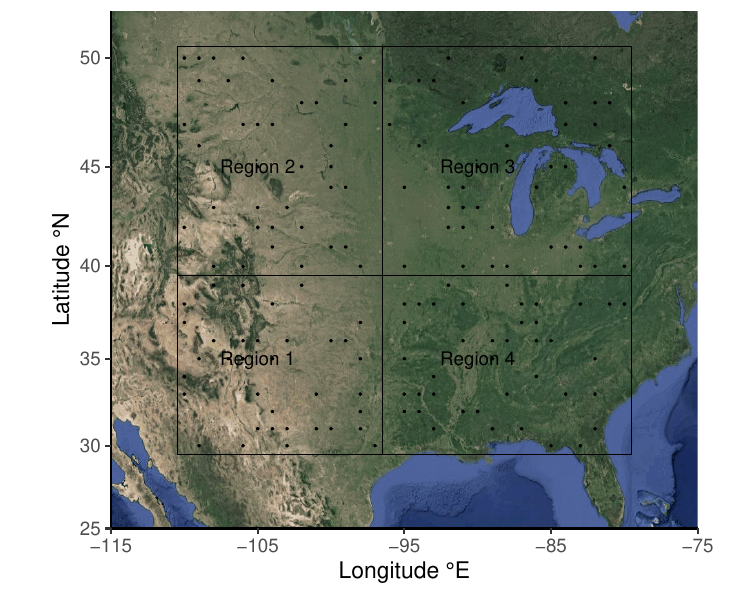}
  \end{subfigure}
  \caption{\REVV{The partition of the study domain into four regions. The black points represent the grid points chosen for the validation set. }}
  \label{fig:regions}
\end{figure}

\REV{
We divide the domain of interest into four regions displayed in Figure~\ref{fig:regions}, and study them separately. These four regions are homogeneous in terms of climate according to the Köppen--Geiger classification \citep[e.g.,][]{beck2018present} and main weather drivers (e.g., in terms of position with respect to the polar and subtropical jet streams). 
Applying our model to each region (rather than to the full spatial domain) allows us both to account partially for the spatial non-stationarity of dependence and to obtain finer meteorological interpretations. Moreover, a similar partition was also obtained by applying \REVV{to} the full domain the clustering method of \citet{Bernard.PAM.2013}, which is suited to maxima and designed to find the groups that are most independent; see Figure~\ref{fig:explore:pam} (Supplementary Material). This supports our split and implies that the four regions can be treated separately for our purposes.} 

\REVV{The use of the GEV   distribution as model for the margins of the monthly maxima considered here was extensively justified in \citet[][Sections 3(a) and (b)]{koch2020trendPROD}.}
\REV{\REVV{Regarding the dependence structure,} diagnostic plots (Figures~\ref{fig:chiplot} and~\ref{fig:condprob}  of the Supplementary Material) showed no evidence that a max-stable model is inappropriate, though these plots use non-parametric estimators that typically give inflated uncertainties near the boundaries. Hence we also applied the max-stability test outlined in Section~\ref{sec:ms_test}. }

\REV{For each variable and each month, we applied our test to each region. Figure~\ref{fig:maxstabtest_data} (Supplementary Material) suggests that a max-stable model is adequate, as the number of p-values below 0.05 remains  reasonable, though Region~4 has a larger number of p-values below $0.2$ for SRH, suggesting that closeness to max-stability varies across space. It also seems to vary with season: small p-values often correspond to summer to early winter (e.g., June to November) for PROD, late summer (e.g., September) for CAPE, and winter and spring (e.g., November to May) for SRH. Nevertheless, note that the power of the test depends heavily on the dependence strength in each region, as shown by simulation in Section~\ref{sec:ms_test}. }


\subsection{Results} 
\label{subsec_results}

\REV{
Regions 1--4 (R1--R4 in the following) respectively contain 140, 154, 164, and 161 grid points. To fit our model, we randomly chose roughly two thirds of the grid points in each region (105, 116, 123, and 121) and validated our models with the rest. The calibration and validation sets are the same for all variables.}

We applied the model described in Section~\ref{sec:model:case_study} separately to \REV{each region}, with $k_1 \in \{2,3,4\}$ knots in the ENSO direction and $k_2 \in \{4,5\}$ knots in the month direction. 
The knots for ENSO are placed evenly between the $10\%$ and $90\%$ quantiles of its values, i.e., between $-1.06^{\circ}$C and $1.16^{\circ}$C, and those for months are positioned evenly between $0.5$ and $12.5$, both of which represent \REV{mid-December.}
We also fitted a range parameter with no covariates. We fitted all models to the calibration set using the truncated pairwise likelihood approach described in \REV{Section~\ref{sec:composite} (Supplementary Material).} 

We assessed the uncertainty of our estimates with basic \REV{parametric bootstrap} confidence intervals, \REV{keeping the covariate information fixed for $200$ replicates; \citet{Gilleland.bootstrap.2020} \REVV{shows} that the parametric bootstrap is suitable in the context of univariate extreme-value theory.} For the range parameter, we used the logarithm as a variance-stabilizing transformation \citep[][p.~195]{davison.hinkley.1997} and derived the basic confidence intervals for $\log$ range before transforming them to the original scale. For model selection, we used $\text{CLIC}^\text{b}$ in~\eqref{eq:CLICb} with the same bootstrap replicates. 

\REV{Table~\ref{table:clic_b} shows that the best models for PROD, CAPE and SRH vary with the region considered: for instance, the best for PROD has three and four knots in the ENSO and month directions in \REVV{R2}, but four and five in R3.}
\REV{These models outperform that with constant range (not shown)}, suggesting that incorporating ENSO and month is valuable. \REV{Models other than those chosen, especially the more complex ones, lead to similar conclusions in terms of influence of EN and LN, but using more knots increases the uncertainty on the parameter estimates (not shown).} Below, by `model' we mean the best model for each of \REVV{PROD, CAPE and SRH}.

\begin{table}[t]
\footnotesize
\centering
\begin{tabular}{lllllllllllll}
  &  \multicolumn{4}{c}{\textbf{PROD}} & \multicolumn{4}{c}{\textbf{CAPE}} & \multicolumn{4}{c}{\textbf{SRH}} \\
   Knots  & R1      & R2      & R3      & R4      & R1      & R2      & R3      & R4      & R1      & R2      & R3      & R4     \\
  \hline
  \hline
    $2\times4$ & 478 & 56 & 3183 & \textbf{0} & 2054 & 601 & 6547 & 1775 & 1029 & 3347 & 1270 & 944\\
  $3\times4$ & 224 & \textbf{0} & 1475 & 444 & 839 & 819 & 1187 & \textbf{0} & 587 & 77 & 3 & 293 \\
$4\times4$ & \textbf{0} & 94 & 1461 & 526 & 945 & \textbf{0} & 1100 & 138 & 184 & 24 & 165 & 395 \\ 
  $2\times5$ & 567 & 222 & 2009 & 680 & 692 & 849 & 5069 & 1896 & 961 & 2800 & 788 & 345 \\ 
  $3\times5$ & 347 & 199 & 136 & 714 & 144 & 1044 & \textbf{0} & 311 & 688 & 213 & \textbf{0} & \textbf{0}\\ 
    $4\times5$  & 352 & 374 & \textbf{0} & 1293 & \textbf{0} & 198 & 304 & 293 & \textbf{0} & \textbf{0} & 76 & 176 \\ 
  \hline
\end{tabular}
\caption{\REVV{Differences of $\mathrm{CLIC^b}$ for R1--R4 and for different configurations $k_1 \times k_2$, where $k_1$ and $k_2$ are the numbers of knots in the ENSO and month directions.} }
\label{table:clic_b}
\end{table}

\begin{figure}[t!]
\centering
   \includegraphics[width=.85\textwidth]{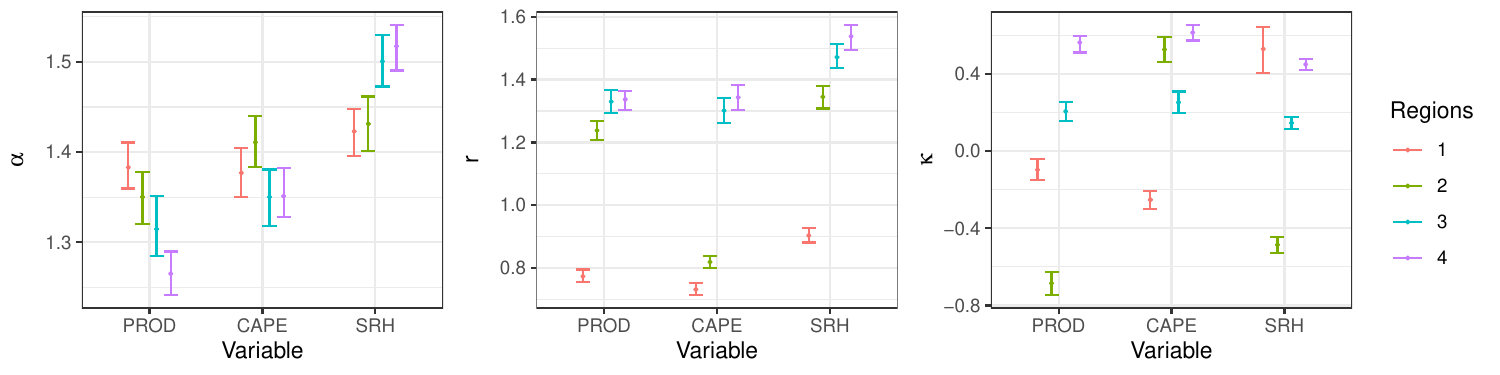} 
   \caption{\REVV{Bootstrap bias-corrected parameter estimates and their $90\%$ bootstrap confidence intervals for each region and variable.}}
   \label{fig:model_parameters}
\end{figure}

The estimates of the smoothness parameter $\alpha$ shown in \REV{Figure~\ref{fig:model_parameters}}
suggest that the models for PROD \REV{are slightly rougher than those for CAPE, which in turn are slightly rougher than those for SRH. A slight systematic downward bias for $\alpha$ was corrected using the bootstrap replicates.} 
\REV{The estimates of the parameters in~\eqref{eq_aniso} show a non-negligible \REVV{dilatation} in the \REVV{vertical} direction ($r$ ranges from \REVV{around $1.25$ to  $1.6$}) for all variables for R3 and R4\REVV{, and a slight compression in the vertical direction ($r$ ranges from around 0.7 to 0.95) for all variables for R1}. \REVV{The rotation parameter} $\kappa$ is positive \REVV{for all variables in R3 and R4, while its sign is not uniform across the variables for R1 and R2}.}



\begin{figure}[t!]
\centering
    \hspace{-1cm}%
  \begin{subfigure}[b]{.33\linewidth}
    \centering
    \includegraphics[width=.99\textwidth]{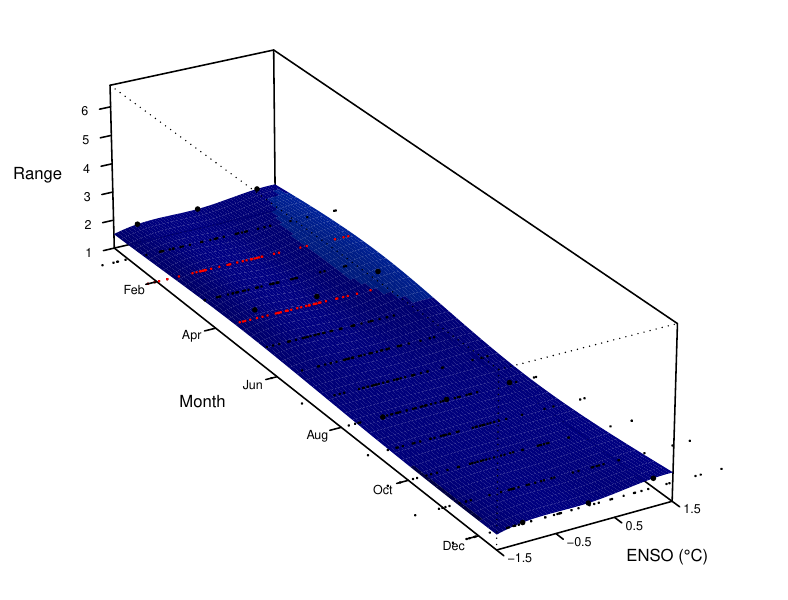}
    \end{subfigure}
    \hspace{-0.5cm}%
  \begin{subfigure}[b]{.17\linewidth}
    \centering
    \includegraphics[width=.99\textwidth]
    {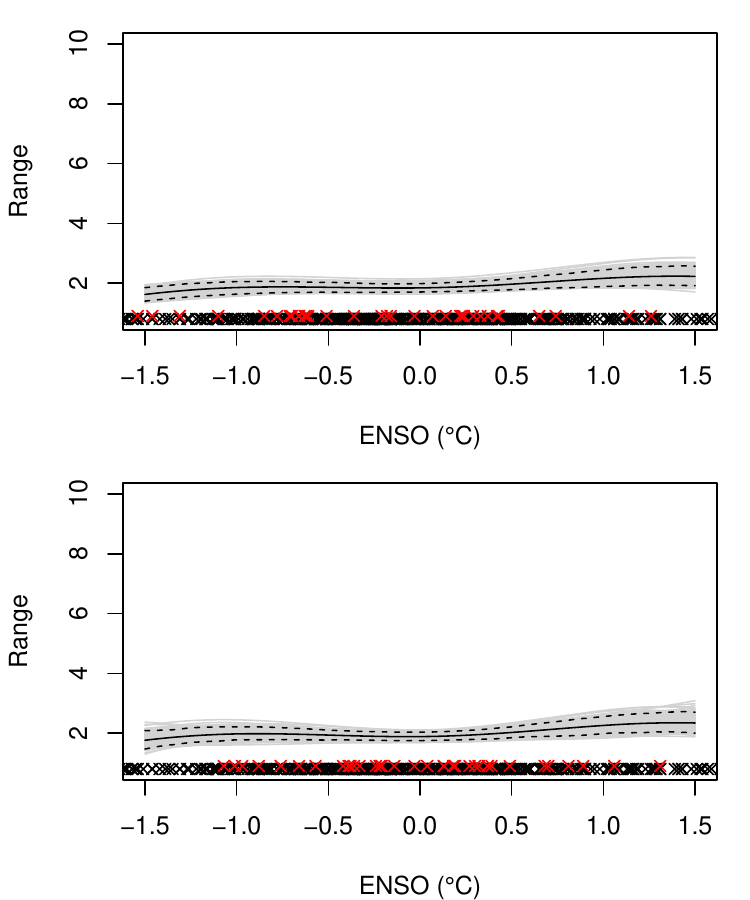} 
    \end{subfigure}
  \begin{subfigure}[b]{.33\linewidth}
    \centering
    \includegraphics[width=.99\textwidth]{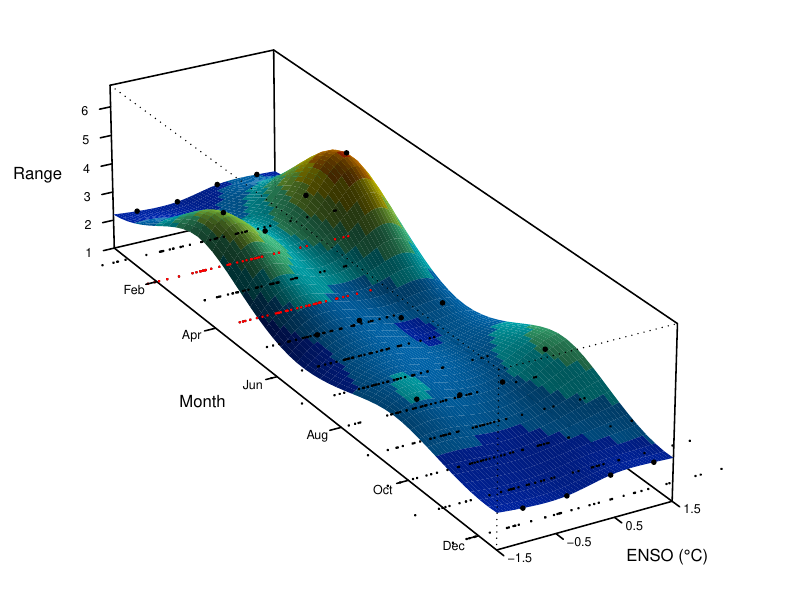}
    \end{subfigure}
    \hspace{-0.5cm}
  \begin{subfigure}[b]{.17\linewidth}
    \centering
    \includegraphics[width=.99\textwidth]
    {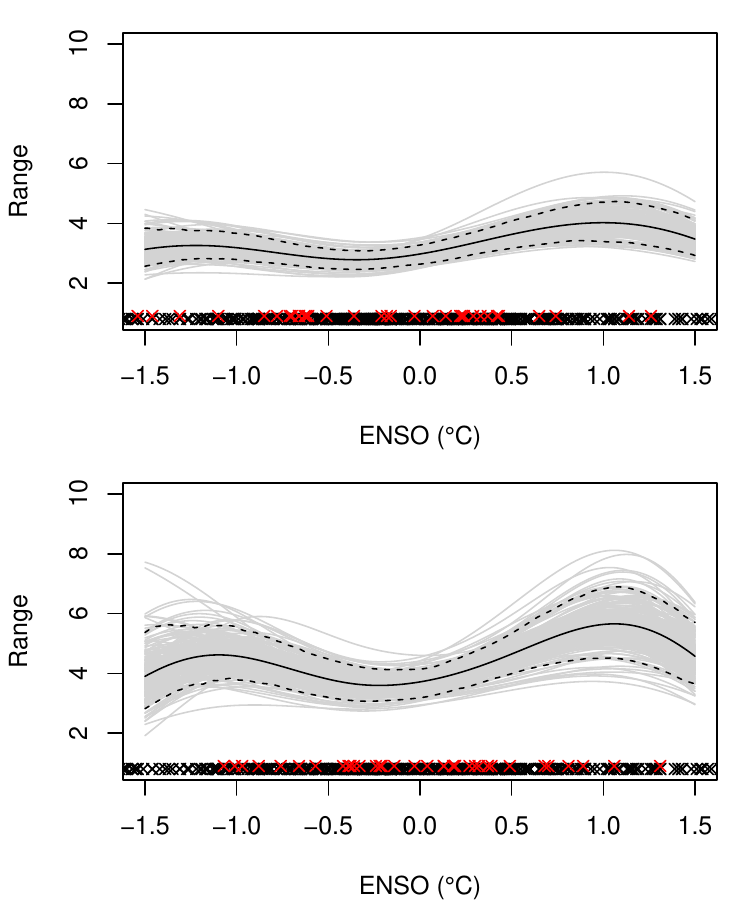}
    \end{subfigure} \\ 
        \hspace{-1cm}%
  \begin{subfigure}[b]{.33\linewidth}
    \centering
    \includegraphics[width=.99\textwidth]{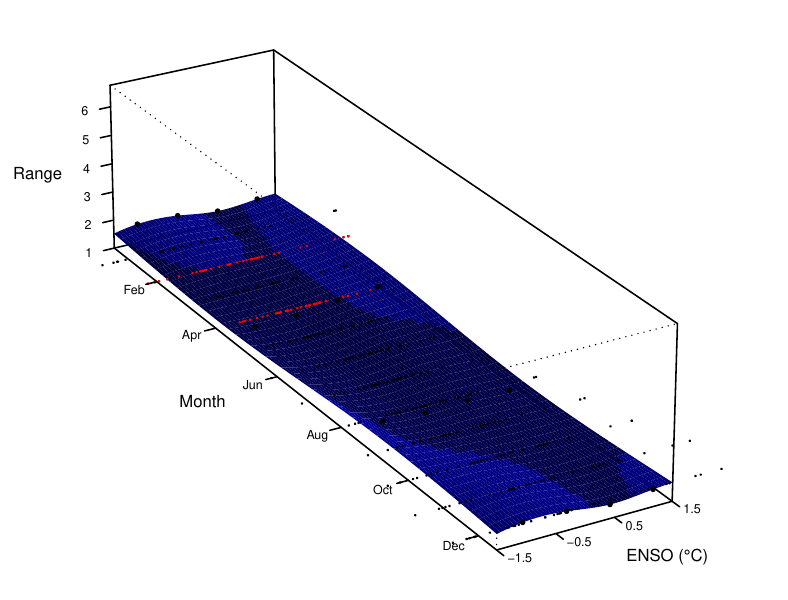}
    \end{subfigure}
    \hspace{-0.5cm}%
  \begin{subfigure}[b]{.17\linewidth}
    \centering
    \includegraphics[width=.99\textwidth]
    {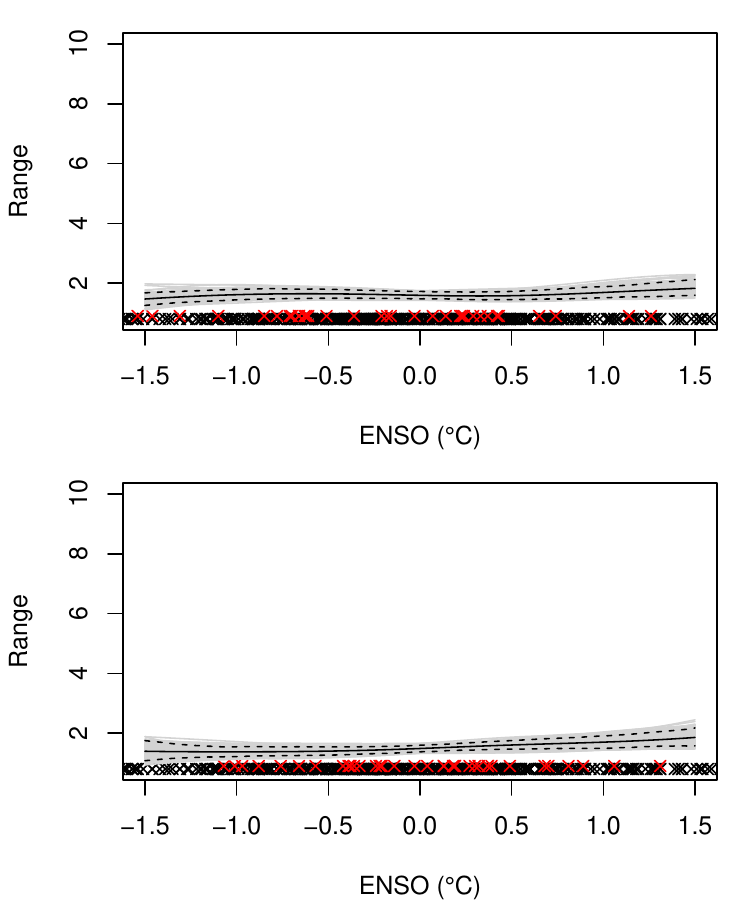} 
    \end{subfigure}
  \begin{subfigure}[b]{.33\linewidth}
    \centering
    \includegraphics[width=.99\textwidth]{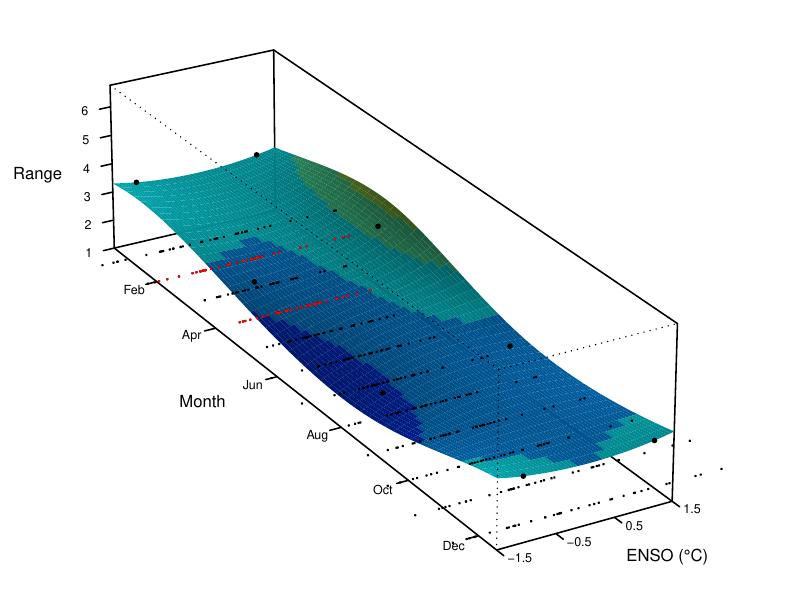}
    \end{subfigure}
    \hspace{-0.5cm}
  \begin{subfigure}[b]{.17\linewidth}
    \centering
    \includegraphics[width=.99\textwidth]
    {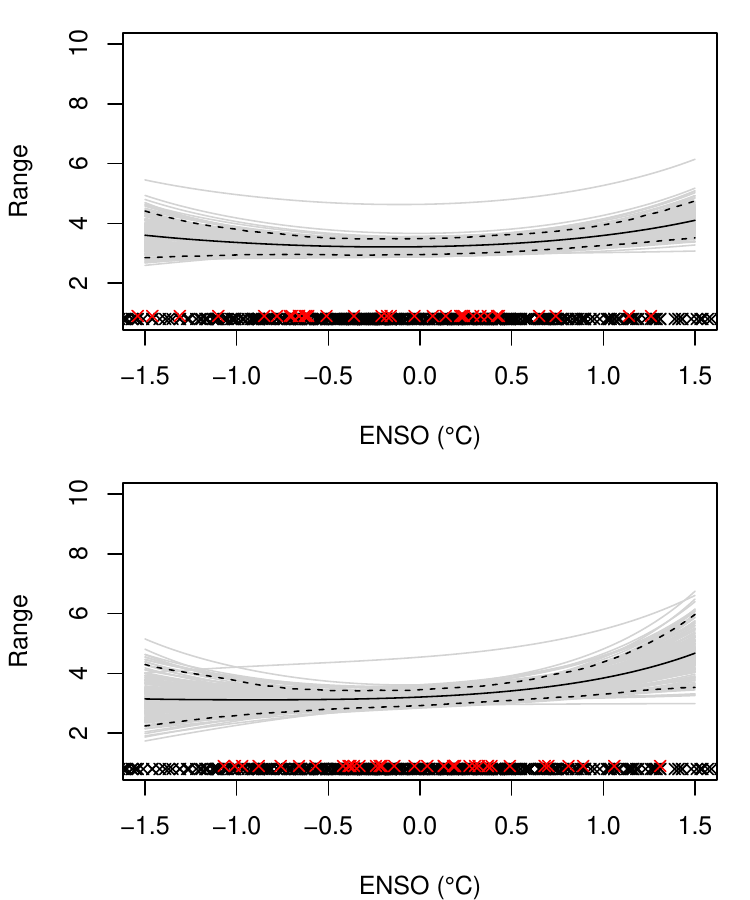}
    \end{subfigure} \\
\caption{\REVV{Trend surfaces for the estimated range parameter $\hat{\rho}$ of the model for PROD in Regions 1 (bottom left), 2 (top left), 3 (top right) and 4 (bottom right). The right-hand panels show slices of the surface in February and April (black) with the values of $\exp\{2\log(\hat{\rho})-\log(\hat{\rho}^{\star}_b)\}$ $(b=1,\dots,200)$ (grey), where $\hat{\rho}^{\star}_b$ is the $b$-th bootstrap estimate, and the $90\%$ bootstrap pointwise confidence limits (dashed). On all plots, the rug represents the ENSO phases for February and April (red) and other months (black). The larger black dots in the left-hand panels indicate the knots.} } 
\label{fig:trend_surface:prod}
\end{figure}

\begin{figure}[t!]
\centering
    \hspace{-1cm}%
  \begin{subfigure}[b]{.33\linewidth}
    \centering
    \includegraphics[width=.99\textwidth]{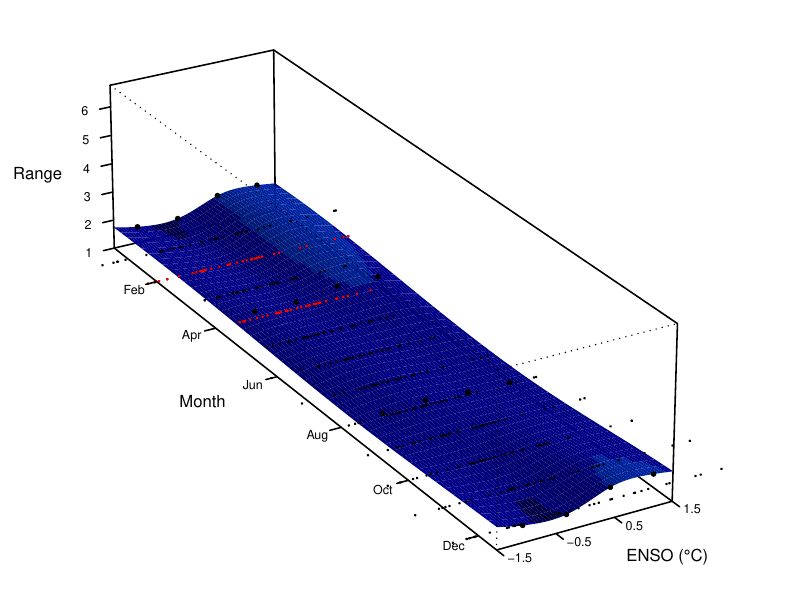}
    \end{subfigure}
    \hspace{-0.5cm}%
  \begin{subfigure}[b]{.17\linewidth}
    \centering
    \includegraphics[width=.99\textwidth]
    {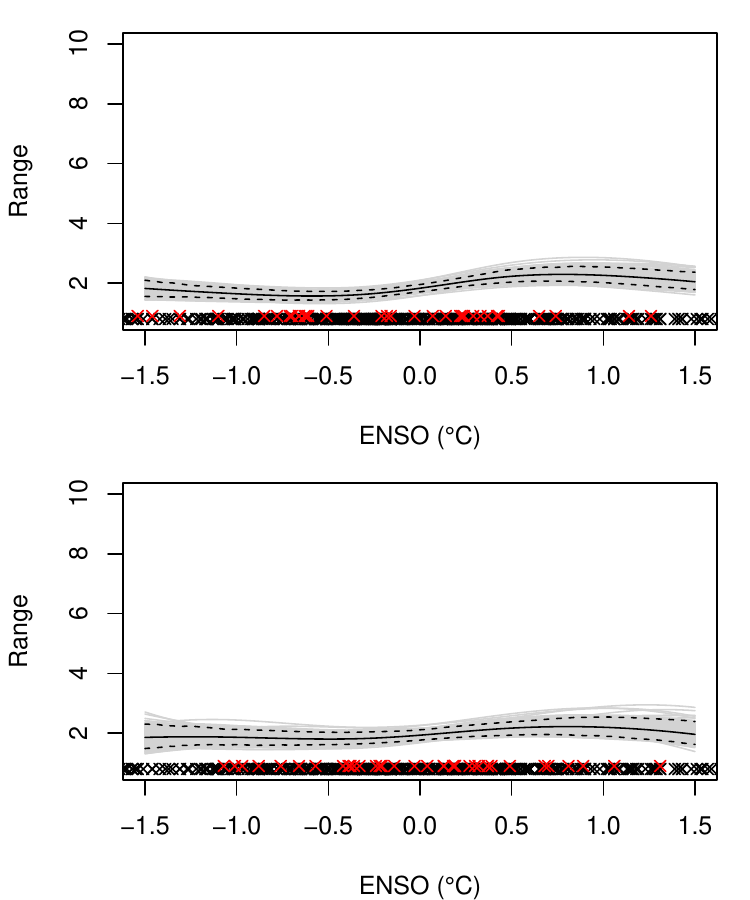} 
    \end{subfigure}
  \begin{subfigure}[b]{.33\linewidth}
    \centering
    \includegraphics[width=.99\textwidth]{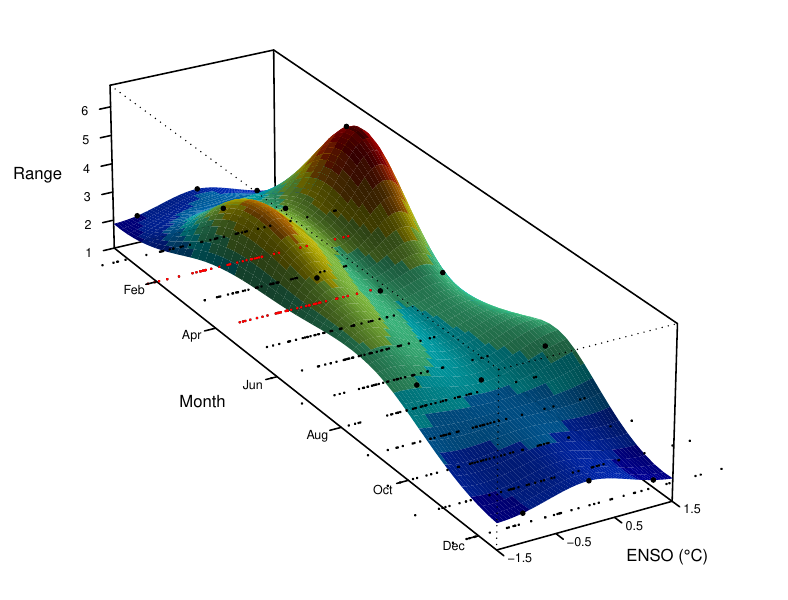}
    \end{subfigure}
    \hspace{-0.5cm}
  \begin{subfigure}[b]{.17\linewidth}
    \centering
    \includegraphics[width=.99\textwidth]
    {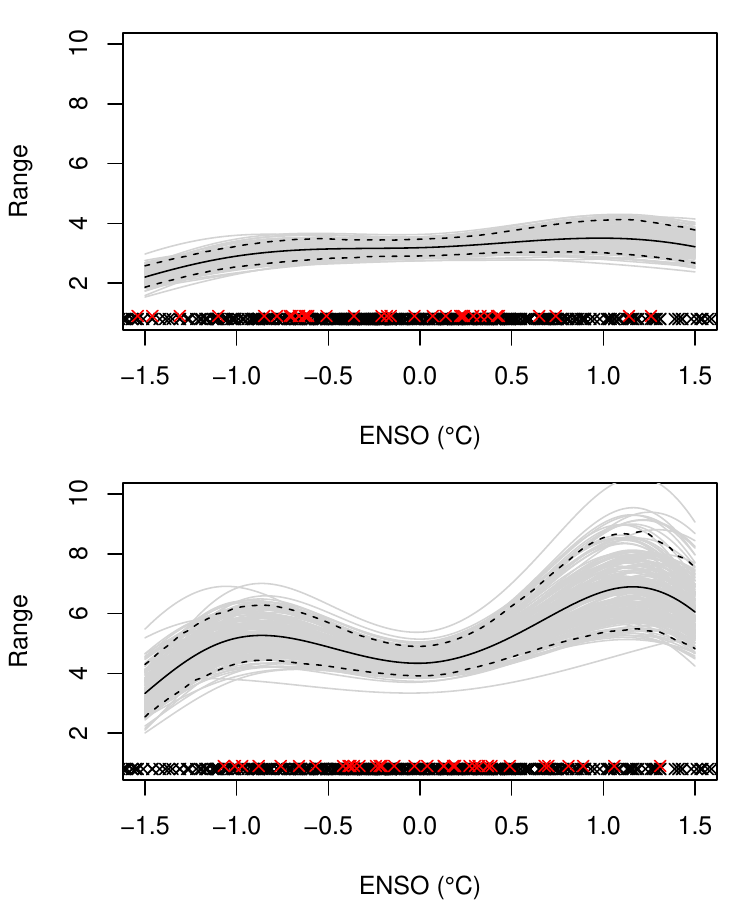}
    \end{subfigure} \\ 
        \hspace{-1cm}%
  \begin{subfigure}[b]{.33\linewidth}
    \centering
    \includegraphics[width=.99\textwidth]{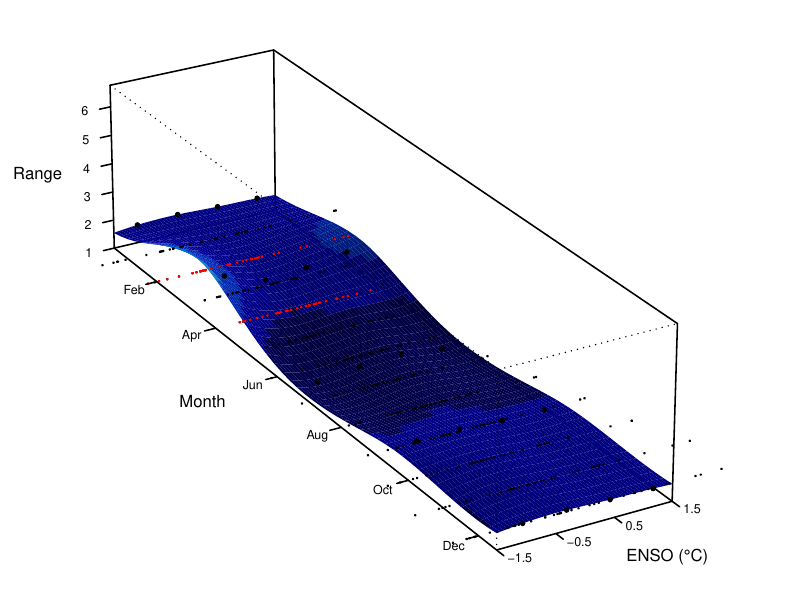}
    \end{subfigure}
    \hspace{-0.5cm}%
  \begin{subfigure}[b]{.17\linewidth}
    \centering
    \includegraphics[width=.99\textwidth]
    {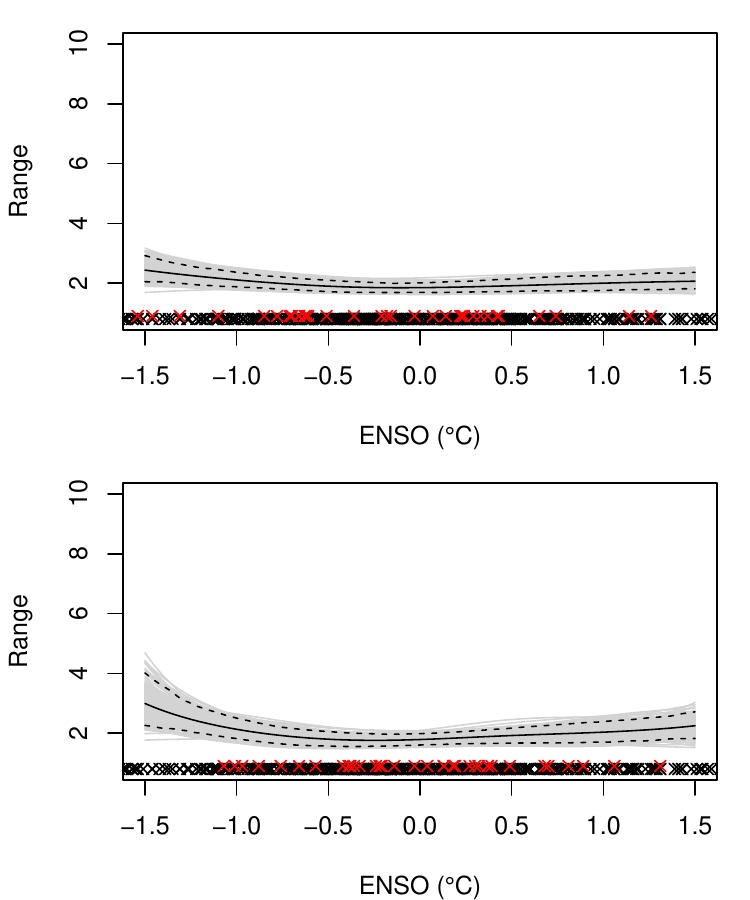} 
    \end{subfigure}
  \begin{subfigure}[b]{.33\linewidth}
    \centering
    \includegraphics[width=.99\textwidth]{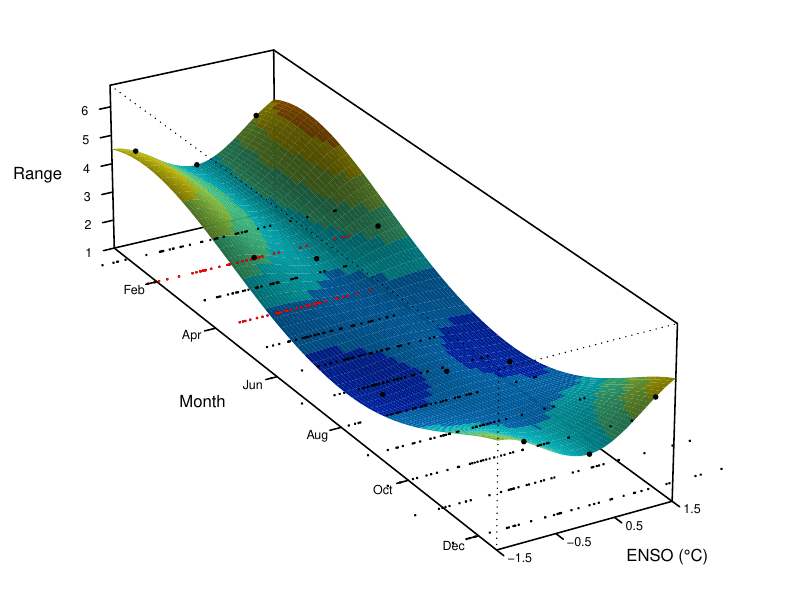}
    \end{subfigure}
    \hspace{-0.5cm}
  \begin{subfigure}[b]{.17\linewidth}
    \centering
    \includegraphics[width=.99\textwidth]
    {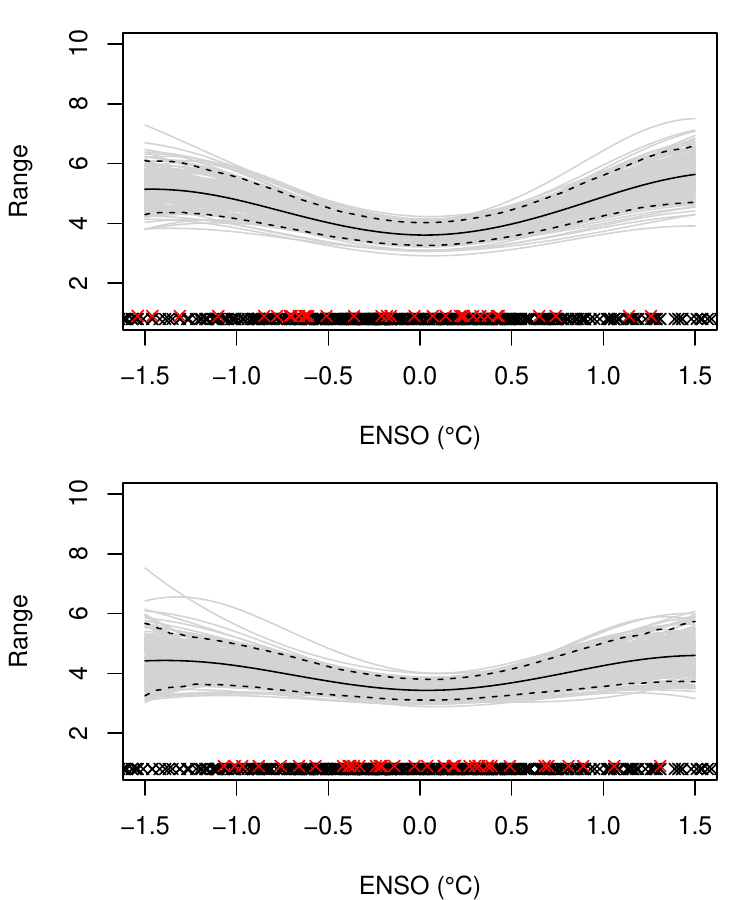}
    \end{subfigure} \\
\caption{\REV{Same as in Figure \ref{fig:trend_surface:prod}, but for CAPE.} }
\label{fig:trend_surface:cape}
\end{figure}

\begin{figure}[t!]
\centering
    \hspace{-1cm}%
  \begin{subfigure}[b]{.33\linewidth}
    \centering
    \includegraphics[width=.99\textwidth]{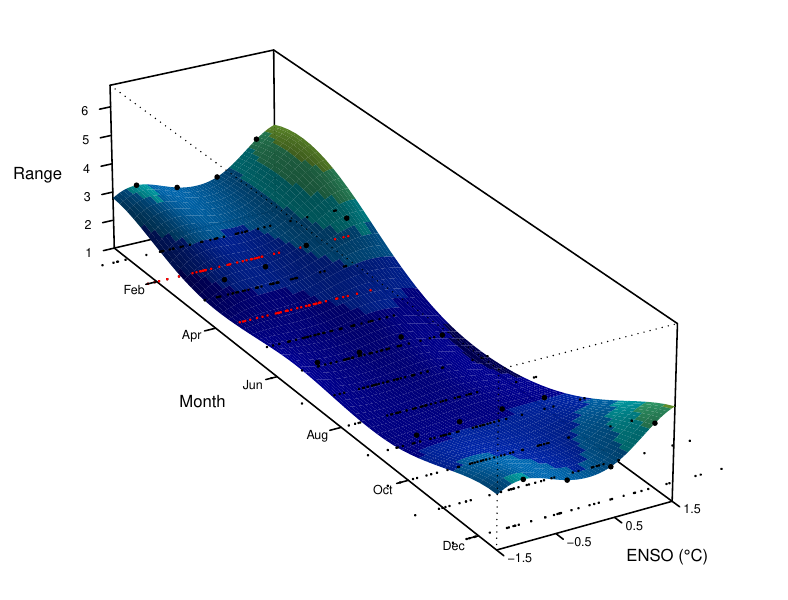}
    \end{subfigure}
    \hspace{-0.5cm}%
  \begin{subfigure}[b]{.17\linewidth}
    \centering
    \includegraphics[width=.99\textwidth]
    {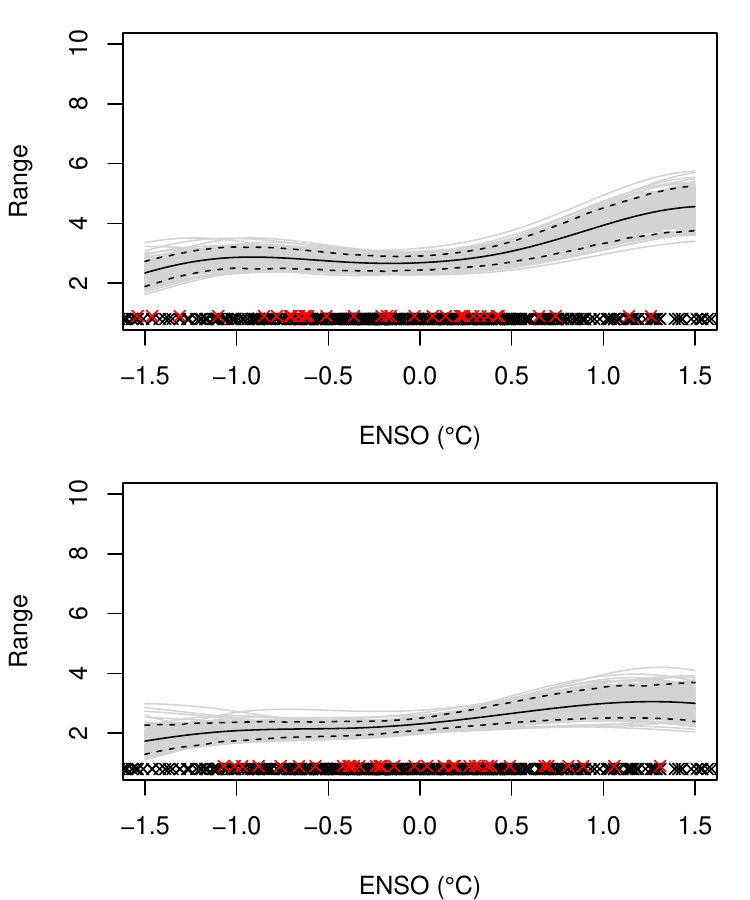} 
    \end{subfigure}
  \begin{subfigure}[b]{.33\linewidth}
    \centering
    \includegraphics[width=.99\textwidth]{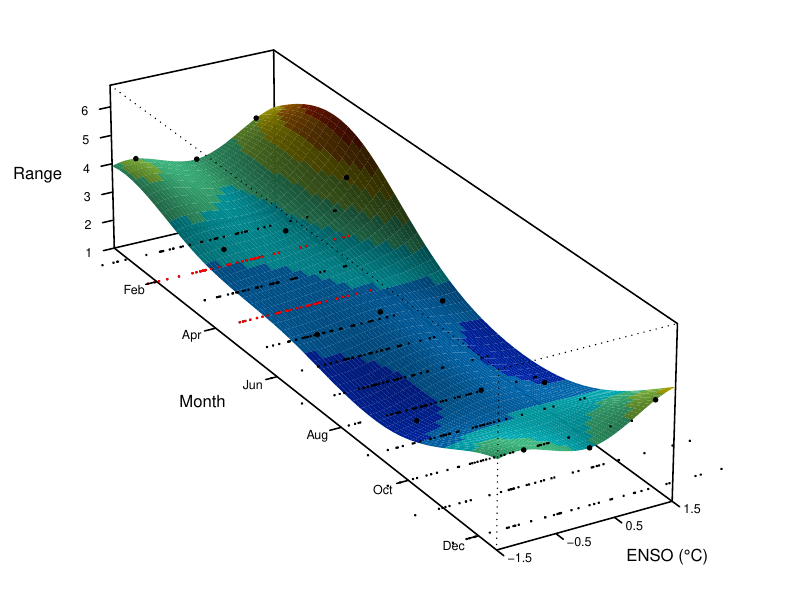}
    \end{subfigure}
    \hspace{-0.5cm}
  \begin{subfigure}[b]{.17\linewidth}
    \centering
    \includegraphics[width=.99\textwidth]
    {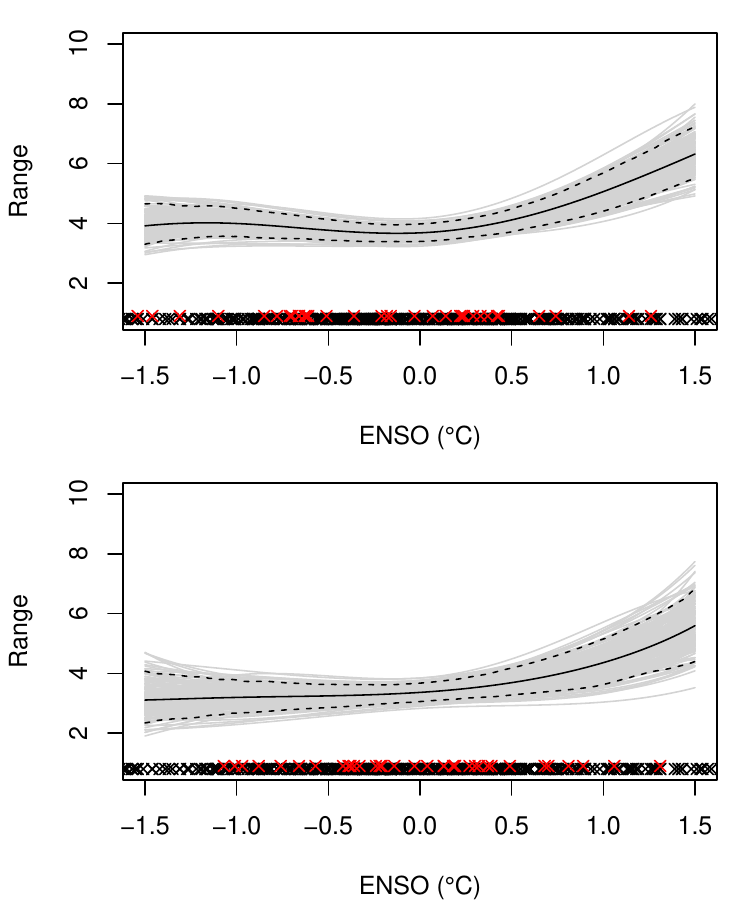}
    \end{subfigure} \\ 
        \hspace{-1cm}%
  \begin{subfigure}[b]{.33\linewidth}
    \centering
    \includegraphics[width=.99\textwidth]{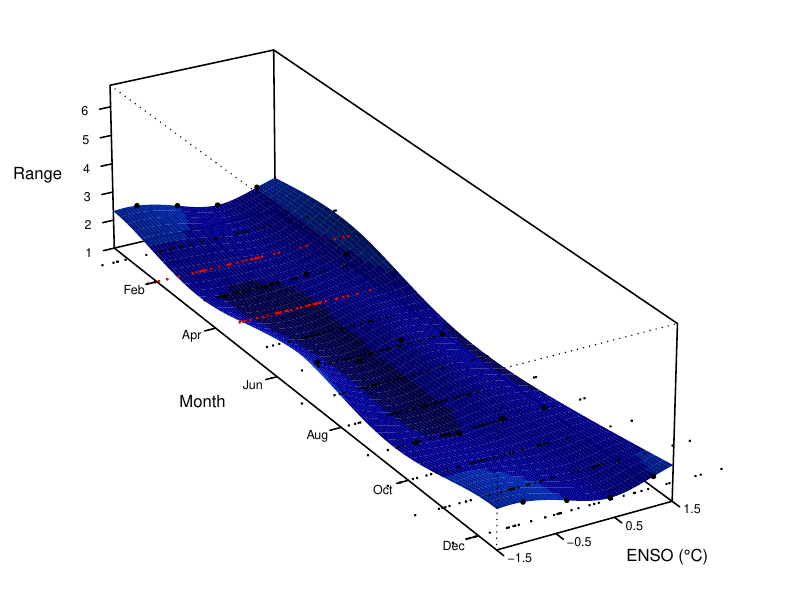}
    \end{subfigure}
    \hspace{-0.5cm}%
  \begin{subfigure}[b]{.17\linewidth}
    \centering
    \includegraphics[width=.99\textwidth]
    {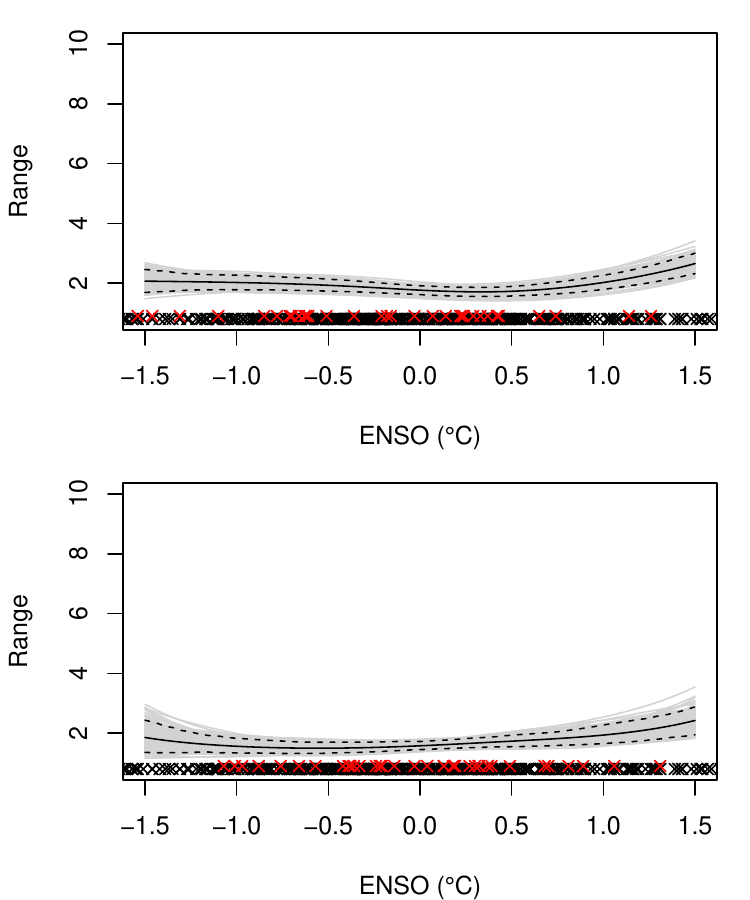} 
    \end{subfigure}
  \begin{subfigure}[b]{.33\linewidth}
    \centering
    \includegraphics[width=.99\textwidth]{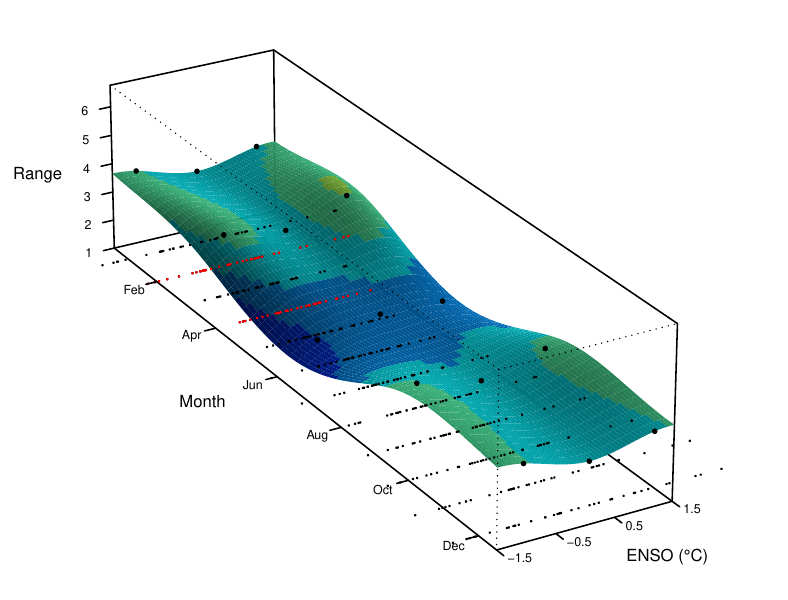}
    \end{subfigure}
    \hspace{-0.5cm}
  \begin{subfigure}[b]{.17\linewidth}
    \centering
    \includegraphics[width=.99\textwidth]
    {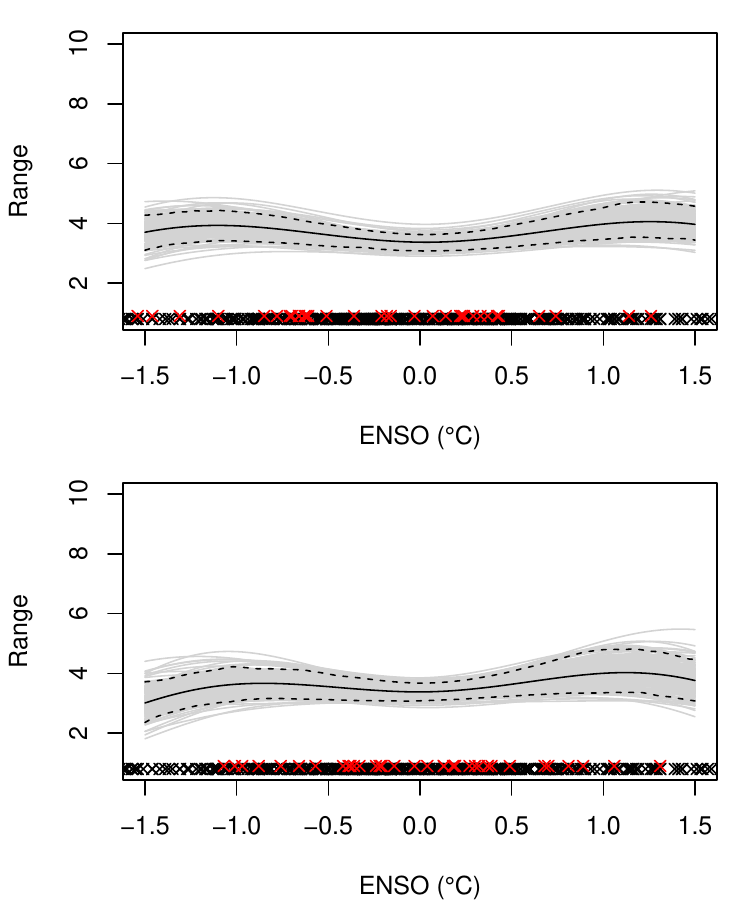}
    \end{subfigure} \\

\caption{\REV{Same as in Figure \ref{fig:trend_surface:prod}, but for SRH.} }
\label{fig:trend_surface:srh}
\end{figure}

\REV{We now discuss the range parameter $\rho$. Figures~\ref{fig:trend_surface:prod}, \ref{fig:trend_surface:cape} and~\ref{fig:trend_surface:srh} show that in nearly all regions} $\rho$ is lowest for all variables in July--September and highest in December--May. \REV{An exception is CAPE in R3, where the lowest $\rho$ is observed in winter; this is unexpected but has no practical implications because CAPE is rather low over that region in winter.}


\REV{The variation of $\rho$ with ENSO is more complex. The main effects occur in late winter and spring so, unless otherwise stated, the statements in this paragraph concern these periods. For PROD, in R3, high absolute values of ENSO are associated with much higher values of $\rho$, the clearest signal being during EN. In R4, increased ENSO is linked to an increase in $\rho$, whereas nothing significant is seen for R1 and R2. Regarding CAPE, in R1, high absolute values of ENSO are associated with \REVV{marginally} larger values of $\rho$, the signal being \REVV{stronger} during LN. Over R2, a slight increase of $\rho$ is noted in winter during EN periods.
In R3 \REVV{and for spring}, large absolute values of ENSO are associated with substantially increased $\rho$, and this effect is accentuated during EN. The same is seen in R4, but rather for winter than spring. For SRH, over R1 \REVV{and R4}, the range parameter increases slightly with the absolute value of ENSO. In R3, \REVV{low} and substantial
increases happen during \REVV{LN and EN}, respectively, and similarly for R2, though the \REVV{EN} effects are more visible in winter. \REVV{To summarize, $\rho$ increases significantly during EN (i) for PROD in R3 and R4 (ii) for CAPE in R3 and R4 (iii) for SRH in \REVV{R2 and} R3. It increases significantly during LN (i) for PROD in R3 (ii) for CAPE in R3 and R4.}
Owing to the interpretation of the range~$\rho$ in Section~\ref{Subsec_MaxStableFields}, the seasonal and ENSO-related variations of $\rho$ for the three variables can be interpreted to some degree as variations of the spatial extent of their extremes.}




For any value of ENSO and each month, we can compute the bivariate extremal coefficient of our model by combining~\eqref{Eq_ExtrCoeffBRField} and~\eqref{Eq_VariogramFinalModelCaseStudy}.  \REV{Figures~\ref{fig:model:extcoeff:prod}--\ref{fig:model:extcoeff:srh} (Supplementary Material) show} the modeled extremal coefficients obtained from the $200$ bootstrap replicates for grid points that are \REVV{$100$ km apart in the horizontal direction}. For $\mathrm{ENSO} = 0^\circ$C, the estimates are \REV{rather stable across months, though they are generally higher in summer. The seasonal variation is more pronounced when $\mathrm{ENSO}=-1^\circ$C and $\mathrm{ENSO}=1^\circ$C, especially for PROD in \REVV{R3}, CAPE in R1, R3--4, and SRH in \REVV{R2--3}. For SRH in these regions, for instance, the estimated extremal coefficient during EN and LN events in July is appreciably higher than in November--April. There is a clear decrease of the extremal coefficient (related to an increase of the range parameter) \REVV{in late winter and spring during EN years.}}

We assessed our model's \REVV{out-sample} performance 
\REVV{on three randomly chosen months-region-variable combinations for which only the validation grid points have been considered. In each case, we pooled together two consecutive months to have enough data to estimate the empirical extremal coefficients.}  
Figure~\ref{fig:extcoeff_models} \REVV{(Supplementary Material)} shows that the theoretical pairwise extremal coefficients computed from our model agree with the empirical ones. For \REVV{PROD} and SRH in the chosen \REVV{month-region} combinations, higher values of ENSO are associated with lower empirical extremal coefficients than when ENSO $\approx 0^{\circ}$C, which is not the case for \REVV{CAPE} in the chosen \REVV{month-region in the middle panel}. \REVV{Our model captures these various effects well (for the middle panel, the gap between the two theoretical curves is not significant as associated confidence bounds overlap), and has satisfactory overall out-sample performance.} The few departures between our model and the data probably stem from the \REV{spatially-constant extremal dependence in our model within each region, though some spatial heterogeneity remains in each case.} 
The fact that the models' extremal coefficient is computed for particular values of  distance, \REVV{month} and ENSO, whereas the boxplots of Figure~\ref{fig:extcoeff_models} are built using a range of distances, \REVV{months and ENSO phases}, may also contribute to the departures. 

\subsection{Meteorological interpretation \REV{and implications}}
\label{Subsec_MeteorologicalExpl}



\REV{Extremes of PROD, CAPE and SRH are generally  more localized in summer than in winter because lower baroclinicity leads to weaker jet stream-related circulation  and thus to a greater importance of local processes and feedbacks. }

\REV{ENSO is known to modulate the mean latitudinal position of the subtropical jet stream across North America \citep[e.g.,][]{cook2008relation}. During the EN phase, the associated changes in the tropics (especially in terms of convection) trigger a southward shift and amplify the subtropical jet stream relative to the \REVV{neutral (N)} and LN phases. This has direct implications for SRH, as the highest values of vertical shear are typically encountered on the path of the jet stream. R2 and R3 lie partially on that path during the N and LN phases, but during the EN phase it is much further south; see Figures 4--6 in \cite{cook2008relation}. 
During EN, we thus expect lower SRH values  for these regions (consistent with what was obtained by \cite{koch2020trendPROD}) but larger homogeneous spatial patterns compared to N and LN phases.  During the latter phases, some portions of R2 and R3 are on the jet path and some are not, leading to smaller patterns than if the jet was present or absent throughout each region. This agrees with the higher range parameter during EN reported for R2 and R3 in Section~\ref{subsec_results}; see Figure~\ref{fig:trend_surface:srh}. As shown by Figures 4--6 of \cite{cook2008relation}, some areas of R1 and R4 lie on the jet path and some lie outside it for all ENSO phases (LN, N and EN). Thus, the size of homogeneous spatial patterns should not vary much with ENSO, consistent with our findings in Section~\ref{subsec_results} (see \REVV{Figure~\ref{fig:trend_surface:srh}}).}

\REV{The meteorological mechanisms linking ENSO and the spatial extent of CAPE events are less explicit than for SRH, since high CAPE values are only indirectly related to the position of the jet. Areas of high CAPE are directly determined by large-scale weather systems such as low-pressure systems and fronts, which themselves are partly driven by the jet stream's position and intensity. For example, synoptic conditions conducive to advection in the lower troposphere of warm and moist air from the Gulf of Mexico  in the presence of rather cold and dry air in the upper troposphere will typically trigger high CAPE. Through their impacts on the jet stream and other potential consequences, variations of ENSO affect the spatial extent of CAPE extremes, but a more detailed meteorological explanation would require deep probing to shed further light on the very complex physical mechanisms involved.}

The evolution of the dependence structure of PROD with respect to ENSO \REVV{mainly} stems from those of CAPE and SRH. 

The link between ENSO and the spatial extent of weather phenomena has received relatively little attention from researchers; see, e.g., \cite{lyon2004strength} and \cite{lyon2005enso} on drought and rainfall extremes.
\REV{In spring and summer and when PROD is high enough for thunderstorms to develop, $\rho$ might be viewed very indirectly as a characteristic dimension of thunderstorm systems (including supercells, multi-cell storms, squall lines, and mesoscale convective complexes). Similarly, if SRH is high enough for cyclones \REVV{to} form in winter and spring, then $\rho$ may be obliquely considered to be a characteristic dimension of cyclones. But care is needed when inferring features of \REVV{thunderstorms or cyclonic storms} from conclusions about their proxies. \cite{koch2020trendPROD} found that the maxima of PROD, CAPE and SRH tend to be larger during LN in late winter and spring. Combined with this, our findings of Section~\ref{subsec_results} entail that, in spring, thunderstorm systems may be larger in the North-East during LN than the N phase, but  the situation for the North-East and South-East during EN remains unclear. Similarly, we cannot draw conclusions about an increase of the extent of cyclonic storms in R2 and R3 (North) during EN, as such a phase is associated with a decrease of SRH values linked to an absence of the jet stream. 
}



\section{Discussion}
\label{Sec_Discussion}

In this paper we propose stochastic models that use covariates such as large-scale atmospheric signals to capture the temporal non-stationarity of the extremal spatial dependence of \REV{convective available potential energy (CAPE), storm relative helicity (SRH) and $\mathrm{PROD}=\sqrt{\mathrm{CAPE}} \times \mathrm{SRH}$, variables that are} 
associated with severe US thunderstorms. We use a fractional Brown--Resnick field whose range parameter depends on \REV{the El Ni\~no-Southern Oscillation (ENSO)} and month to assess how these variables affect the spatial extents of these phenomena.


One novel methodological contribution is a max-stability test, based on empirical likelihood and the bootstrap, that accounts for unknown margins and provides a further diagnostic to assess whether a sub-asymptotic model should replace a max-stable \REV{model. Another,} a bootstrap-based estimator of the composite Kullback--Leibler divergence, enables better model selection than does the CLIC, especially when one first estimates marginal distributions and then uses composite likelihood separately to estimate a dependence structure. 
 
 
Max-stability appears to be an appropriate assumption for our data. 
\REV{Our findings, combined with those of \citet{koch2020trendPROD}, suggest that, in the North-East and in spring, thunderstorm systems during \REVV{La Ni\~na} may be more intense and  larger than during neutral ENSO phases or \REVV{El Ni\~no} events. Drawing conclusions about cyclones and thunderstorms for other regions is difficult, because translating our findings on their proxies requires caution \REVV{and because the ENSO effects on the marginal levels and spatial extent often differ}. Moreover, the reliability of our conclusions depends on the quality of the reanalysis data, in which vertical wind shear (and thus SRH) is typically well-represented, but thermodynamic variables like CAPE are less reliable \citep[see][and references therein]{taszarek2021global}.}

Our model accounts for variation of extremal dependence across time but not over space, a drawback for regions with heterogeneous weather influences. \REV{We partially overcome this  by applying our model to regions that are rather homogeneous in terms of climate and weather drivers. Allowing smooth spatial variation in the dependence structure, such as in \citet{Huser2016nonstat} or \citet{Zhong.2022}, could be useful, especially in applications where we are agnostic about the different climatic regions of the domain considered. A local likelihood approach such as that of  \citet{hector2022distributed} could be used for the fitting. Studying the effects of other atmospheric signals, such as the North Atlantic Oscillation and the Madden--Julian Oscillation, would also be worthwhile, but would require suitable data. Finally, introducing time as a covariate could help in detecting climate-change related influences ``orthogonal'' to those of the atmospheric signals included in the model.}

Using our model for risk assessment or forecasting would require simulated vectors of covariates $\bm{x}_t$ ($t=1, \ldots, T$), entailing treating them as realizations of random vectors $\bm{X}_t$. Point 2 in the definition of our general model in Section~\ref{Subsubsec_GeneralVersion} would then be: the spatial fields $\{ Z(\bm{s}, 1) : \bm{s} \in \mathcal{X}\}, \ldots, \{ Z(\bm{s}, T) : \bm{s} \in \mathcal{X}\}$ are independent conditionally on $\bm{X}_1,\ldots, \bm{X}_T$. Unconditionally, any temporal dependence would be driven by the dynamics of the random process $\{ \bm{X}_t: t =1, \ldots, T \}$, but without specific assumptions on that dynamics, the resulting space-time model $\{ Z(\bm{s}, t) : \bm{s} \in \mathcal{X}, t  = 1, \ldots, T \}$ 
will not be a space-time max-stable field,
unlike models developed by \cite{davis2013max}, \cite{huser2014space} or \cite{embrechts2016space}.


\spacingset{0.97}

\bibliographystyle{agsm_jon.bst}


\bibliography{Biblio_Final.bib}       

\newpage

\spacingset{1.7}
\bigskip
\begin{center}
{\large\bf SUPPLEMENTARY MATERIAL}
\end{center}

\section{Supplement}

\subsection{\REV{Estimation of max-stable fields} }\label{sec:composite}

The $D$-dimensional multivariate density of a max-stable random field can be intractable, as the exponent measure in~\eqref{Eq_MultivariateDfMaxStable} can be difficult to characterize unless $D$ is small and the exponential leads to a combinatorial explosion of the number of terms in the density. Full likelihood inference is thus typically out of reach and so composite likelihood techniques \citep[e.g.,][]{Varin2011} have been extensively used. Pairwise composite likelihoods are most common \citep[e.g.,][]{padoan2010likelihood, blanchet2011, davison2012}, but higher order composite likelihoods have also been investigated \citep[e.g.,][]{HuserDavison,Castruccio2016}. Under mild regularity conditions, the maximum pairwise likelihood estimator is strongly consistent and asymptotically normal, with a larger variance than the maximum likelihood estimator. \cite{padoan2010likelihood} and \cite{sang2014} showed that truncating the pairwise likelihood by ignoring pairs of sites that are far apart can improve its statistical efficiency; for similar findings in other settings, see the references in \cite{sang2014}. Ignoring some pairs also decreases the computational burden, which is valuable for large values of $D$, such as our $D=619$. \cite{Castruccio2016} showed that truncation increases statistical efficiency by more for pairwise or triplewise likelihoods than for higher order composite likelihoods.  

Let $z_{\bm{s},t}$ denote the maximum at $\bm{s}$ during the $t$-th period, transformed marginally to standard Fréchet, let $\bm{\psi}$ denote the vector of dependence parameters of the max-stable model, and let $f_{\bm{s}_d, \bm{s}_{d'}; \bm{\psi}}$ denote the corresponding pairwise density for grid points $\bm{s}_{d}, \bm{s}_{d'} \in \mathcal{X}$.
The truncated pairwise log-likelihood is 
\begin{equation}
\label{Eq_TruncLogLik}
    l(\boldsymbol{\psi}) = \sum_{t=1}^{T} \sum_{d = 1}^{D-1} \sum_{d'=d+1}^{D} \mathbb{I}_{ \left \{ || \bm s_d -  \bm s_{d'} ||\leq \sqrt{2  c^2} \right \}} \log  f_{\bm{s}_d, \bm{s}_{d'}; \bm{\psi}}(z_{\bm{s}_d,t}, z_{\bm{s}_{d'},t}),
\end{equation}
where, \REVV{in this formula, the distance corresponds to the Euclidean distance in the longitude-latitude coordinate system, $c$ is the number of neighbours we want to consider in the longitude direction,}
and $\mathbb{I}_{\{\cdot\}}$ is the indicator function. Choosing truncation distance $\sqrt{2  c^2}$ rather than $c$ allows us to take \REV{all first adjacent neighbours into account, including those on the diagonals.} 

\begin{figure}[!ht]
\centering
  \begin{subfigure}[b]{.92\linewidth}
    \centering
    \includegraphics[width=.99\textwidth]{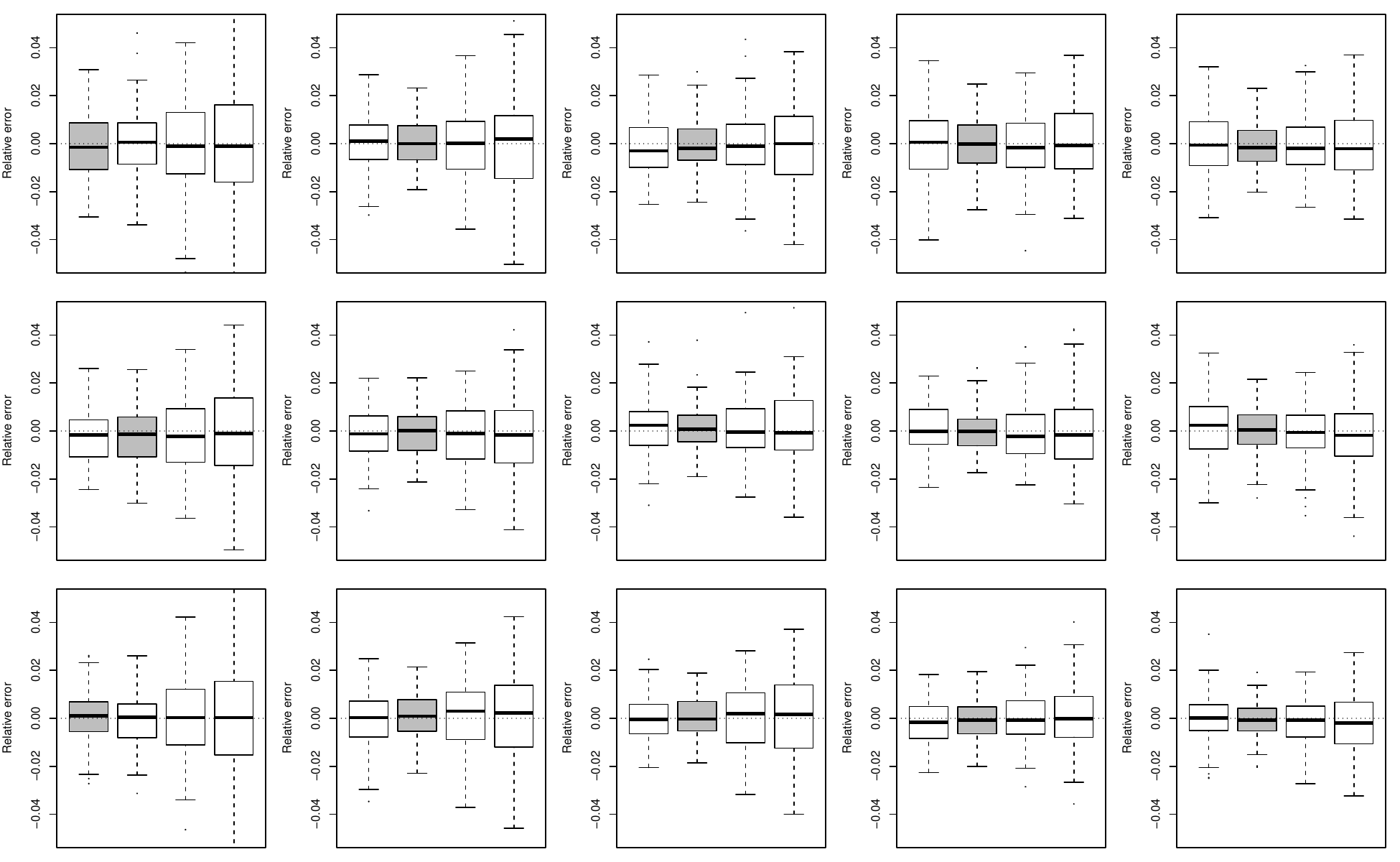}
  \end{subfigure}
  \caption{Boxplots of the relative errors of the estimates of \REVV{$\alpha$} for $c=1, 2, 3, 4$ (from left to right). The boxplot associated with the lowest relative root mean squared error is highlighted in grey. The rows correspond to $D=121$ (top), $144$ (middle), $169$ (bottom) and each column corresponds to a value of $\rho$: $1, 1.5, 2, 3, 4$ (from left to right).} \label{fig:taper_simulation}
\end{figure}

We chose a value of $c$ adapted to the present context by simulating Brown--Resnick fields having semivariogram~\eqref{Eq_ExpressionAnisotropicVariogram} with $r=1$, $\kappa=0$, $\alpha=1$ and \REVV{$\rho \in \{1, 1.5, 2, 3, 4 \}$}, on squares containing \REVV{$121$, $144$, and $169$ }grid points. In each of these $15$ settings, we simulated $444$ (the number of months in our data) independent realizations of the field $400$ times independently. For each of these $400$ experiments, we estimated $\rho$ using the truncated pairwise log-likelihood~\eqref{Eq_TruncLogLik} with $c \in \{ 1, 2, 3, 4 \}$. The smoothness parameter \REVV{$\alpha=1.4$} and scaling and rotation parameters $r=1$ and $\kappa=0$ are close to the estimates from our data. We let the range $\rho$ and the spatial domain vary, as the optimal truncation distance depends on both. Our results (Figure~\ref{fig:taper_simulation}; \REVV{the results for $\rho$ are similar}) show that estimation becomes more precise when $\rho$ decreases and $D$ increases.  In most settings $c=2$ leads to more precise estimation, and we use this value for our case study.


\newpage 
\subsection{Simulation study for Section~\ref{sec:model:case_study}}

\begin{figure}[!ht]
\centering
\begin{subfigure}{.95\linewidth}
    \includegraphics[width=.99\textwidth]{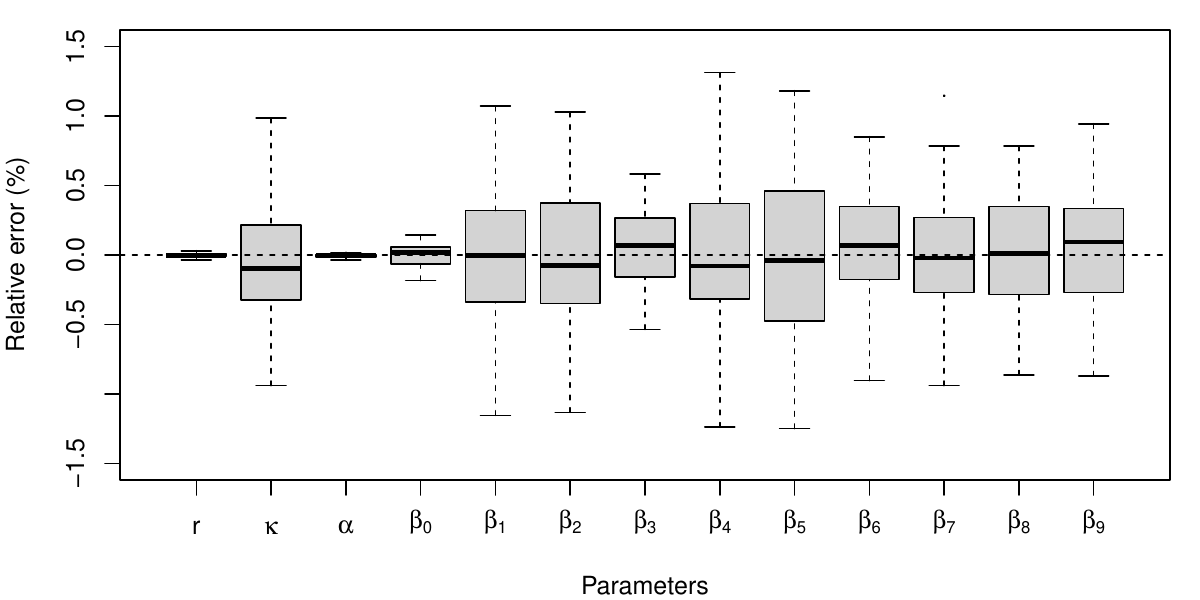}  
    \end{subfigure} 
   \caption{\REV{Relative error for the parameters in the simulation study.}}
   \label{fig:simulation_actual}
\end{figure}
\begin{figure}[!ht]
\centering
\begin{subfigure}{.95\linewidth}
   \includegraphics[width=.5\textwidth]{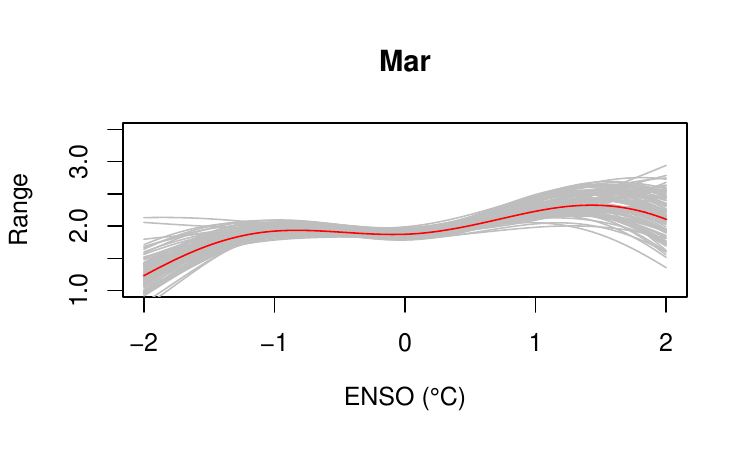} 
   \includegraphics[width=.5\textwidth]{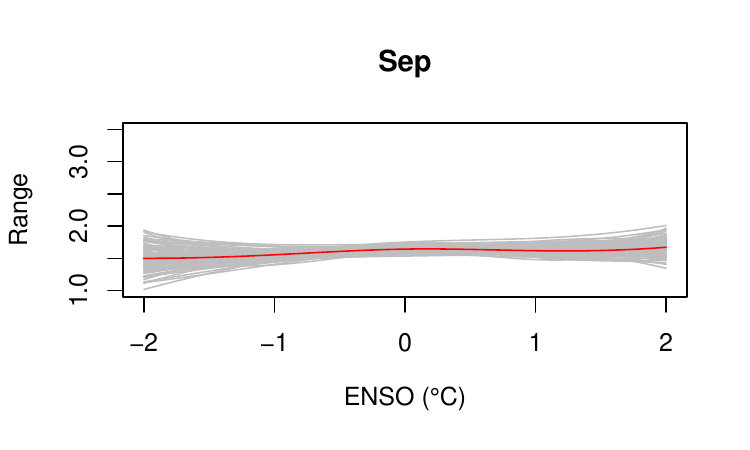} 
    \end{subfigure}
   \caption{\REV{The estimated range parameter with form given in (\ref{EqLinearBasisExpansonRange}), based on the same simulation study as that in Figure~\ref{fig:simulation_actual}, for $t$ mod $12$ = 3 (March, left) and $t$ mod $12$ = 9 (September, right).} }
   \label{fig:simulation_actual_1}
\end{figure}

We consider~\eqref{Eq_VariogramFinalModelCaseStudy} with $T=444$ on a square containing \REV{100} grid points\REV{, corresponding to the lowest number of grid points considered in our case study,} and with parameters $r = 0.72 $, $\kappa= -0.08$, and $\alpha = 1.26$, three knots  with coordinates $-1.06$, $0.05$, $1.16$ in the ENSO \REV{direction}, 
and knots at $0.5$, $4.5$, $8.5$, $12.5$ in the month direction. As we use a circular P spline basis for months, the values of the spline are the same at $0.5$ and $12.5$, giving three distinct knots in both the ENSO and month directions, i.e., nine knots over the space of covariates. The corresponding coefficients, $\beta_0=0.52$ (intercept), $\beta_1=-0.03$, $\beta_2=0.02$, $\beta_3=0.07$, $\beta_4=0.11$, $\beta_5=-0.07$, $\beta_6=-0.23$, $\beta_7=-0.03$, $\beta_8=0.02$ and $\beta_9=0.04$, correspond to those obtained when fitting this model to the data described in Section~\ref{sec:data_explore}. We estimate all the parameters for $100$ independent replicates of such data. \REV{Figure~\ref{fig:simulation_actual}} suggests that all the estimators are essentially unbiased and that all parameters are recovered well, but that the estimators of  the parameters associated with the spline basis tend to be  more variable than are the others. The smoothness parameter $\alpha$ and the scaling factor $r$ are very well estimated. \REV{Figure~\ref{fig:simulation_actual_1}} shows that signals and non-signals are equally well-identified.

\newpage 

\subsection{For Section~\ref{sec:ms_test}}

\subsubsection{Algorithms}

Let $\bm {{y}}_j = ({y}_{1,j},\dots ,{y}_{D,j})^{\prime}$ denote the $j$-th underlying \REV{(here, three-hourly)} observation, empirically transformed to be standard Fr\'echet-distributed, where $D$ is the number of grid points. Given \REV{$n_M$} underlying \REV{(here again, three-hourly)} observations \REV{over all years, i.e., $n_M=n$ (block size) $\times M$ (number of years)}, let $({\bm {y}_1}, \dots, {\bm {y}_{n_M} } )^{\prime}\in \mathbb{R}^{n_M\times D}$ denote the transformed dataset. The procedure for generating the max-stable vector is 
    \begin{enumerate}
    \item Compute $\tilde{R}_j=||y_{1,j},\dots, y_{D,j}||_1$, known as the radial coordinates, where $||\cdot||_1$ is the $L_1$ norm. Then calculate $\bm{\tilde{W}_j}= \bm{y}_{j}/\tilde{R}_j$, $j=1,\dots,n_M$, commonly called the angular coordinates. Keep those $\bm{\tilde{W}}_j$ for which $\tilde{R}_j>r_0$, with $r_0$ fixed to be the empirical $p$-quantile of $\tilde{R}_1,\dots,\tilde{R}_n$, so the number of observations retained is $n_0=(1-p)n_M$, where $0<p<1$. Let $R_i$ and $\bm{W}_i$ ($i=1,\dots,n_0$) denote the coordinates retained.
    \item Following the empirical likelihood approach of \citet{Einmahl-Segers.2009}, but extended to $D>2$, we find the estimated angular probability measure 
    $$
    \hat{Q}(\bm w) = \sum_{i=1}^{n_0} q_i \mathbb{I}_{\{\bm{W}_i<\bm w\}}, \quad \bm w\in[0,1]^D,
    $$
    where $\mathbb{I}_{\{\cdot\}}$ is the indicator function and the tilting probabilities $\{q_i\}_{i=1}^{n_0}$ satisfy
    \begin{equation}\label{eq:lagrange}
        q_i = \argmax_{q_i} \prod_{i=1}^{n_0} q_i, \text{ such that } \sum_{i=1}^{n_0} q_i = 1 \text{ and } \sum_i^{n_0} q_i \bm{W}_i/R_i = \bm D^{-1},
    \end{equation}
    where $\bm D^{-1}= (1/D, \dots, 1/D)^{\prime}$, and (\ref{eq:lagrange}) is solved using Lagrange multipliers. 
    \item With $\{q_i\}_{i=1}^{n_0}$ and $\{\bm{W}_i\}_{i=1}^{n_0}$, generate a simple max-stable vector using Algorithm \ref{alg:ms}, based on \citet{dombry.2016}. 
    
    \end{enumerate}

\begin{algorithm}[t]
\SetAlgoLined
    \KwInput{Tilting weights $\{q_i\}_{i=1}^{n_0}$ and vectors $ \{\bm W_i\}_{i=1}^{n_0}$}
\KwOutput{Max-stable $D$-dimensional vector with standard Frech\'et margins}
 Generate $E^{\star} \sim \text{Exp}(1)$ \;
    Set $R^{\star} = D/E^{\star}$\;
    Set $\bm Z = ( Z_1, \dots,  Z_D )= (0, \dots, 0) $\;
 \While{$R^{\star}> \min \{Z_1, \dots, Z_D \} $}{
     Draw $\bm W^{'}=(W_1^{'},\dots,W_D^{'})$ from the set $\{ \bm W_1, \dots, \bm W_{n_0}\}$ with sampling probabilities $\{q_1, \dots, q_{n_0}\}$\;
 \For{$j\gets1$ \KwTo $D$}{
    Set $Z_j = \max (Z_j, R^{\star} W_j^{'})$\;
    }
    Generate $E^{\star}\sim \text{Exp}(1)$\;
    Set $R^{\star}=\frac{1}{(1/R^{\star} + E^{\star}/D)}$\;
 }
 \Return{} $\bm Z$ \;
 \caption{Simulate max-stable vector using tilting weights}
 \label{alg:ms}
\end{algorithm}

\subsubsection{Simulation study}

 We first generate from a multivariate logistic extreme-value distribution 
$$
G(y_1, \ldots, y_D) = \exp \left \{-\left(\sum_{d=1}^{D} y_d^{-1/\lambda}\right)^{\lambda} \right \}, \quad y_1, \ldots, y_D >0,
$$
with dependence parameter $\lambda\in \{0.1,0.5,0.9\}$, and perform our max-stability test on these observations with $B=200$. We then repeat this experiment with data from a multivariate Gaussian distribution with common pairwise correlation $\zeta\in\{0.1,0.5,0.9,0.99\}$. To assess the empirical size and power of the test, we replicate both experiments \REV{$500$} times. Pointwise maxima of Gaussian fields converge in~\eqref{eq:maxima_convergence} to the degenerate independent max-stable field, but a block size of $240$ is insufficient for convergence, so using the multivariate normal distribution is an approximate but reasonable way to assess the power of our test.  

\begin{figure}[!t]
\centering
    \includegraphics[width=.8\textwidth]{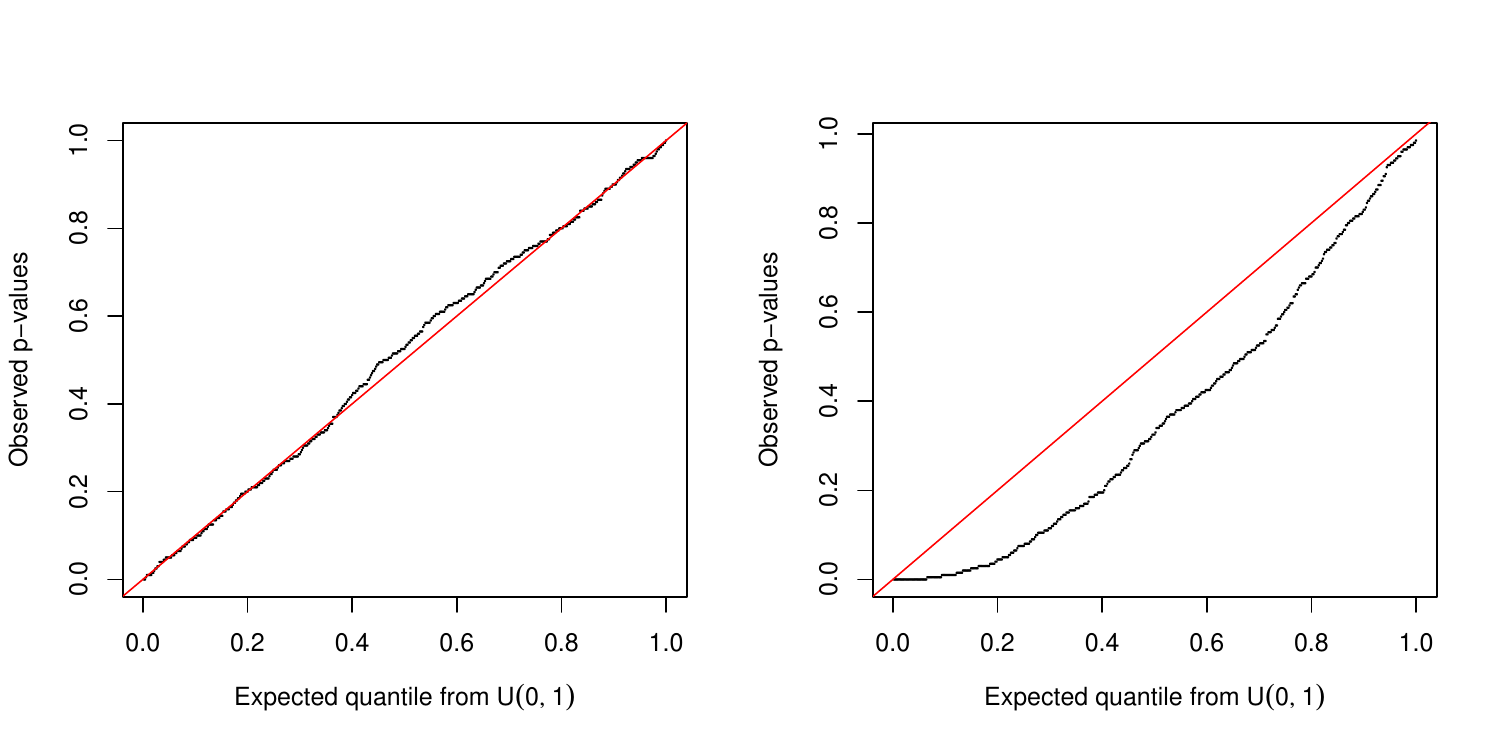} 
   \caption{Quantile-quantile plots for the $500$ p-values with the simulation setting involving the max-stable logistic, $\lambda=0.1$ (left), and normal, $\REV{\zeta}=0.9$ (right), distributions.}
   \label{fig:ms_test_1}
\end{figure}

\begin{figure}[!ht]
\centering
  \begin{subfigure}[b]{.45\linewidth}
    \centering
    \includegraphics[width=.99\textwidth]{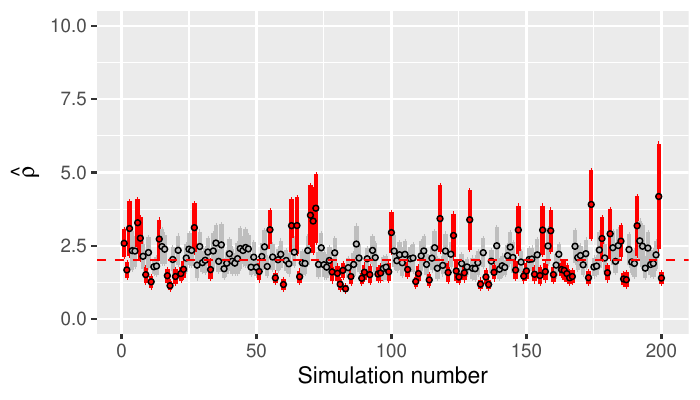}
  \end{subfigure}
  \begin{subfigure}[b]{.45\linewidth}
    \centering
    \includegraphics[width=.99\textwidth]{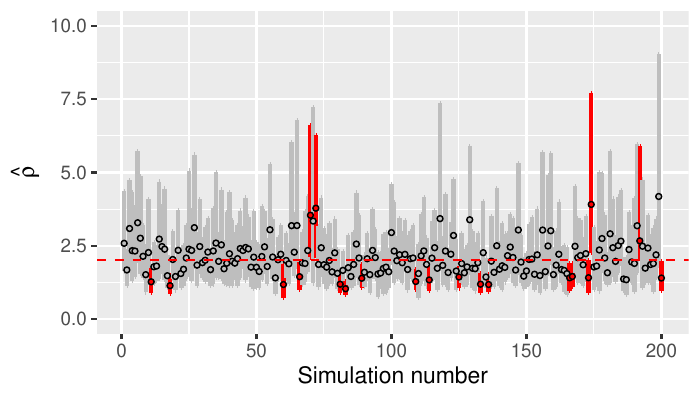}
  \end{subfigure} \\
  \begin{subfigure}[b]{.45\linewidth}
    \centering
    \includegraphics[width=.99\textwidth]{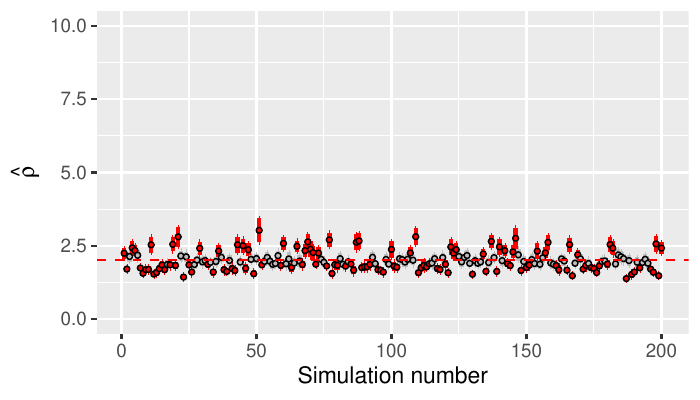}
  \end{subfigure}
  \begin{subfigure}[b]{.45\linewidth}
    \centering
    \includegraphics[width=.99\textwidth]{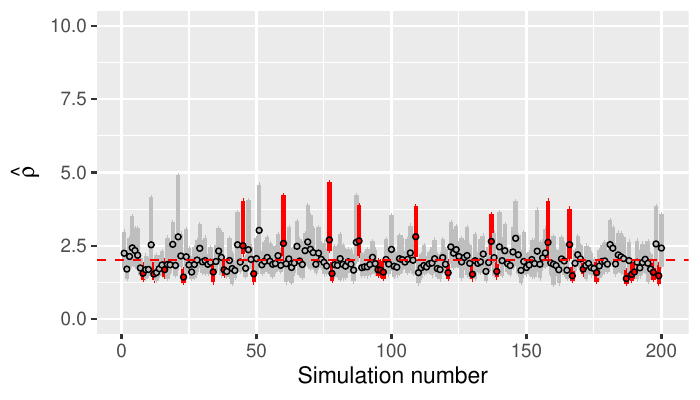}
  \end{subfigure} 
  \caption{Nominal $95\%$ sandwich (left) and bootstrap (right) confidence intervals for the range parameter $\rho$ (with logarithm as variance-stabilizing transform) in the model BR$_1$ using the two-step estimation procedure, from 200 simulations with 40 replications, $D=25$ (top) and $D=225$ (bottom). The red confidence intervals represent those that do not cover the true value represented by the red dashed line.}
  \label{fig:simulation:sandwichCI}
\end{figure}

Table~\ref{table:ms_test} shows that the empirical size of the test is controlled reasonably well in this setting, with the 500 A-D and Kolmogorov--Smirnov p-values correctly showing no departure from uniformity at the $5\%$ level for the logistic models, which are max-stable. \REV{Uniformity of the p-values for the multivariate normal model is rejected at the $5\%$ level except when $\zeta=0.1$, 
perhaps because the 
block maxima are already close to the independent max-stable distribution as this convergence becomes quicker with decreasing $\zeta$. As expected, the power rises as dependence increases. }Figure~\ref{fig:ms_test_1}, which shows quantile-quantile plots of the p-values in two simulation settings, illustrates the non-uniformity in the Gaussian case.


\begin{table}[t]
\centering
\begin{tabular}{lrrrr}
  \hline
 & $5\%$ & $20\%$ & p-val AD & p-val KS \\ 
  \hline
Max-stable logistic, $\lambda=0.1$ &  5.40 & 18.80 & 0.20 & 0.18 \\ 
Max-stable logistic, $\lambda=0.5$  & 4.80 & 19.00 & 0.71 & 0.83  \\ 
Max-stable logistic, $\lambda=0.9$ & 5.40 & 20.00 & 0.40 & 0.22  \\ 
  \hline
Normal, $\zeta=0.1$  & 4.80 & 21.40 & 0.48 & 0.65\\
Normal, $\zeta=0.5$ & 14.20 & 31.40 & 0.00 & 0.00   \\ 
Normal, $\zeta=0.9$ & 22.00 & 40.40 & 0.00 & 0.00  \\ 
   \hline
\end{tabular}
\caption{\REV{Empirical size $(\%)$ (top three lines) and power $(\%)$ (bottom three lines) for tests at the $5\%$ and $20\%$ nominal levels with $(B,p)=(200,0.9)$, for $500$ samples of size $n_M=240\times 40$, and different levels of dependence. The last two columns show the p-values for the Anderson--Darling and Kolmogorov--Smirnov tests of uniformity for the 500 p-values.} 
}
\label{table:ms_test}
\end{table}

\newpage

\subsection{For Section~\ref{sec:bootstrap}}


\newpage

\subsection{For Section \ref{Subsec_sub-region}}

\begin{figure}[t!]
\centering
    \begin{subfigure}[b]{.33\linewidth}
    \centering
    \includegraphics[width=.99\textwidth]{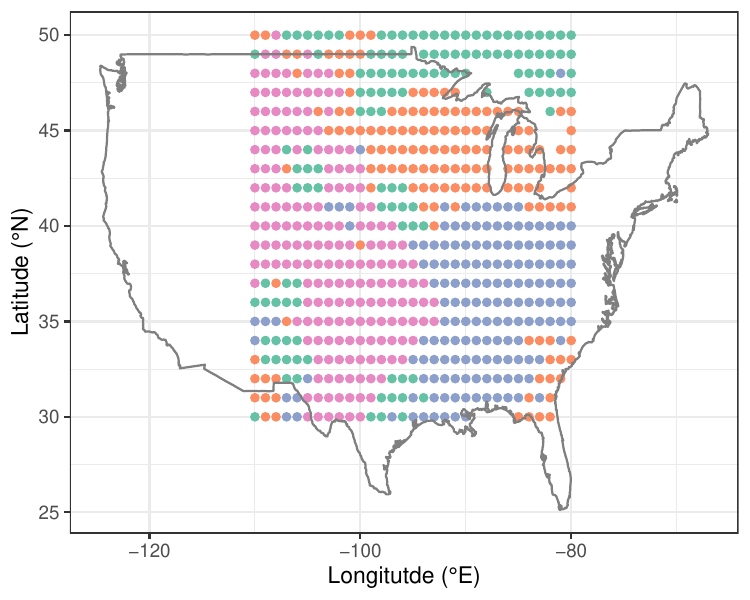}
  \end{subfigure}%
  \begin{subfigure}[b]{.33\linewidth}
    \centering
    \includegraphics[width=.99\textwidth]{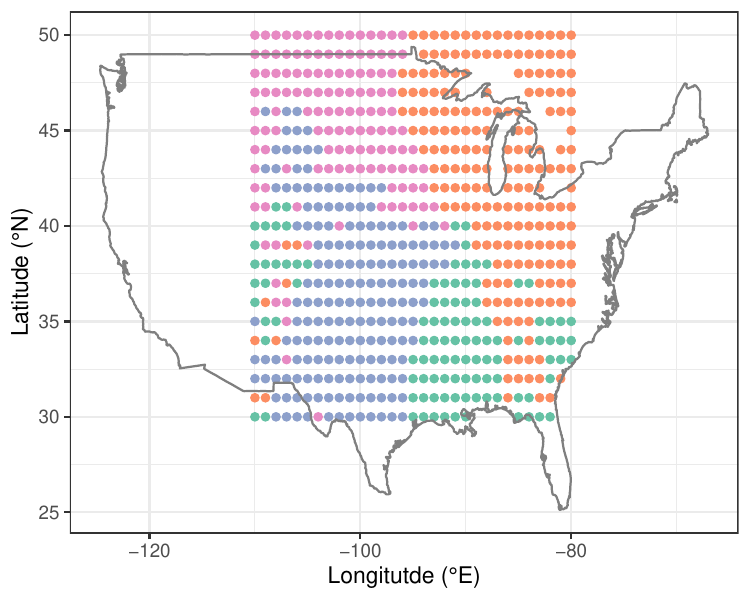}
  \end{subfigure} 
  \begin{subfigure}[b]{.33\linewidth}
    \centering
    \includegraphics[width=.99\textwidth]{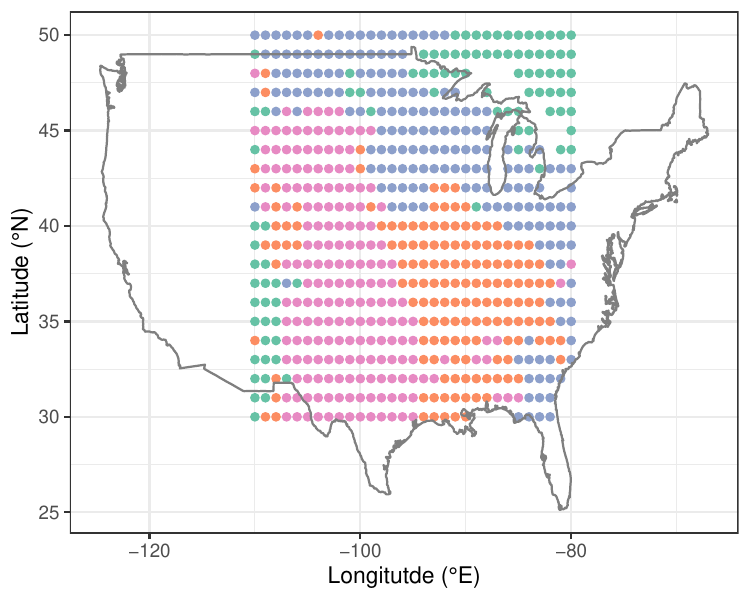}
  \end{subfigure} 
   \caption{\REV{Resulting clusters (indicated with different colours) from applying the algorithm of \citet{Bernard.PAM.2013} with four clusters on our July maxima for PROD (left), CAPE (centre) and SRH (right).} }
   \label{fig:explore:pam}
\end{figure}

Figure \ref{fig:explore:pam} shows the resulting clusters from applying the PAM to chosen monthly maxima. 

\begin{figure}[t!]
\centering
  \begin{subfigure}[b]{.32\linewidth} 
    \centering
    \includegraphics[width=.99\textwidth]{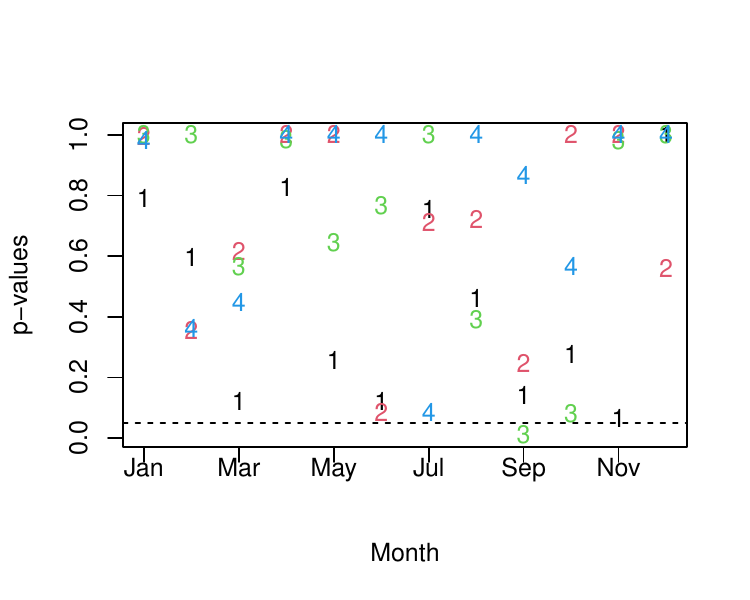}
  \end{subfigure}
  \begin{subfigure}[b]{.32\linewidth} 
    \centering
    \includegraphics[width=.99\textwidth]{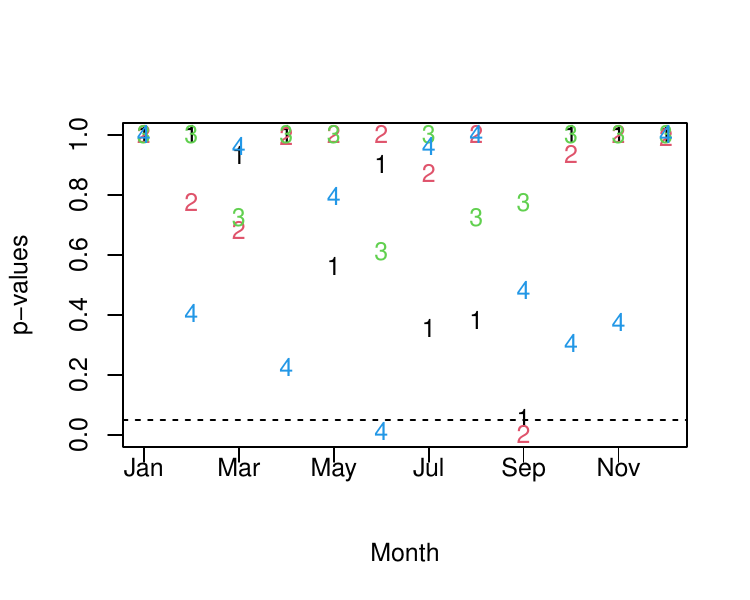}
  \end{subfigure}
  \begin{subfigure}[b]{.32\linewidth} 
    \centering
    \includegraphics[width=.99\textwidth]{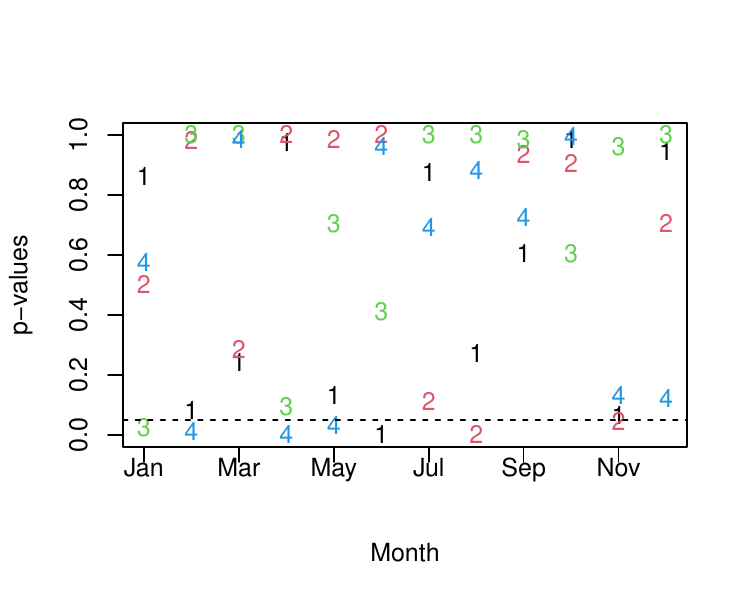}
  \end{subfigure}
  \caption{\REV{P-values from applying our max-stability test on the region (in numbers from 1-4) month (x-axes) combinations for PROD (left), CAPE (centre) and SRH (right). The dashed horizontal lines indicate the 0.05 level.} }
  \label{fig:maxstabtest_data}
\end{figure}

Figure \ref{fig:maxstabtest_data} shows the p-values from applying our max-stability test to the data.

Here we describe the two types of diagnostic plots implemented in our case study. For two random variables $X_1$ and $X_2$ with bivariate distribution function $F$ and marginal distribution functions $F_1$ and $F_2$, two measures of dependence are 
\begin{equation*}
\chi(u) = 2 - \frac{\log \Pr(F_1(X_1) < u, F_2(X_2)<u)}{\log \Pr(F_2(X_2)<u)}, \quad 0 \leq u \leq 1.
\end{equation*}
and 
\begin{equation*}
\chi^{\prime}(u)= \mathrm{Pr}\{F_1(X_1) > u \mid F_2(X_2) > u \}.
\end{equation*}
It is straightforward to see that $\lim_{u \to 1} \chi(u) = \lim_{u \to 1}\chi^{\prime}(u).$ If $F$ is a bivariate max-stable distribution, then $\chi^{\prime}(u)$ stabilises as $u\rightarrow 1$, and $\chi(u)$ is stable for all $u\in[0,1]$  \citep{coles1999dependence}. In practice, due to the lack of data when $u$ is close to 0 and 1, the two measures are only evaluated in the regions between the two values. 

\begin{figure}[t!]
\centering
\begin{subfigure}{.95\linewidth}
   \includegraphics[width=.31\textwidth]{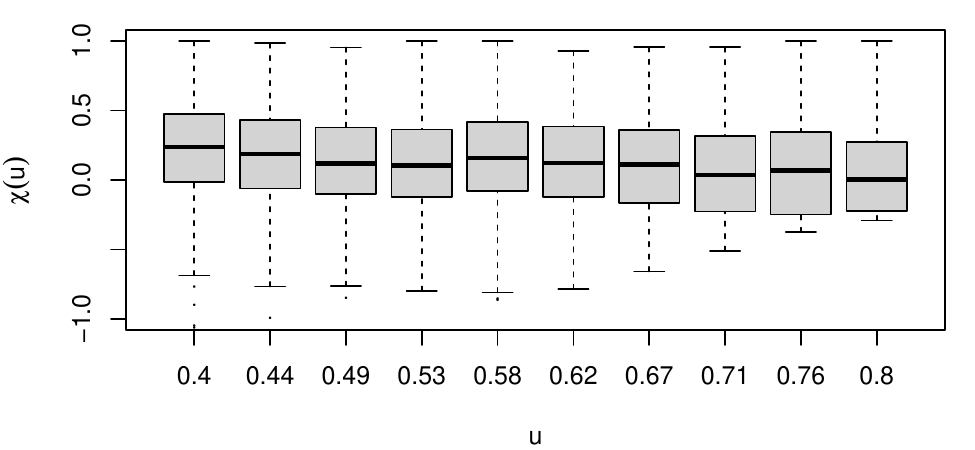}   
      \includegraphics[width=.31\textwidth]{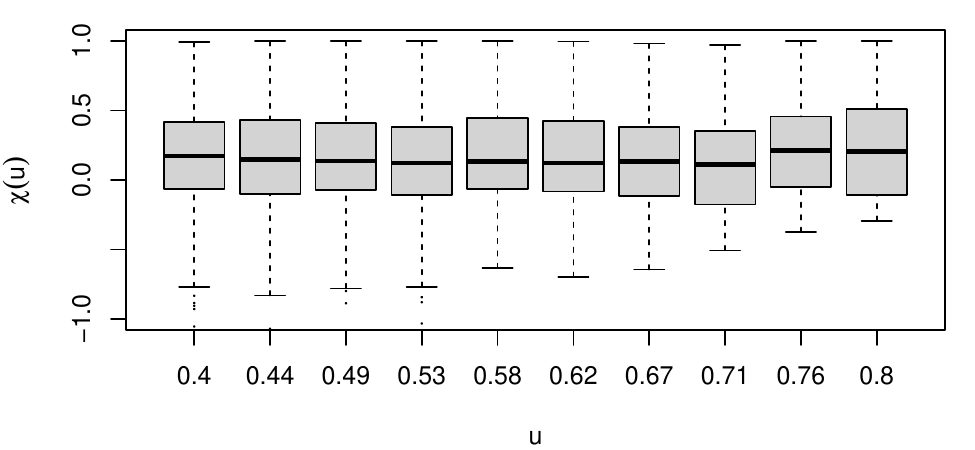}  
         \includegraphics[width=.31\textwidth]{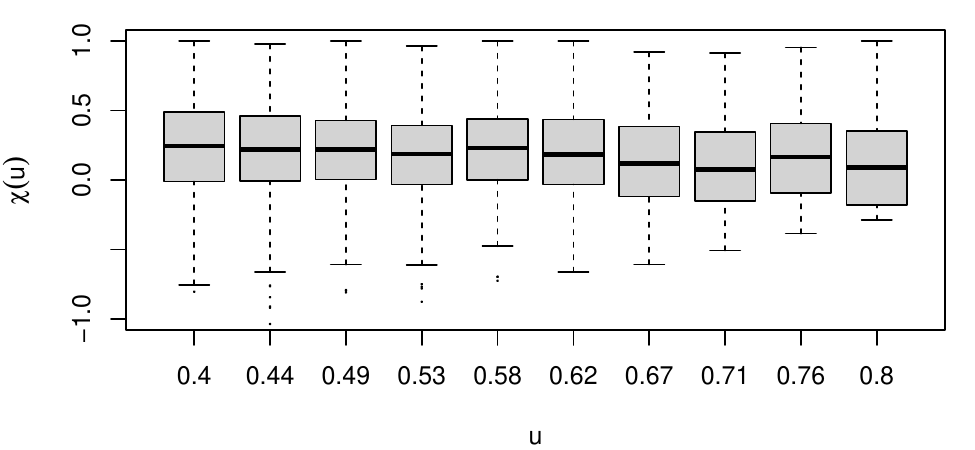}  
    \end{subfigure} \\
    \begin{subfigure}{.95\linewidth}
   \includegraphics[width=.31\textwidth]{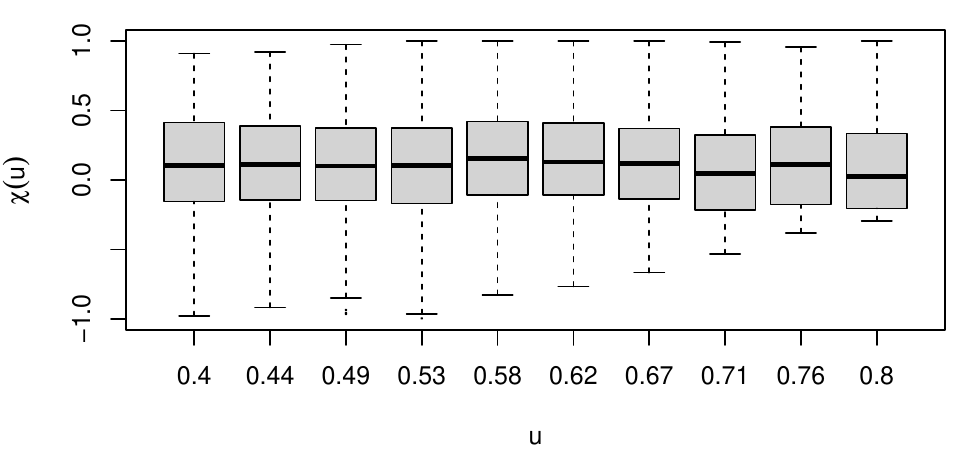}   
      \includegraphics[width=.31\textwidth]{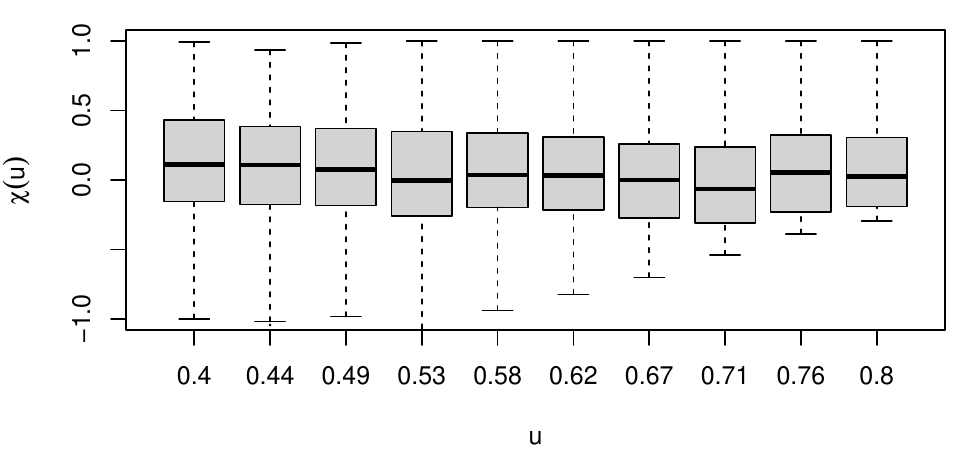}  
         \includegraphics[width=.31\textwidth]{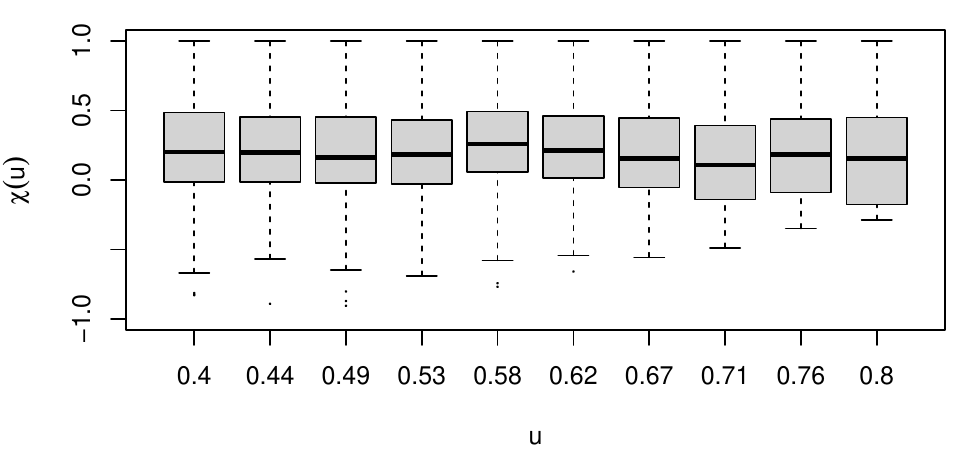}  
    \end{subfigure} 
   \caption{\REVV{Boxplots of empirical estimates of $\chi(u)$ for all neighbouring pairs (i.e., pairs with $c=1$ in Section~\ref{sec:composite}) with $u\in[0.4,0.8]$, 
   in region-month combinations 2-August (top) and 3-February (bottom), for monthly maxima of (left to right) PROD, CAPE and SRH.}}
   \label{fig:chiplot}
\end{figure}

\REVV{Figure \ref{fig:chiplot} finds no evidence that $\chi(u)$ is unstable in the two chosen regions for all variables. There is also little evidence that $\chi^ \prime(u)$ is decreasing for increasingly high thresholds, though these plots display high uncertainty}. 

\begin{figure}[t!]
\centering
\begin{subfigure}{.95\linewidth}
   \includegraphics[width=.31\textwidth]{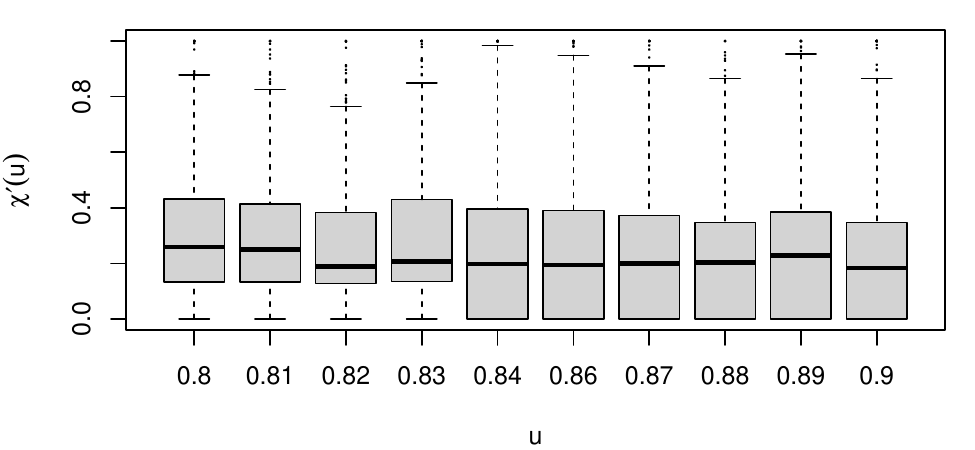}     
      \includegraphics[width=.31\textwidth]{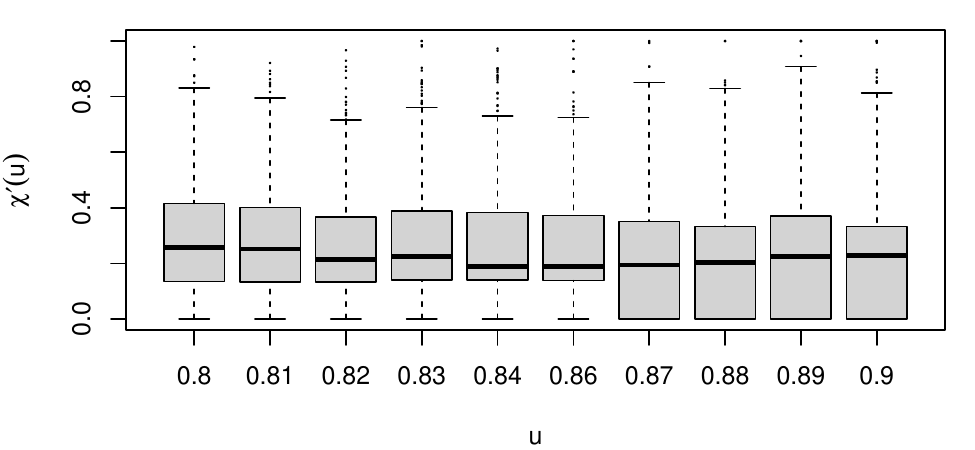} 
         \includegraphics[width=.31\textwidth]{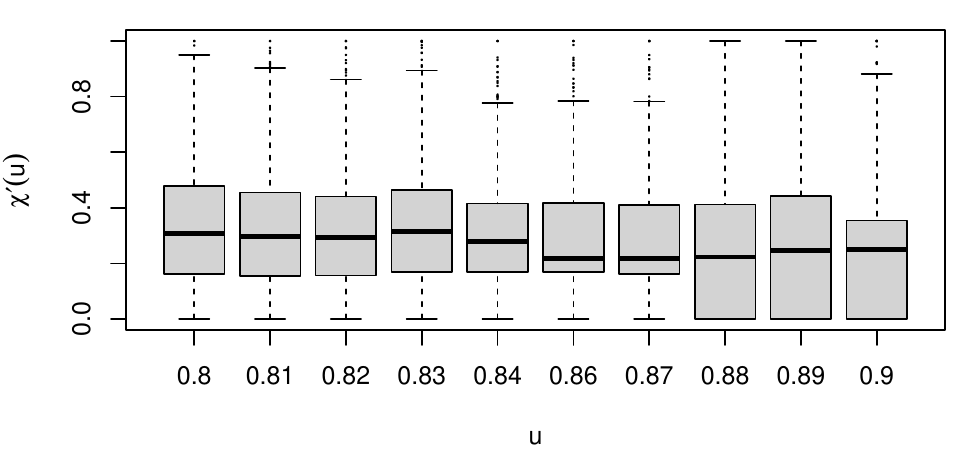} 
    \end{subfigure} \\
    \begin{subfigure}{.95\linewidth}
   \includegraphics[width=.31\textwidth]{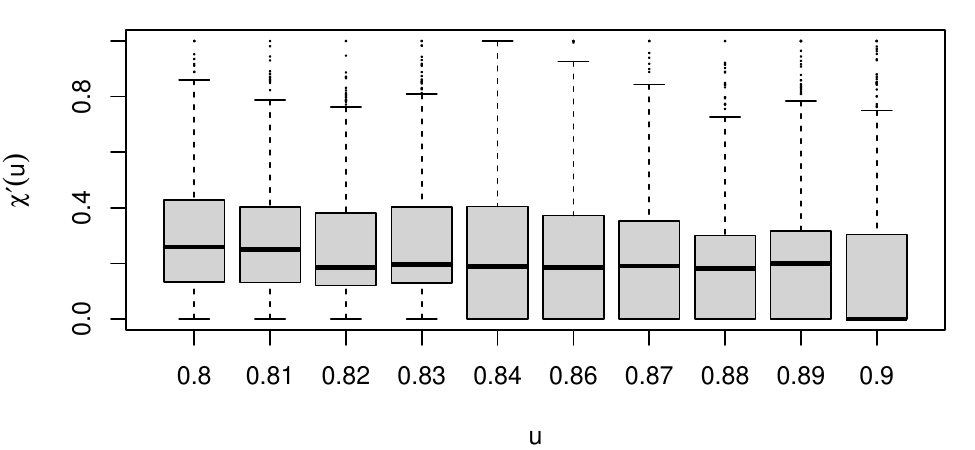} 
      \includegraphics[width=.31\textwidth]{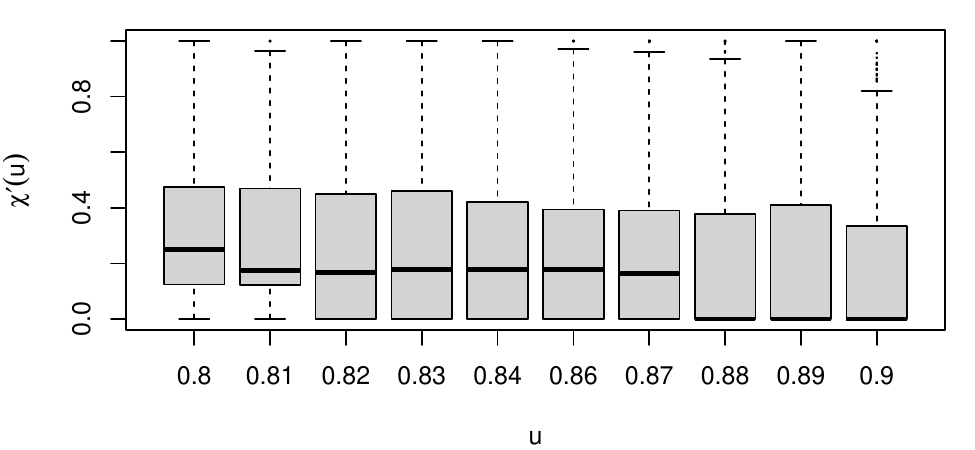} 
         \includegraphics[width=.31\textwidth]{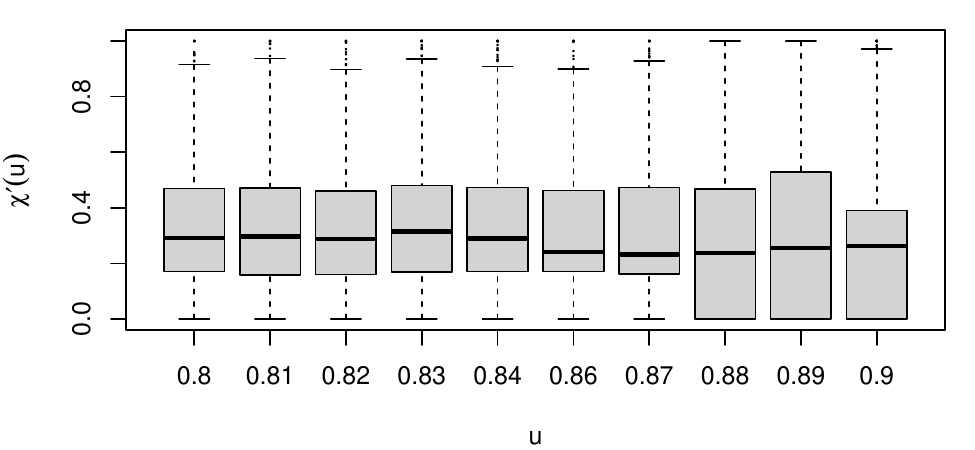} 
    \end{subfigure}
   \caption{\REVV{Same as Figure \ref{fig:chiplot} but for $\chi^\prime(u)$ with $u\in[0.8,0.9]$.}  }
   \label{fig:condprob}
\end{figure}

\subsection{For Section~\ref{subsec_results}}\label{sec:supplement:trend_surfaces}

\begin{figure}[t!]
\centering
    \begin{subfigure}[b]{.47\linewidth}
    \centering
    \includegraphics[width=.99\textwidth]{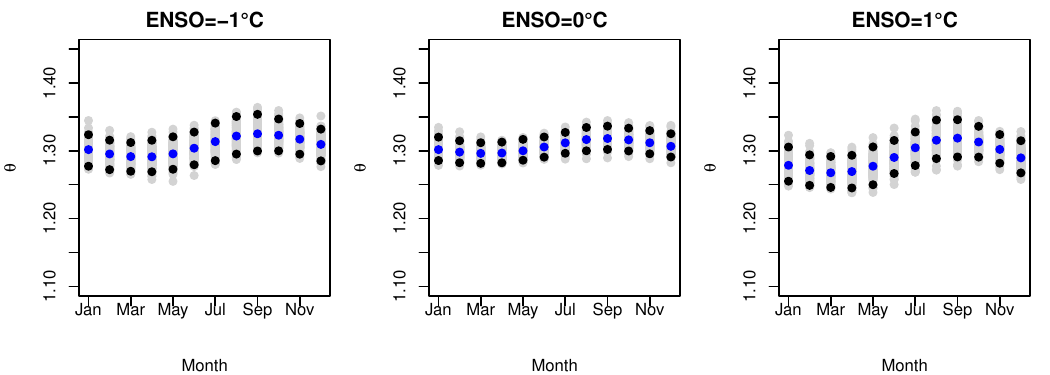}
  \end{subfigure}
  \hspace{0.4cm}
      \begin{subfigure}[b]{.47\linewidth}
    \centering
    \includegraphics[width=.99\textwidth]{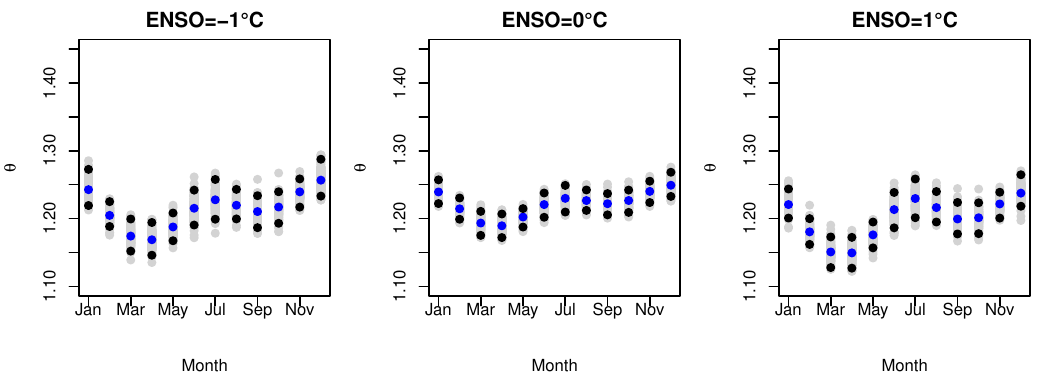}
  \end{subfigure}
  \begin{subfigure}[b]{.47\linewidth}
    \centering
    \includegraphics[width=.99\textwidth]{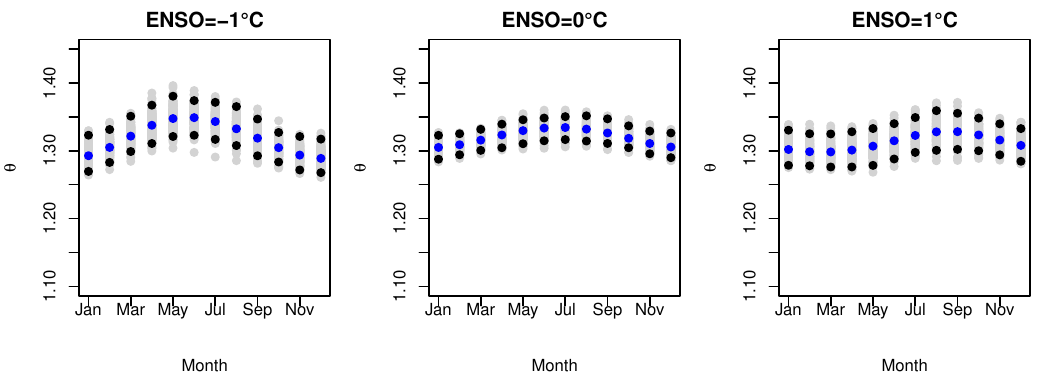}
  \end{subfigure}
    \hspace{0.4cm}
    \begin{subfigure}[b]{.47\linewidth}
    \centering
    \includegraphics[width=.99\textwidth]{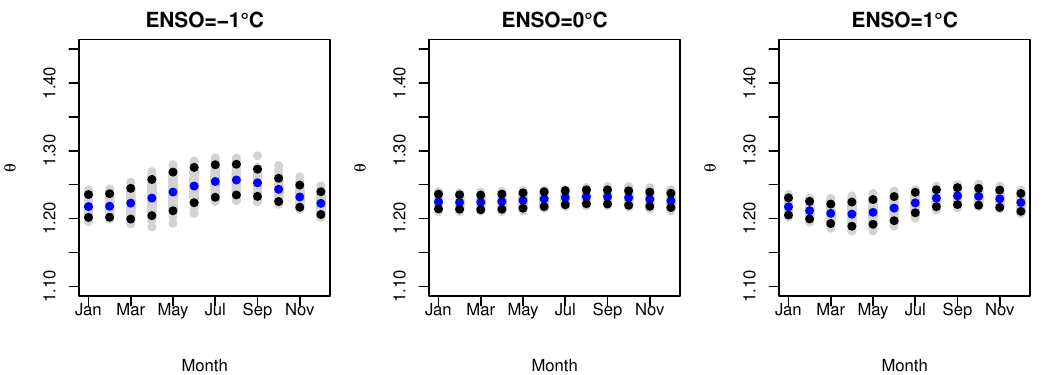}
  \end{subfigure} \\
   \caption{\REV{Bias-corrected estimates (blue) of the modeled bivariate extremal coefficient for PROD in regions 1 (bottom left), 2 (top left), 3 (top right) and 4 (bottom right), when ENSO equals $-1^\circ$C (left), $0^\circ$C (center) and $1^\circ$C (right), for two grid points \REVV{$100$ km apart in the horizontal direction}. The grey dots are the $200$ values of $h^{-1}\{2h(\hat{\theta})-h(\hat{\theta}^\star_b)\}$, $b=1,\dots,200$, where $\hat{\theta}^\star_b$ is the $b$-th bootstrap estimate and $h(x)=\log\{(x-1)/(2-x)\}$, for $x\in[1,2]$. The black dots indicate the lower and upper $90\%$ bootstrap pointwise confidence limits.
   }}
   \label{fig:model:extcoeff:prod}
\end{figure}

\begin{figure}[t!]
\centering
    \begin{subfigure}[b]{.47\linewidth}
    \centering
    \includegraphics[width=.99\textwidth]{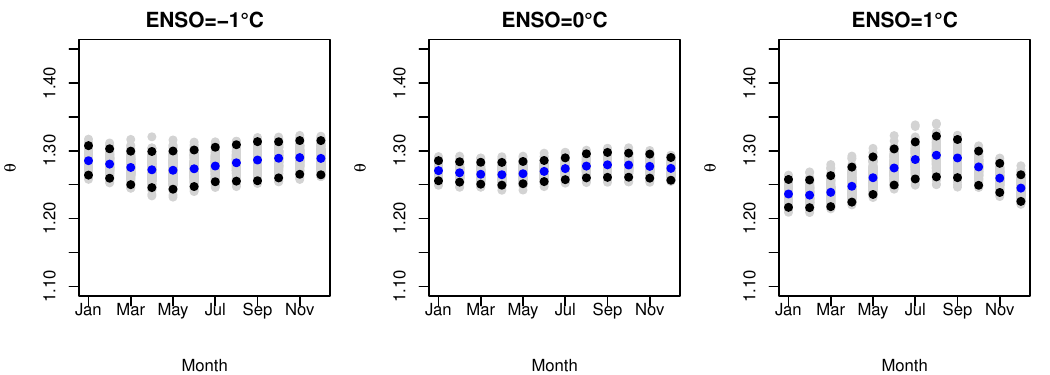}
  \end{subfigure}
  \hspace{0.4cm}
      \begin{subfigure}[b]{.47\linewidth}
    \centering
    \includegraphics[width=.99\textwidth]{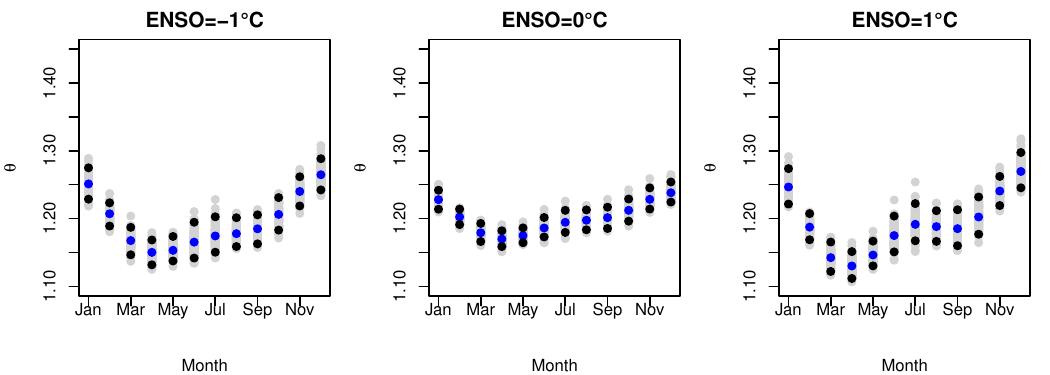}
  \end{subfigure}
  \begin{subfigure}[b]{.47\linewidth}
    \centering
    \includegraphics[width=.99\textwidth]{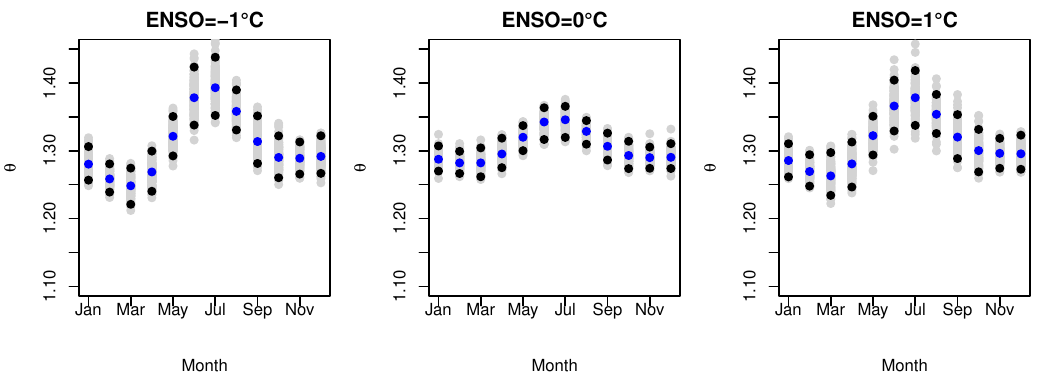}
  \end{subfigure}
    \hspace{0.4cm}
    \begin{subfigure}[b]{.47\linewidth}
    \centering
    \includegraphics[width=.99\textwidth]{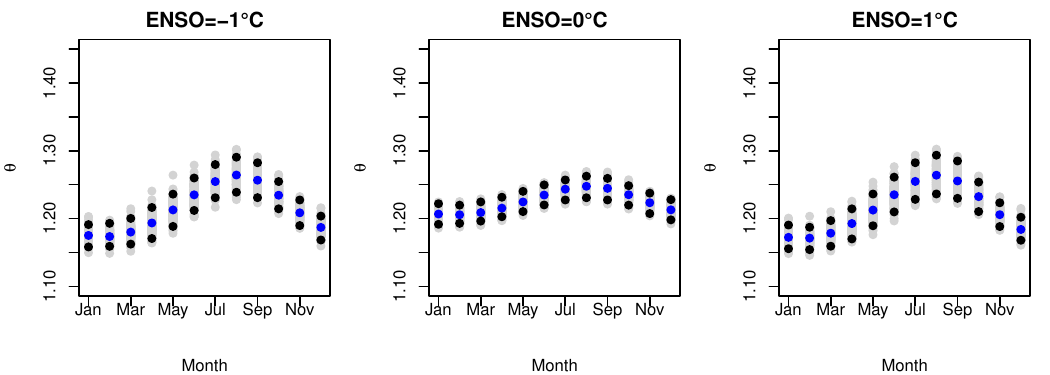}
  \end{subfigure} \\
 
   \caption{\REV{Same as Figure \ref{fig:model:extcoeff:prod} but for CAPE.
   }}
   \label{fig:model:extcoeff:cape}
\end{figure}

\begin{figure}[t!]
\centering
    \begin{subfigure}[b]{.47\linewidth}
    \centering
    \includegraphics[width=.99\textwidth]{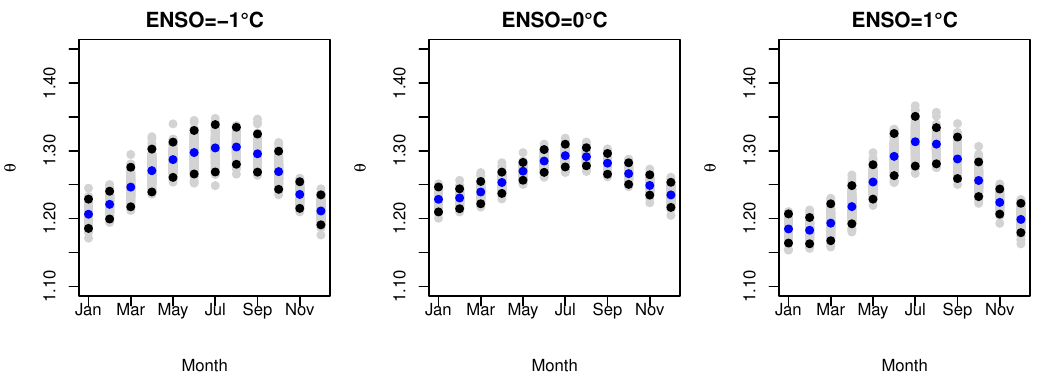}
  \end{subfigure}
  \hspace{0.4cm}
      \begin{subfigure}[b]{.47\linewidth}
    \centering
    \includegraphics[width=.99\textwidth]{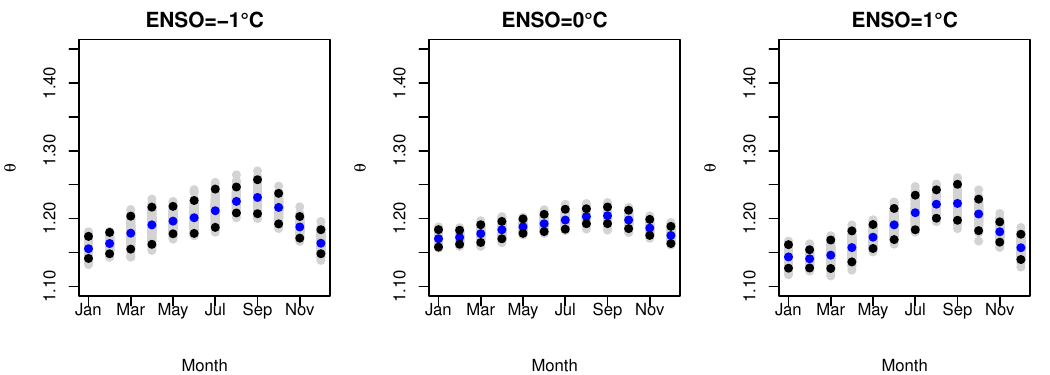}
  \end{subfigure}
  \begin{subfigure}[b]{.47\linewidth}
    \centering
    \includegraphics[width=.99\textwidth]{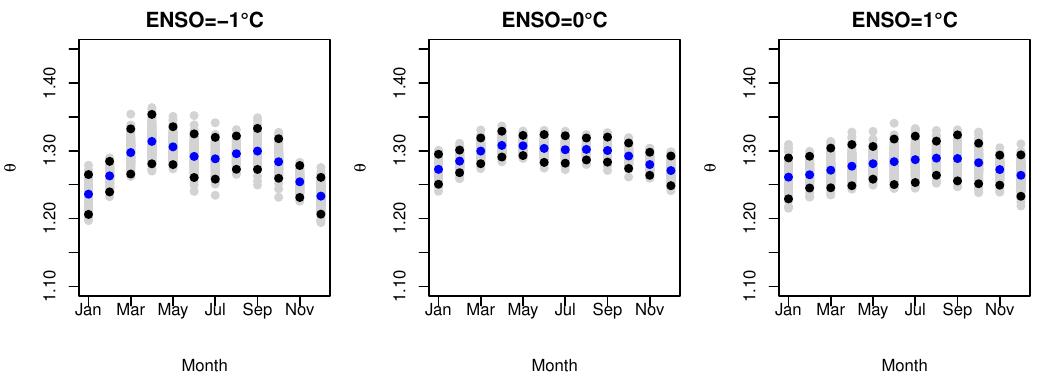}
  \end{subfigure}
    \hspace{0.4cm}
    \begin{subfigure}[b]{.47\linewidth}
    \centering
    \includegraphics[width=.99\textwidth]{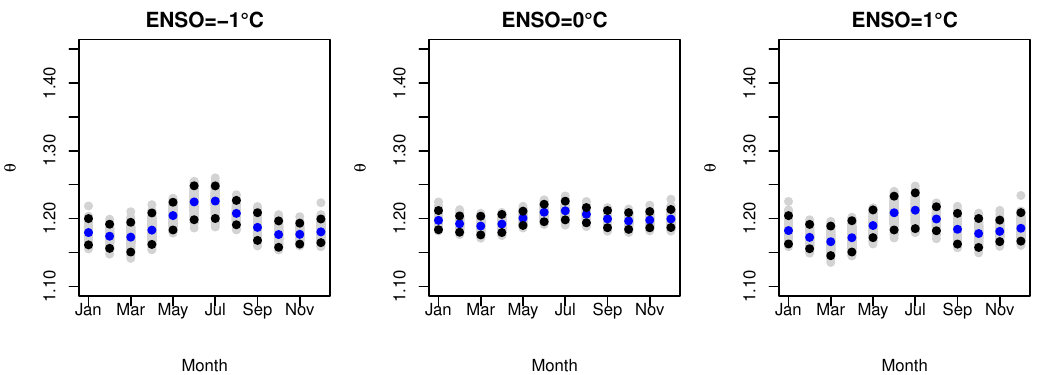}
  \end{subfigure} \\
 
   \caption{\REV{Same as Figure \ref{fig:model:extcoeff:prod} but for SRH.
   }}
   \label{fig:model:extcoeff:srh}
\end{figure}

\begin{figure}[t!]
\centering
\begin{subfigure}{.99\linewidth}
   \includegraphics[width=.32\textwidth]{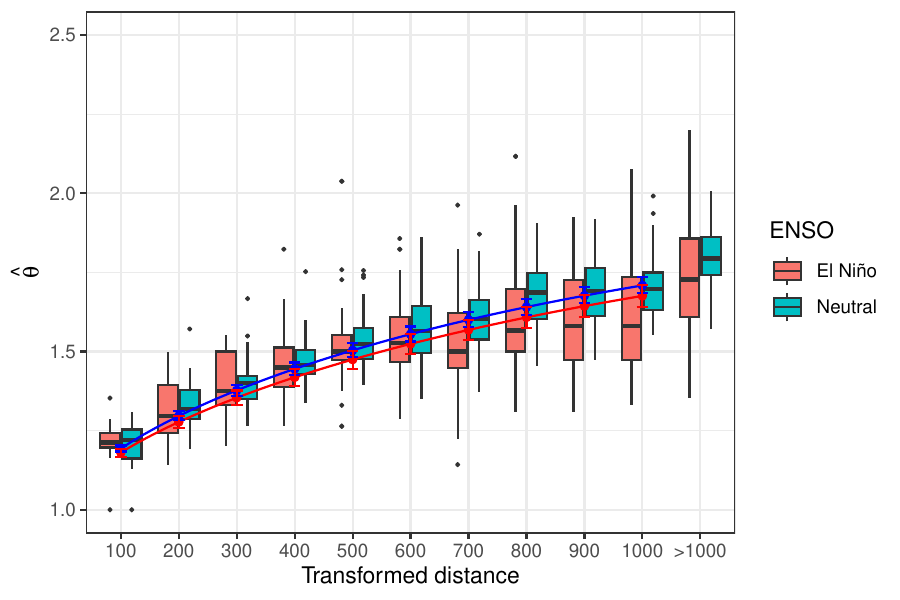} 
   \includegraphics[width=.32\textwidth]{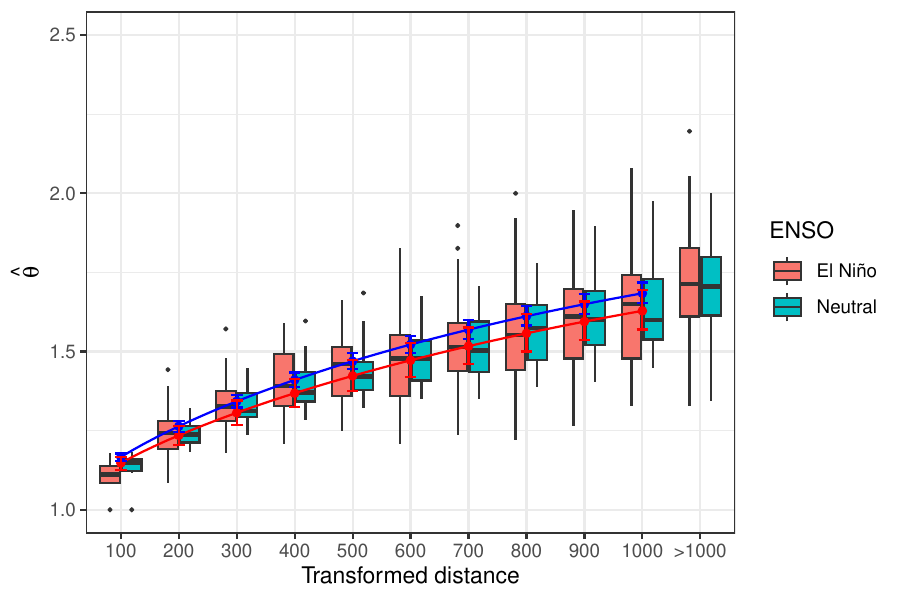} 
   \includegraphics[width=.32\textwidth]{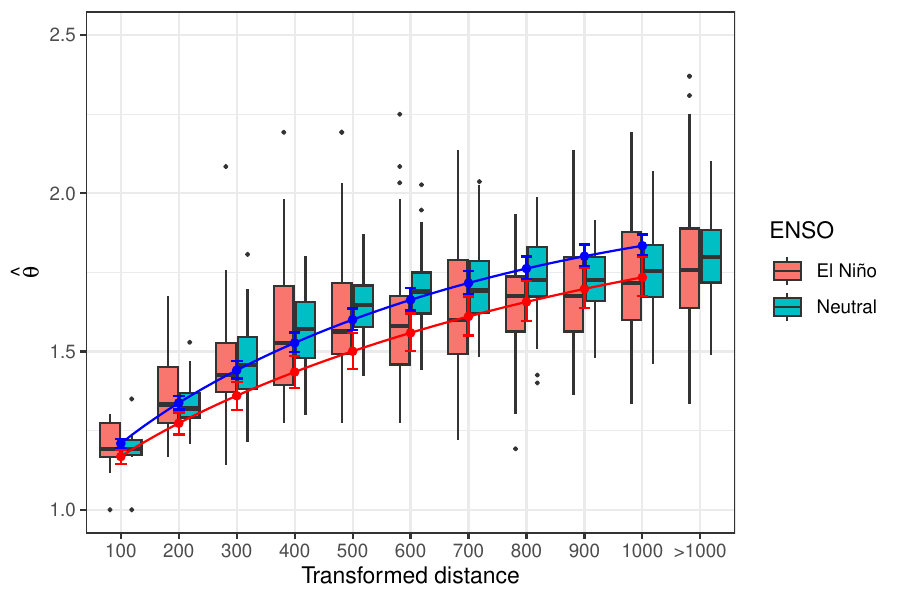}  
    \end{subfigure}
   \caption{
   \REVV{Out-sample performance of the models for two months pooled together, for months-region-variable combinations (June/July-$4$-PROD), (August/September-$3$-CAPE) and (March/April-$2$-SRH), respectively. The lines depict the modeled extremal coefficients computed at the average value between the months (e.g., 3.5 for March/April), with ENSO equal to its average value in those months where ENSO$\geq0.5^\circ$C (red), and $0^\circ$C (blue), as a function of distance in the transformed space, computed using (\ref{eq_aniso}) with the estimated $\kappa$ and $r$. The whiskers indicate pointwise basic bootstrap $90\%$ confidence limits. The pairs of boxplots summarize the empirical estimates of the extremal coefficient for pairs of grid points whose distance in the transformed space lies in the ranges $[0, 100)$, $[100, 200)$, \dots, $>1000$. The red and blue boxes correspond to ENSO$\geq0.5^\circ$C and $|\text{ENSO}|<0.5^\circ$C, respectively. } }
   \label{fig:extcoeff_models}
\end{figure}

\REV{Figures \ref{fig:model:extcoeff:prod}, \ref{fig:model:extcoeff:cape} and \ref{fig:model:extcoeff:srh}} shows the modeled bivariate extremal coefficient estimates obtained from $200$ bootstrap replicates for grid points that are \REVV{$100$ km apart in the horizontal direction}. \REVV{Figure \ref{fig:extcoeff_models} shows the out-sample performance of the models.} 

\end{document}